\documentclass[fleqn,usenatbib]{aa}
\usepackage{graphicx, color}
\usepackage{natbib}
\usepackage{hyperref}
\usepackage{threeparttable}

\usepackage{txfonts}
\usepackage{array}
\usepackage{subfigure}
\usepackage{graphicx}	
\usepackage{amsmath}	
\usepackage{multicol}        
\usepackage{bm}		
\usepackage{pdflscape}	

\newcommand{\kms}{\,km\,s$^{-1}$} 

\usepackage{xcolor}
\usepackage{siunitx}

\newcommand{\cm}{\ensuremath{\,{\rm cm}}}

\renewcommand{\kms}{\ensuremath{\,{\rm km\,s^{-1}}}} 

\newcommand{\K}{\ensuremath{\, {\rm K}}}

\newcommand{\MHz}{\ensuremath{\, {\rm MHz}}}

\newcommand{\mJy}{\ensuremath{\,{\rm mJy}}}

\renewcommand{\arcmin}{\ensuremath{\,{\rm arcmin}}}

\usepackage[normalem]{ulem}

\def\hi{\textsc{Hi~}}

\makeatletter
\renewcommand*\aa@pageof{, page \thepage{} of \pageref*{LastPage}}
\makeatother

\DeclareRobustCommand{\VAN}[3]{#2}
\let\VANthebibliography\thebibliography
\def\thebibliography{\DeclareRobustCommand{\VAN}[3]{##3}\VANthebibliography}

\begin{document} 

   \title{Detections of 21-cm absorption with a blind FAST survey at z $\leqslant$ 0.09}

   \subtitle{}

   \author{Wenkai Hu
   \inst{1,2,3}\thanks{wenkai.hu@lam.fr}, Yougang Wang\inst{2}\thanks{wangyg@bao.ac.cn}, Yichao Li\inst{4}, Yidong Xu\inst{2}, Wenxiu Yang\inst{2,5}, Guilaine Lagache\inst{1}, Ue-Li Pen\inst{6,7,8,9,10}, Zheng Zheng\inst{2,11}, Shuanghao Shu\inst{2,5}, Yinghui Zheng\inst{2,5}, Di Li\inst{2,5,11}, Tao-Chung Ching\inst{2,11}, Xuelei Chen\inst{2,4,5}\thanks{xuelei@cosmology.bao.ac.cn}
           }

   \institute{Aix Marseille Univ, CNRS, CNES, LAM, Marseille 13388, France
        \and
            National Astronomical Observatories, Chinese Academy of Sciences, Beijing 100101, China
        \and
            ARC Centre of Excellence for All Sky Astrophysics in 3 Dimensions (ASTRO 3D), Australia
        \and
            Department of Physics, College of Sciences, Northeastern University, Shenyang 110819, China
        \and
            School of Astronomy and Space Science, University of Chinese Academy of Sciences, Beijing 100049, China
        \and
            Institute of Astronomy and Astrophysics, Academia Sinica, Astronomy-Mathematics Building, No. 1, Sec. 4, Roosevelt Road, Taipei 10617, Taiwan
        \and
            Canadian Institute for Theoretical Astrophysics, University of Toronto, 60 Saint George Street, Toronto, ON M5S 3H8, Canada
        \and
            Canadian Institute for Advanced Research, 180 Dundas St West, Toronto, ON M5G 1Z8, Canada
        \and 
            Dunlap Institute for Astronomy and Astrophysics, University of Toronto, 50 St George Street, Toronto, ON M5S 3H4, Canada
        \and
            Perimeter Institute of Theoretical Physics, 31 Caroline Street North, Waterloo, ON N2L 2Y5, Canada
        \and 
            Research Center for Intelligent Computing Platforms, Zhejiang Laboratory, Hangzhou 311100, China
            }

   \date{Received November 24, 2022; accepted January 11, 2023}

  \abstract{
  We present the early science results from a blind search of the extragalactic \hi 21-cm absorption lines at $z \leqslant 0.09$ with the drift-scan observation of the Five-hundred-meter Aperture Spherical radio Telescope (FAST). We carried out the search using the data collected in 643.8\, hours by the ongoing Commensal Radio Astronomy FasT Survey (CRAFTS), which spans a sky area of 3155\,deg$^{2}$ ($\sim$81 per cent of CRAFTS sky coverage up to January 2022) and covers 44827 radio sources with a flux density greater than 12 mJy. Due to the radio frequency interference (RFI), only the relatively clean data in the frequency range of 1.3--1.45 GHz are used in the present work. Under the assumption of $T_{s}/c_{f}$ = 100 K, the total completeness-corrected comoving absorption path length spanned by our data and sensitive to Damped Lyman $\alpha$ Absorbers (DLAs; $N_{\hi} \geqslant 2\times10^{20} \cm^{-2}$) are $\Delta X^{\mathrm{inv}}$ = 8.33$\times10^3$ ($\Delta z^{\mathrm{inv}} = 7.81\times10^{3}$) for intervening absorption. For associated absorption, the corresponding values are $\Delta X^{\mathrm{asc}}$ = 1.28$\times10^1$ ($\Delta z^{\mathrm{asc}} = 1.19\times10^{1}$). At each time point of the drift scan, a matched-filtering approach is used to search \hi absorbers. 
  Three known \hi absorbers (UGC\,00613, 3C\,293 and 4C\,+27.14) and two new \hi absorbers (towards the direction of NVSS\,J231240-052547 and NVSS\,J053118+315412) are detected blindly. We fit the \hi profiles with multi-components Gaussian functions and calculate the redshift (0.063, 0.066), width, flux density, optical depth and \hi column densities for each absorption. Our results demonstrate the power of FAST in blindly searching \hi absorbers. For absorption towards NVSS\,J231240-052547, the optical counterparts are faint and currently lack existing spectra. The most likely interpretation is that a radio-loud active galactic nucleus (AGN) is faint in the optical as the background source, with a faint optical absorber in between. 
  NVSS\,J053118+315412 exhibits an associated absorption with a complex profile, which may suggest unsettled gas structures or gas accretion onto the supermassive black hole (SMBH). The expanding collection of blind radio detections in the ongoing CRAFTS survey offers a valuable opportunity to study AGN and associated interstellar medium (ISM) interaction, and intervening absorbers optically without overwhelming quasi-stellar object (QSO) background light.
  }

   \titlerunning{Detections of 21-cm absorption with a FAST blind survey}
   \authorrunning{Wenkai Hu et al}

   \maketitle
%

\section{Introduction}

Hydrogen is the most abundant and widely distributed element in the
Universe. The 21-cm hyperfine transition of neutral hydrogen (\hi) has
been used as an important tool to study a wide range of astrophysical
processes, including galaxy dynamics, galaxy mergers and galaxy
interaction, star formation history, as well as tracing the cosmic
large-scale structure \citep{2008PhRvL.100i1303C}. 


Substantial effort has been dedicated to measuring the \hi\ emission line, e.g. the \hi Parkes All Sky Survey (HIPASS) \citep{2004MNRAS.350.1210Z,2004MNRAS.350.1195M}, and the Arecibo Legacy Fast Arecibo L-Band Feed Array (ALFALFA) survey \citep{2005AJ....130.2598G,2007AJ....133.2569G,2007AJ....133.2087S}, and the Jansky Very Large Array (JVLA) deep survey \citep{2014arXiv1401.4018J}. However, due to the limited sensitivity of the telescopes, wide-field \hi emission survey has been limited to the local Universe, while individual 21-cm deep field surveys have only detected \hi galaxies up to $z \approx 0.3$ \citep{2008ApJ...685L..13C,2001Sci...293.1800Z,1538-4357-668-1-L9,2016ApJ...824L...1F}. Combined with the observations of damped DLAs at high-z, the \hi mass density ($\Omega_{\hi}$) of the Universe is well-constrained at z < 0.2 and 2 < z < 5, with the \hi content poorly constrained at $0.2 < z < 2$ (c.f. \citealt{2019MNRAS.489.1619H} and references therein).

The \hi 21-cm absorption lines, arising from \hi gas absorbing the flux of the bright background source, could be a complement to the observation of the \hi emission at higher redshifts. Unlike \hi flux in emission which is flux-limited, the detectability of the \hi absorption depends only on the column density of foreground gas and the strength of the background sources. \hi absorption may be a good alternative tool to measure the \hi content and constrain the redshift-evolution of $\Omega_{\hi}$ at intermediate redshifts $0.2 < z < 2$ \citep{2018A&ARv..26....4M,2020MNRAS.498..883G}. 

Two types of \hi absorption systems exist, with one being the associated absorption system where the absorbing gas is situated in the same extragalactic object that emits the background bright continuum. Studies of associated absorption systems typically concentrate on the extragalactic object (usually an AGN) and its interaction with the ISM within the same object. This includes exploring the kinematics and structure of the ISM in galaxies that host AGNs and investigating the potential mechanisms by which \hi gas fuels supermassive black holes.

Significant progress has been made in associated absorption studies in the past two decades (see \citealt{2018A&ARv..26....4M} for a review). Some physical pictures have been established and revealed by observations, including the detection of circumnuclear absorbing structures containing \hi \citep{1996ApJ...470..394T,2010A&A...513A..10S}, the identification of \hi outflows potentially driven by the jet \citep{2003ApJ...593L..69M,2021A&A...647A..63S} and the indication of fuelling of SMBH through the accretion of small gas clouds \citep{2010AJ....139...17A,2014A&A...571A..67M}. Although many discoveries have been made, the statistics are still limited by small sample sizes, and there is a lack of conclusive observational evidence for \hi playing a role in fuelling the central SMBH. To further the understanding of \hi associated absorption, a larger and more diverse sample from different physical environments is necessary.

The other type of \hi absorption system is referred to as the intervening absorption system that results from gas in a foreground Galactic or extragalactic object absorbing the flux of a bright background source that is not related. The intervening absorptions are usually employed to investigate the ISM characteristics in both our Galaxy and distant galaxies. A major fraction of the neutral hydrogen of the post-reionization universe has been found in DLAs, which are found in the optical spectrum of QSO as strong \hi absorption lines with $N_{\hi} \geqslant 2\times10^{20} \cm^{-2}$ \citep{2005ARA&A..43..861W}. Despite numerous studies, the physical nature of the DLAs remains an unsolved problem (for recent work, see e.g. \citealt{Bordoloi2022}). The QSOs used for such search are usually bright, making it difficult to observe the optical emission from the DLA system itself, in analogy to the practical challenge of detecting planets in reflected light. Besides, the optical surveys have additional biases from dust extinction, which could be systematically correlated to the self-shielding required for the existence of DLAs, to the gravitational magnification bias.


Most radio-loud QSOs are optically faint, opening the possibility of
uncovering a sample of absorption systems without blinding the optical
background. To our knowledge, only two 21cm absorbers (towards UGC 6081 and 3C
286) have been discovered in untargeted blind radio searches. Absorber
in UGC 6081 is an associated system \citep{2011ApJ...742...60D} and 3C
286 is a bright optical QSO \citep{1973ApJ...184L...7B}. Only 18 21cm
absorbers have been discovered in bright radio sources targeted blind
radio searches
\citep{2004ApJ...613L.101D,2015MNRAS.453.1249A,2020MNRAS.494.3627A,2020MNRAS.498..883G,2020MNRAS.499.4293S,2021ApJS..255...28G,2021ApJ...907...11G,2022MNRAS.509.1690M}. Of these, 16 still had a bright optical QSO, one had a radio background with unknown type \citep{2020MNRAS.494.3627A} and one had powerful
radio galaxy (PKS\,0409-75) with 3000 \kms and should be considered an
associated absorber \citep{2022MNRAS.509.1690M}. 

Blind radio searches offer the opportunity to detect DLAs in front of optically faint QSOs, enabling a detailed study of DLA properties without interference from a bright background source. A rare opportunity was the observation of the afterglow of GRB030323, which showed both a DLA and Lyman $\alpha$ emission \citep{vreeswijk} in the gamma-ray burst (GRB) host galaxy. However, at present, the number of DLAs detected from GRB afterglow is still very few.

To achieve sufficient sensitivity, detecting the relatively weak 21-cm absorption technically requires a long integration time, and the observation is also affected by technical limitations such as unstable bandpass and RFI. Known that the compact sources have the higher rate of \hi absorption detection \citep{2010MNRAS.406..987E,2011MNRAS.418.1787C} and the majority of DLAs show strong Mg$_{\rm II}$ absorption, most \hi absorption searches have been pointed at selected targets with these features. However, these pre-selection methods could introduce biases in studying the intrinsic physics of \hi absorptions, which have not been well-understood.
 
An unbiased radio survey would address the DLA question directly, help overcome these limitations and also produce a more comprehensive depiction of the interaction between the ISM and host galaxy. \citet{2011ApJ...742...60D} carried out a pilot search of \hi 21-cm absorption spanning -650 \kms < cz < 17,500 \kms and covering 517.0 deg$^{2}$ with the ALFALFA Survey. The intrinsic absorption of UGC 6081 was re-detected, but no new absorption line systems were detected. \citet{2020MNRAS.494.3627A} reported a spectroscopically blind search for 21-cm absorption lines in the First Large Absorption Survey in \hi (FLASH \citep{2022PASA...39...10A}) using the Australian Square Kilometre Array Pathfinder (ASKAP). No absorptions were found in a purely blind search in the GAMA 23 field. Still, a 21-cm absorption at $z=0.3562$ towards NVSS\,J224500-343030 was detected after cross-matching the radio sources with optical spectroscopic identifications of galaxies. Besides, \citet{2022MNRAS.516.2947S} presented the results of a search for 21 cm \hi absorption from three Galaxy And Mass Assembly (GAMA) survey fields. Two associated \hi absorption systems in SDSS J090331+010847 and SDSS J113622+004852 are detected from a sample of 326 radio sources with 855.5 MHz peak flux density over 10 mJy. Recently, \citet{2021ApJ...907...11G} developed the Automated Radio Telescope Imaging Pipeline (ARTIP),  which is designed to process large volumes of radio interferometric data from the MeerKAT Absorption Line Survey (MALS). Applying this pipeline to the observations of a field centred on PKS 1830-211, they re-detected the known \hi 21-cm absorption at z = 0.19, and OH 1665 and 1667 MHz absorption at z = 0.89 using MeerKat L-band \citep{2021ApJ...907...11G} and UHF-bands \citep{2021A&A...648A.116C}. 

In the not-too-distant future, an unbiased census of intrinsic \hi absorption could be established by large blind \hi absorption line surveys planned with the next-generation telescopes, i.e. the Square Kilometre Array or its pathfinders, including the MeerKAT Absorption Line Survey (MALS; \citealt{Gupta:2018A7}), the Widefield ASKAP L-band Legacy All-sky Blind surveY (WALLABY; \citealt{2020Ap&SS.365..118K}), the First Large Absorption Survey in \hi (FLASH; \citealt{2020MNRAS.494.3627A}), and the Search for \hi Absorption with AperTIF (SHARP; \citealt{2017A&A...604A..43M}).

As a complement in the northern hemisphere, the FAST \citep{2011IJMPD..20..989N,2020RAA....20...64J} can play an important role in blind survey of \hi absorption. FAST is the largest filled-aperture single-dish radio antenna in the world, with high sensitivity and large sky coverage. Equipped with 19-beams L-band receiver, covering a wide band from 1050 MHz to 1450 MHz \citep{8073111,2020RAA....20...64J}, corresponding to redshift $0 <z< 0.35$ for \hi observations, FAST can effectively search for \hi absorption lines in the above redshift range. According to the forecast provided by \citet{2017RAA....17...49Y}, FAST should be able to detect $\sim$ 200 \hi absorption systems, including associated and intervening systems, in a one-month survey around the celestial equator. \citet{2021MNRAS.503.5385Z} presented a study of extragalactic \hi 21-cm absorption lines using FAST. They observed 5 \hi absorption systems which were previously identified in the 40$\%$ data release of the ALFALFA survey. The \hi absorption profiles given by FAST have much higher spectral resolution and higher $\mathrm{S/N}$ ratio than obtained in the previous searches, demonstrating the power of FAST in revealing detailed structures of \hi absorption lines. Also, \citet{2020MNRAS.499.3085Z} carried out an OH absorption survey towards 8 associated and 1 intervening \hi absorbers at redshifts z $\in$ [0.1919, 0.2241] using FAST and constrained the OH to \hi relative abundance ([OH]/[HI]) to be $\leqslant$ 5.45 $\times 10^{-8}$.

In this paper, we report the early science results from a purely blind search for \hi absorption lines at $z\leqslant$0.09 in $\sim 81$ per cent (643.8 hours and 3155 deg$^{2}$) of the sky covered by the CRAFTS survey up to January 2022. Three known \hi absorbers (UGC\,00613, 3C\,293 and 4C\,+27.14) and two new \hi absorbers (NVSS\,J231240-052547 and NVSS\,J053118+315412) are found using our search pipeline. The signal of 4C\,+27.14 has also been re-detected in follow-up observations with FAST.

This paper is organized as follows: Section~\ref{sec:data} describes the CRAFTS data and the follow-up observation used in this work. We then summarize the data processing procedure, including candidate searching by the matched-filtering approach, candidates selection methods and \hi absorption measuring method in Section~\ref{sec:data_analysis}. The 10 candidates selected from the CRAFTS data and the confirmed \hi absorbers are presented in Section~\ref{sec:Results}. We discuss the implications of our methods for the blind \hi absorption survey in the future in Section~\ref{sec:Discussion}. In Section~\ref{sec:summary}, a summary of this work is presented. Throughout this paper we assume H$_{0} = 70$ km s$^{-1}$ Mpc$^{-1}$, $\Omega_{\rm m} = 0.3$ and $\Omega_{\Lambda} = 0.7$, but the results are not sensitive to these parameters.

\section{Data}
\label{sec:data}

\subsection{CRAFTS}

\begin{figure*}
    \centering
    \includegraphics[width=0.9\textwidth]{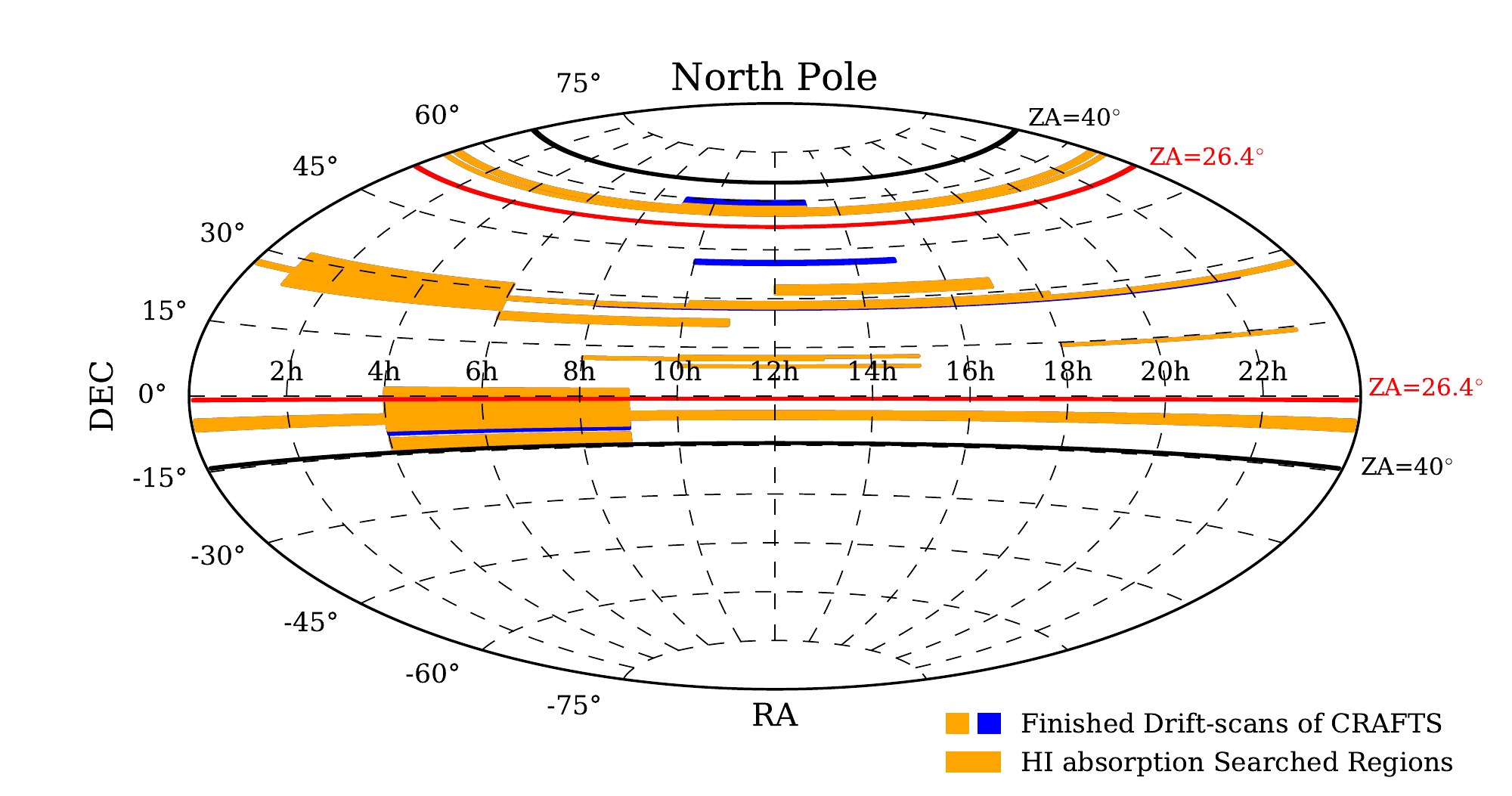}
    \caption{The CRAFTS sky coverage in Equatorial coordinates, as up to January 7th, 2022. We have carried out our search in the orange regions, and the orange and blue regions have been surveyed with CRAFTS. The zenith angle of 40° (maximum zenith angle for FAST) and 26.4° (zenith angle within which FAST has full gain) are shown as black circles and red circles, respectively.}
    \label{surveyed_sky}
\end{figure*}
The Commensal Radio Astronomy FasT Survey (CRAFTS; \citealt{2018IMMag..19..112L}) is a multi-purpose drift-scan survey using the FAST L-band Array of 19 feed-horns (FLAN, \citealt{8105012}), which aims to observe the galactic and extragalactic \hi emission and to search for pulsars and radio transients. According to the CRAFTS plan, the survey region will be observed by two passes of drift scans, with the 19-beam feed rotated by 23.4° to achieve a super-Nyquist sampling of the surveyed area. The full survey will cover over 20000 deg$^{2}$ within a declination (DEC) range between $-14^\circ$ and $66^\circ$, and reach a redshift of 0.35. Limited by the allocated time, there is only one survey pass at present. 

In Figure~\ref{surveyed_sky}, we show the CRAFTS sky coverage up to 2022-01-07. Data have been taken for both the orange and blue regions, while the orange regions have been used for \hi absorption search in the present work, which is $\sim$81\% of the total surveyed region.

The raw CRAFTS data set has a time resolution of 0.2 s, and a frequency resolution of 7.63 kHz ($\sim 1.61$ km s$^{-1}$ at $z=0$), covering a frequency range from 1000 MHz to 1500 MHz. The receiver response drops at the two ends of the frequency band, so only the data in the frequency range of 1.05–1.45 GHz is utilized. Additionally, in consideration of our scientific objectives and computing capacity, we further re-bin the data into a frequency resolution of 15.26 kHz and a time resolution of  12$/\cos\delta_{\rm{dec}}$ s ($\sim$ the transit time in drift scan), where $\delta_{\rm{dec}}$ is the declination of the pointing. The instantaneous sensitivity of each beam of the FAST system will be 2.15 mJy beam$^{-1}$, assuming a bandwidth of $\Delta\nu$ = 15.26 kHz, and 12 s integration time per beam \citep{2020MNRAS.493.5854H}. According to the calculation presented in Section~\ref{sec:completeness}, a velocity-integrated signal-to-noise ratio ($(\mathrm{S/N})_{\rm{int}}$) of $\sim$ 12 corresponds to the 90\% completeness limit. Considering the expected final RMS noise of 2.15 mJy beam$^{-1}$ and assuming source coverage of 100 per cent, a source with a flux density greater than 12 mJy is sufficient to detect \hi absorption with a peak signal-to-noise ratio ($(\mathrm{S/N})_{\rm{peak}}$) better than 5.5 (corresponding to a velocity-integrated signal-to-noise ratio ($(\mathrm{S/N})_{\rm{int}}$) of $\sim$ 12, for the Gaussian narrow line template with a full width at half maximum (FWHM) of 30 \kms and the frequency resolution of 15.26 kHz). According to the NRAO VLA Sky Survey (NVSS) \citep{1998AJ....115.1693C} Catalogue, whose completeness limit is about 2.5 mJy, 44827 radio sources with a flux density greater than 12 mJy are covered in our search region.

\begin{figure*}
    \centering
    \includegraphics[width=0.85\textwidth]{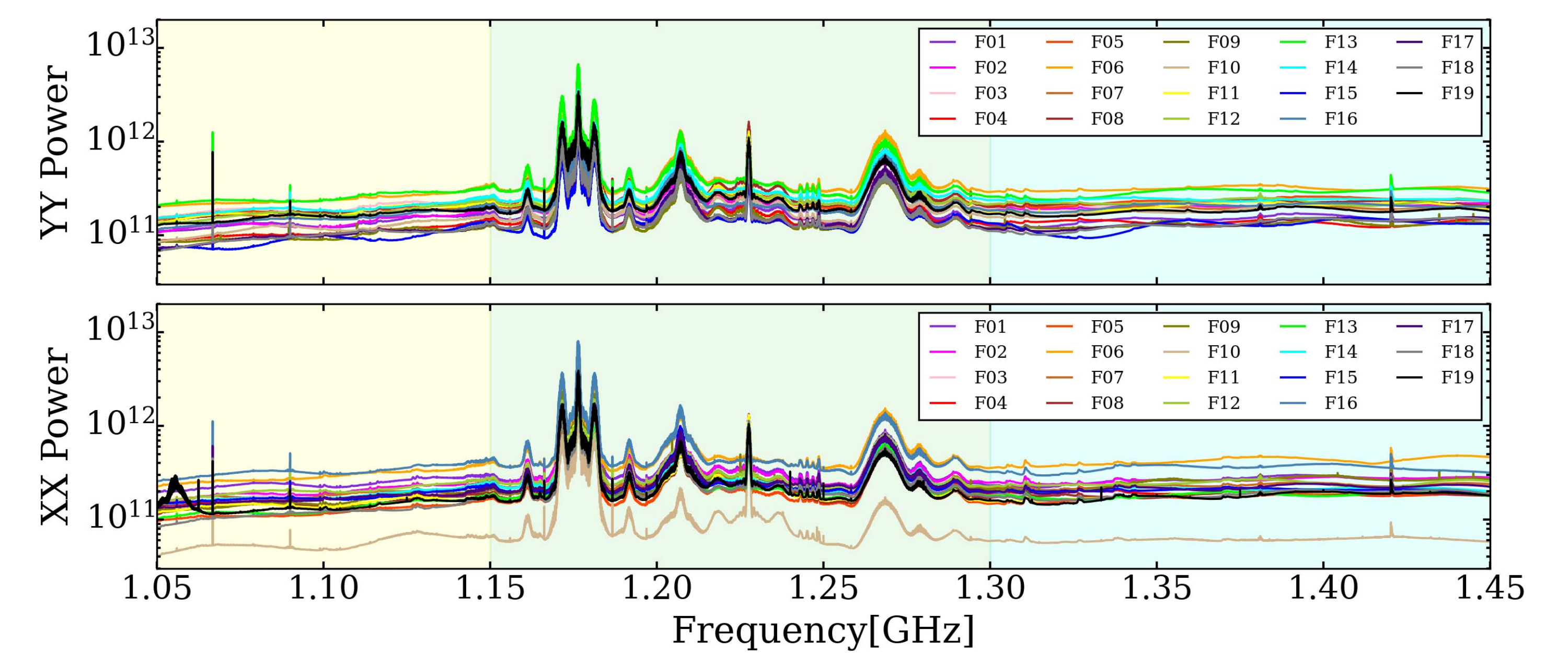}
    \caption{The raw frequency spectra for the 19 feeds of the FAST L-band multi-beam receiver feeds. These spectra are obtained by taking the time average of the raw data throughout the whole time of observation. The XX and YY polarizations are shown in the top and bottom panels, respectively. }
    \label{freq_spec}
\end{figure*}

Figure~\ref{freq_spec} shows a sample of the raw time-averaged frequency spectra of the CRAFTS data for the 19 beams, each with one curve. RFI dominates the frequency range from 1.15 GHz to 1.3 GHz (green shaded region in Figure~\ref{freq_spec}). In the range of 1.05 GHz to 1.15 GHz (yellow shaded region in Figure~\ref{freq_spec}), there are also many smaller peaks due to chronically present RFI \citep{2021MNRAS.508.2897H}. Considering the data quality, the data in the frequency range of 1.3 GHz to 1.45 GHz (blue shaded region in Figure~\ref{freq_spec}) which is relatively clean, is used in the present study. We are also searching the \hi absorption in the range of 1.05 GHz to 1.15 GHz, this will be presented in the future. The peaks at 1420 MHz are from the Galactic neutral hydrogen.

\begin{figure}
    \centering
    \includegraphics[width=0.42\textwidth]{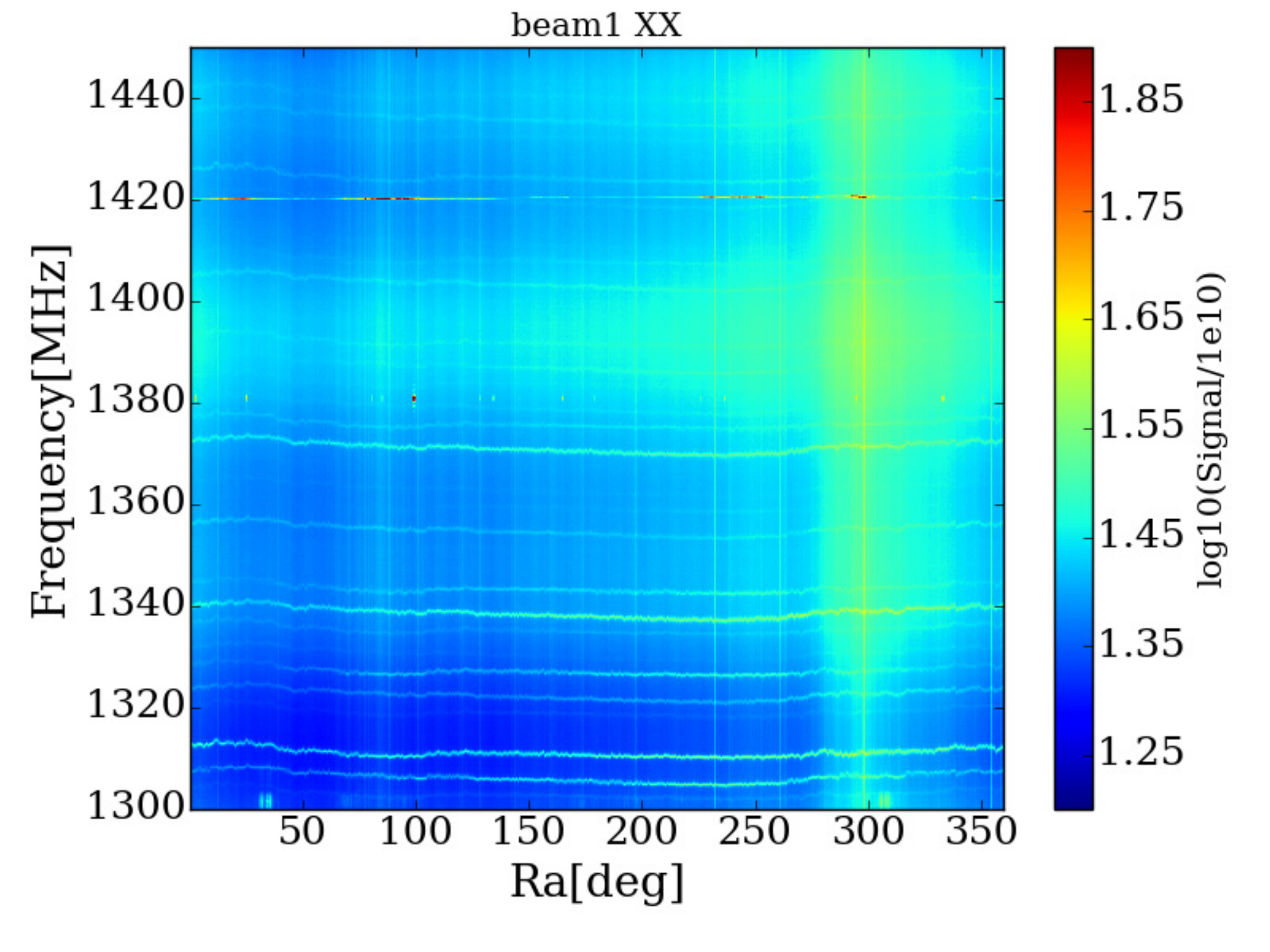}\\
    \includegraphics[width=0.42\textwidth]{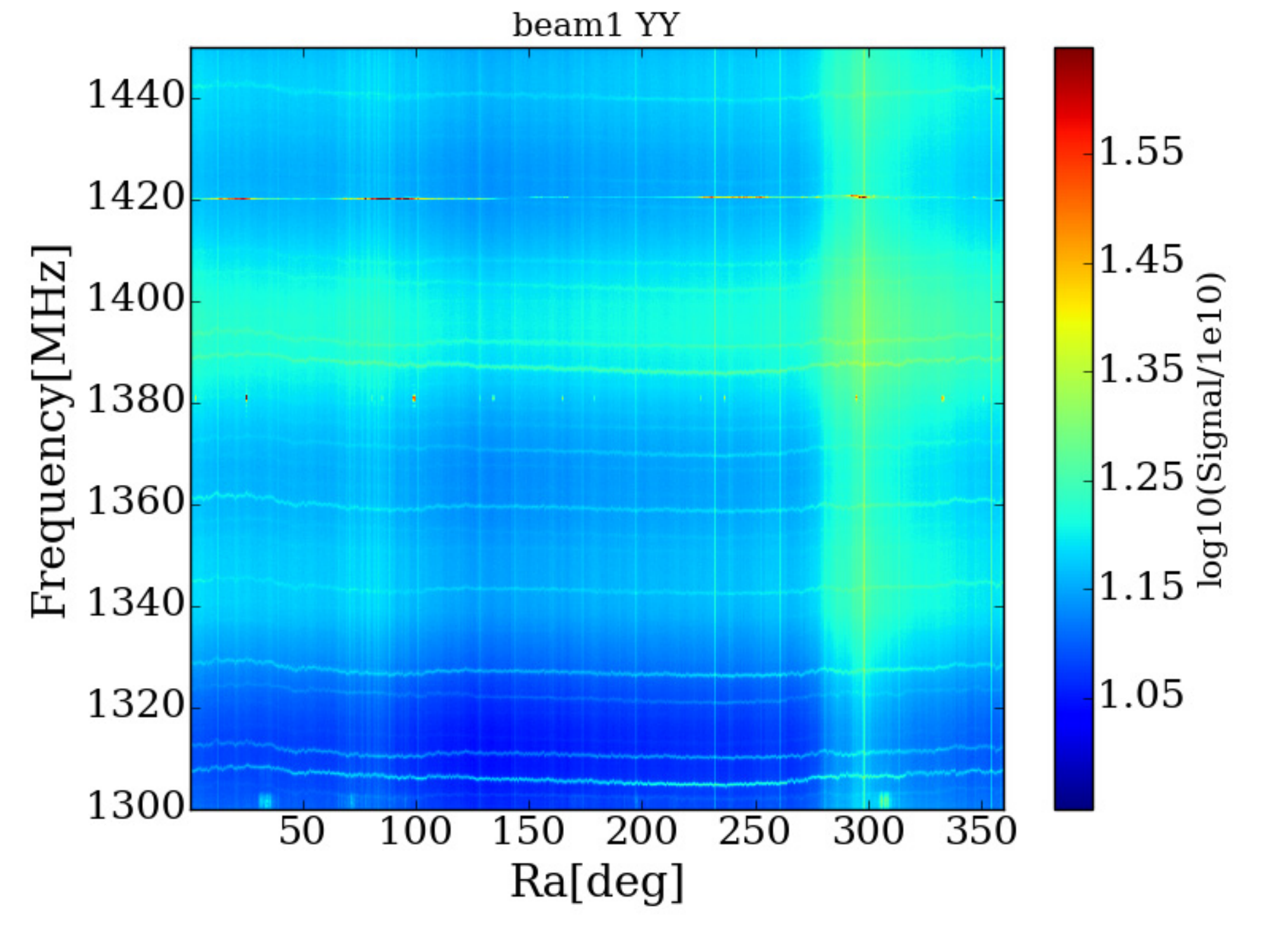}
    \caption{The waterfall plot of the raw data for feed 1 of the FAST L-band multi-beam receiver feeds. The two polarizations are shown in the top and bottom subpanels, respectively.}
    \label{waterfall}
\end{figure}

As an example, in Figure~\ref{waterfall}, we show the waterfall plot of the raw data of feed 1 from the observation on February 13th 2020, which scanned a stripe at the Declination of +27d15m0.0s. The horizontal linear features at some frequency points are chronically RFI.

\subsection{Follow-up Observation}
In the blind searching in CRAFTS data (described in the next section), 10 candidates, including 3 previously known \hi absorption systems (UGC\,00613, 3C\,293 and 4C\,+27.14) are detected. Two known absorbers (UGC 00613 and 3C 293) and all new absorption systems are further confirmed by the CRAFTS data of several neighbouring beams with a high signal-to-noise ratio (see Section~\ref{sec:selection} and Section~\ref{sec:Results}). In order to confirm the signal of the candidates and obtain a more significant \hi absorption feature for 4C\,+27.14, we also made a follow-up observation with FAST. It was carried out at night from 18:55:00 UTC to 23:17:00 UTC on 2021-08-21.

To suppress the fluctuations in the bandpass and remove the sky signal, the follow-up observation used the ON-OFF tracking mode. The integration time for both source-on and source-off observations is 990 s. The cycle of 330s source-on followed by 330 s source-off observation is repeated 3 times for a target. The follow-up data set has a time resolution of 1 s, and a frequency resolution of 7.63 kHz, covering the frequency from 1000 MHz to 1500 MHz. Under the condition that the frequency resolution is 7.63 kHz and the integration time is 990 s, the spectral RMS of 0.33 mJy can be obtained. The source 3C 48 was also observed for flux calibration. 

\section{Analysis}
\label{sec:data_analysis}

To begin, we extract spectra at each time point from the data cubes. We process the XX and YY spectra separately to search for \hi absorption and combine them to create the final spectra if selected as candidates. For each XX and YY polarization, we mask the RFI and obtain the band baseline by smoothing with a low-pass filter using the {\tt medfilt} function \citep{van1995python,2020SciPy-NMeth}. The median filter window size is set to 0.65 MHz, carefully selected to be smaller than the chronically present RFI and bandpass fluctuations yet larger than the typical size of \hi absorption profiles.

Note that we flag only strong RFI ($\geqslant 5 \sigma$) in our data. This is because some true \hi absorbers may reside in the location of chronically present RFI. Furthermore, the spurious candidates produced by RFI will be excluded later in our blind search pipeline.

We then search for the \hi absorbers blindly by cross-correlating the flux spectrum with templates. This matched-filtering approach was developed for identifying emission signals in the Arecibo Legacy Fast ALFA survey \citep{2007AJ....133.2087S} and is more sensitive and faster than the usual peak-finding algorithms. Here, we provide a brief summary of the key steps. Spectra at each time point were extracted from the data cubes. After the removal of the baseline of the bandpass, the matched-filtering approach was applied to find the absorption profiles. The final candidates are selected by use of the transit information recorded by the 19-beams of FAST.

\subsection{Searching Algorithms}
\label{sec:algorithms}

The matched filter is applied to the bandpass baseline-removed data. The signal $g(x)$ is assumed to be represented as:
\begin{eqnarray}
    g(x) \simeq \alpha t(x - \mu; \sigma),
    \label{model}
\end{eqnarray}
where $t$ is a template function over frequency channel $x$, $\alpha$, $\mu$ and $\sigma$ are the amplitude, the central position and the width, respectively. 
The Gaussian function with negative amplitude is used as the template, 
\begin{eqnarray}
t(x) &=& -\exp(-\frac{(x-\mu)^2}{2\sigma^{2}}).
\end{eqnarray}

The best-fit template is found by minimising the $\chi^{2}$:
\begin{eqnarray}
    \chi^{2} = \sum_{x=1}^{N_{\rm f}}[\alpha t(x - \mu; \sigma) - g(x)]^{2},
    \label{chi}
\end{eqnarray}
where $N_{\rm f}$ is the number of the spectral channel. Expanding the $\chi^{2}$ gives:
\begin{eqnarray}
    \chi^{2} = \alpha^{2}N_{\rm f}\sigma_{\rm t}^{2} + N_{\rm f}\sigma_{\rm g}^{2} - 2\alpha N_{\rm f}\sigma_{\rm g}\sigma_{\rm t}c(\mu),
    \label{chi_expanding}
\end{eqnarray}
where $\sigma_{\rm g}^{2}$ and $\sigma_{\rm t}^{2}$ are the variance of the signal and the template, $c(x)$ is the normalized cross-correlation function:
\begin{eqnarray}
    c(x) = g(x) * t(x - \mu; \sigma) = \frac{1}{N_{\rm f}\sigma_{\rm g}\sigma_{\rm t}}\sum_{n}[g(n)t(n-x)].
    \label{crosscorrelation}
\end{eqnarray}
Minimizing the $\chi^{2}$ with respect to $\alpha$ gives:
\begin{eqnarray}
    \alpha = \frac{\sigma_{\rm g}}{\sigma_{\rm t}}c(\mu).
    \label{alpha}
\end{eqnarray}
Replacing the $\alpha$ in $\chi^{2}$ using Eq.~(\ref{alpha}), we have: 
\begin{eqnarray}
    \chi^{2} = N_{\rm f}\sigma_{\rm g}^{2}[1 - c(\mu)^{2}],
    \label{chi_reshape}
\end{eqnarray}
$\chi^{2}$ is minimized by maximizing the value of the cross-correlation between the observational data and the template. 

The template is parameterized by the peak amplitude $\alpha_{\rm max}$, the central channel position $\mu_{\rm max}$ and the width $\sigma_{\rm max}$.
In realistic data processing, the best fit parameters ($\alpha$, $\mu$ and $\sigma$) which minimize the $\chi^{2}$ are obtained as follows:
\begin{enumerate}
\item Generating a set of templates covering a physical range (from 3.22 \kms to 325.22 \kms with an increment of 0.322\kms, the range is selected to ensure comprehensive coverage of all detected absorptions in terms of their width \citep{2015A&A...575A..44G}) of width $\sigma$;
\item Calculating the convolution function $c(x)$ for each template;
\item For each convolution function, obtain $\mu_{\rm max}$ by finding the spectral channel at which $c(x)$ is maximized;
\item For these values of $c(\mu_{\rm max})$ from templates with different $\sigma$, obtaining the $\sigma_{\rm max}$ by finding the largest value of $c(\mu_{\rm max})$;
\item Obtaining the peak amplitude $\alpha_{\rm max}$ using Eq.\,(\ref{alpha}).
\end{enumerate}

The velocity-integrated signal-to-noise ratio of the absorption profile can be calculated following \cite{2007AJ....133.2087S} :
\begin{eqnarray}
  \mathrm{S/N} =
    \begin{cases}
      \frac{F_{\rm{int}}/W}{\sigma_{\mathrm{rms}}} \times \left(\frac{W/2}{dv}\right)^{1/2}& \text{W < 400 \kms}\\
      \frac{F_{\rm{int}}/W}{\sigma_{\mathrm{rms}}} \times \left(\frac{400/2}{dv}\right)^{1/2}& \text{W $\geqslant$ 400 \kms,}\\
    \end{cases}
    \label{signal_to_noise}
\end{eqnarray}
where $F_{\rm{int}}$ is the velocity-integrated flux, $\sigma_{\mathrm{rms}}$ is the RMS noise, $W$ is the width of the signal in \kms and $dv$ is the velocity resolution. For the rest of the content, all signal-to-noise ratios are velocity-integrated unless stated otherwise.

The \hi absorption signal is searched in both the XX and YY polarizations of each beam. Only those candidates with a total $\mathrm{S/N}$ $ > 5.5$ and can be found at nearly the same frequency ($\Delta \nu < 0.05$MHz) with $\mathrm{S/N} > 3.5$ in both XX and YY polarizations are selected.

\subsection{Completeness}
\label{sec:completeness}
The absorption signal may be diluted either due to the presence of emission \citep{2022MNRAS.516.2050M} at low redshift or poor baseline subtraction as a result of fluctuations of the bandpass. In order to quantify how this affects our \hi absorption search, we estimate completeness by adding 1.67$\times10^{8}$ mock absorption lines to real spectra and calculating the fraction detected using our search method and selection threshold. The mock absorption signals are simulated using a Gaussian function template. The peak flux of mock absorption ($F_{\rm{peak}}$) is calculated using Eq.\,(\ref{signal_to_noise}) and $F_{\rm{int}} = \sigma_{\rm{abs}}\sqrt{2\pi}F_{\rm{peak}}$, where $\sigma_{\rm{abs}}$ refers to the standard deviation of the Gaussian template for absorption signals.

\begin{figure}
    \centering
    \includegraphics[width=0.45\textwidth]{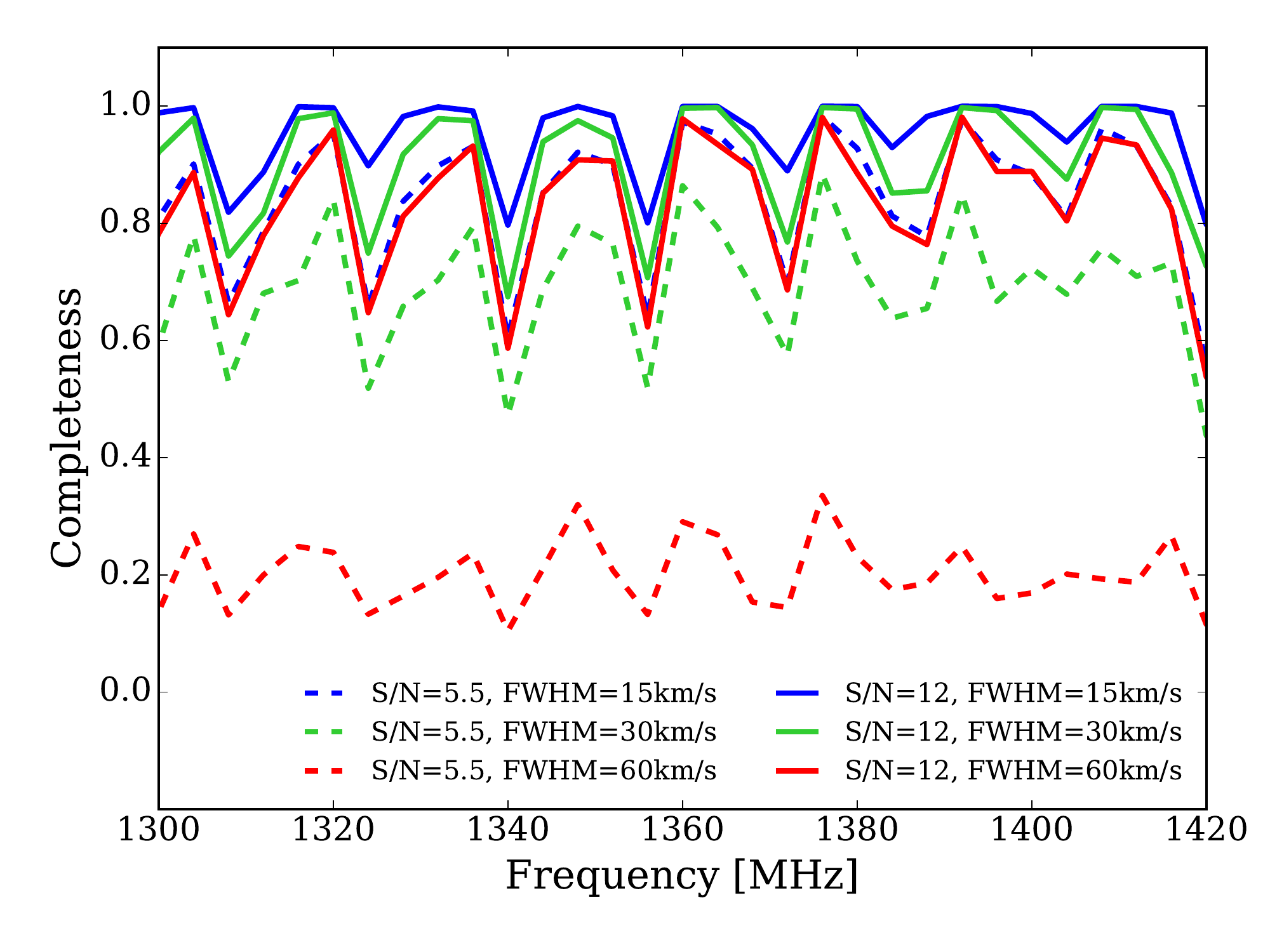}
    \caption{The estimated completeness of mock absorption for various combinations of parameters, including velocity widths of 15 $\kms$ (blue), 30 $\kms$ (green), and 60 $\kms$ (red), as well as velocity-integrated $\mathrm{S/N}$ of 5.5 (dashed lines) and 12 (solid lines).}
    \label{completeness_moreparameters}
\end{figure}

\begin{figure}
    \centering
    \includegraphics[width=0.45\textwidth]{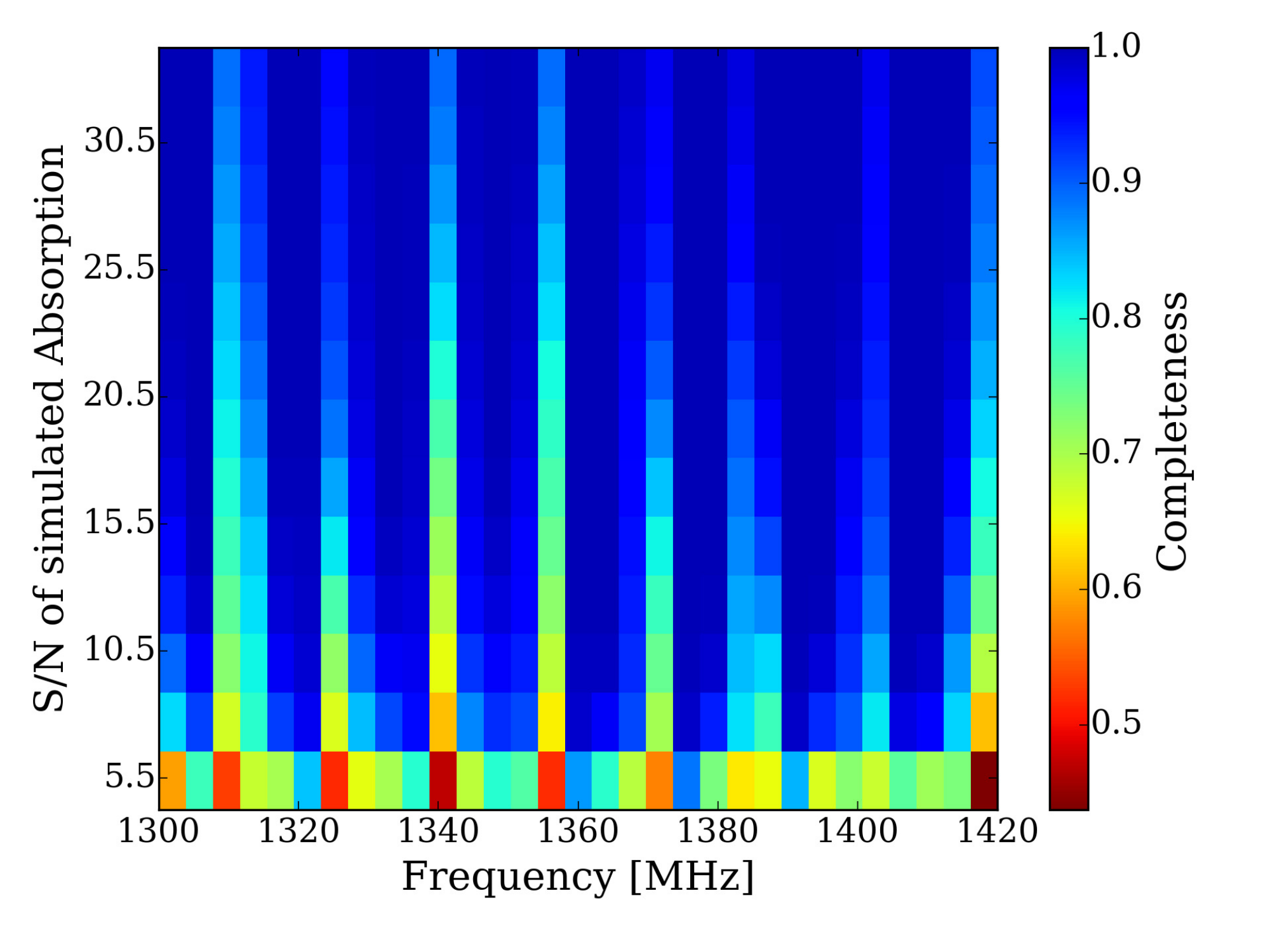}
    \caption{The estimated completeness for mock absorption with velocity-integrated $\mathrm{S/N}$ ranges from 5.5 to 33.0 at frequencies 1.30 - 1.42 GHz.}
    \label{completeness}
\end{figure}

In Figure\,\ref{completeness_moreparameters}, we show the estimated completeness of mock absorption profiles under different conditions, including velocity widths FWHM of 15 $\kms$, 30 $\kms$, and 60 $\kms$, as well as velocity-integrated $\mathrm{S/N}$ of 5.5 (the threshold we used for search in real data) and 12. For a fixed velocity-integrated sensitivity, narrower lines are more complete compared to broader lines, as the latter is more susceptible to being diluted by bandpass fluctuations. Recognizing that the sensitivity is affected by the selection of velocity width, we have chosen a value of FWHM = 30 $\kms$, which corresponds to the mean velocity width for all intervening 21-cm detected absorbers \citep{2016MNRAS.462.4197C,2016MNRAS.462.1341A}. Figure\,\ref{completeness} shows the average completeness for mock absorption with velocity width of FWHM = 30 $\kms$ and velocity-integrated $\mathrm{S/N}$ ranging from 5.5 to 33.0 at frequency 1.30 - 1.42 GHz. Frequency-dependent structures exhibited in the completeness result from the chronically present RFI at specific frequencies (as shown in Figure\,\ref{waterfall}). As expected, our search is less complete for absorption with lower $\mathrm{S/N}$. As shown in Figure\,\ref{completeness}, the velocity-integrated $\mathrm{S/N}$ cut of $\sim$ 12 corresponds to the estimate of the 90$\%$ completeness limit. 
As a consequence, the signal-to-noise ratio of 12 is adopted in the comoving absorption path calculation in Section~\ref{sec:expected_detection}.

\subsection{Standing Waves}
However, even with the threshold shown in Section~\ref{sec:algorithms}, tens of thousands of candidates have passed this screening. Nearly all of the spurious candidates originate from the fluctuations of the bandpass and chronically present RFI. The instability of the bandpass, which is mainly the result of standing waves, makes it difficult to remove the bandpass baseline accurately and find the true absorption signals. Most of these are excluded by using the multi-beam information in the drift scans, but the variation of the bandpass can be a severe source of error in such searches.  

In the \hi absorption feature search above, we ran our pipeline with the raw data before bandpass calibration. We have also tried to carry out the analysis of the bandpass-calibrated data. The built-in noise diode of the FAST receiver is used as a real-time calibrator to estimate the bandpass variations.  For CRAFTS high-cadence calibration mode, the noise diode is injected for $81.92\, {\rm \mu s}$ in every $196.608\, {\rm \mu s}$, with an amplitude of $\sim 1\,$K, which is a small fraction of the system temperature ( $\sim 20\,$K). Data with or without noise collected by the pulsar backend are averaged for every $0.2\,$s (equals to the sampling time of the spectral line backend) respectively, and their difference can be converted to spectrum backend units, relying on the fact that the bandpass shape and system temperature are determined mostly by the front end, and the two backends produce similar bandpasses. The 30-minutes averaged noise spectra are used to calibrate the variations of the bandpass. We then carry out the same analysis using the calibrated data. However, the final result is not affected much. To ensure no absorption is left out because of the imperfect baseline fitting (due to the presence of standing waves), we also tried a smaller median filter window size (0.45 MHz) and lower $\mathrm{S/N}$ threshold (4 $\sigma$). We find that it has little effect on the final candidate samples after multi-beam cross-correlation selection.

The results show that our calibration can only correct the variations over large frequency scales ($\Delta \nu > 10 \MHz$), the fluctuations at small frequency scales ($\Delta \nu \sim 1$ MHz) remain unimproved, and it is the small-scale fluctuations which generate the spurious signals. The precision of the calibration procedure is therefore limited, due to noise or fluctuations in the noise-diode signal. So further selection methods are needed to screen for spurious candidates. 

\subsection{Candidates Selection}
\label{sec:selection}
Figure\,\ref{freq_spec} and Figure\,\ref{waterfall} demonstrate the chronically present RFI and features produced by standing waves, which presents a challenge in accurately modelling the bandpass baseline and results in numerous false candidates. It is not feasible to determine the authenticity of a candidate using only one scan. However, with the drift-scan strategy and multi-beam system data, it becomes possible to differentiate the genuine \hi absorption signal from the spurious signal.

In CRAFTS, the feed array is rotated by 23.4 degrees to achieve super-Nyquist sampling. Perpendicular to the scanning direction, the spacing between the inner 17 beams is less than half FWHM. During the drift scan, the true \hi absorber will be scanned by multiple beams distributed along the scanning path, resulting in detection in different beams but the same sky location. 
In contrast, the spurious signals caused by RFI may appear in data from multiple beams, but do not necessarily originate from the same location in the sky. On the other hand, the transit time for each beam in drift scan is $\sim$ 12$/\cos\theta_{\rm{za}}$ seconds. In the drift-scan data with a time resolution of 12$/\cos\theta_{\rm{za}}$ seconds, the true signal can be detected in $<3$ continuous time points (corresponding to 9 $\arcmin$ in size, a conservative estimate of the angular size of extragalactic radio source), while the spurious signals can be detected in $\geqslant 5$ continuous or discontinuous time points.

We apply our selection pipeline to the preliminary \hi absorption candidates that passed the matched-filtering search:
\begin{enumerate}
\item Excluding the candidates if they can be detected at the same time point in more than 5 beams that are far apart,
\item Excluding the candidates if they can be detected in more than 6 continuous or discontinuous time points in the data from every single beam,
\item Giving priority to the candidates in the remaining catalogue if they can be detected in more than one beams which distribute along the direction of the scanning.
\item Checking the candidates survived after step 1 and step 2 by eye and comparing automated candidates from step 3 with the visual candidates.
\end{enumerate}

\begin{figure*}
    \centering
    \begin{multicols}{2}
    \includegraphics[width=8.8cm]{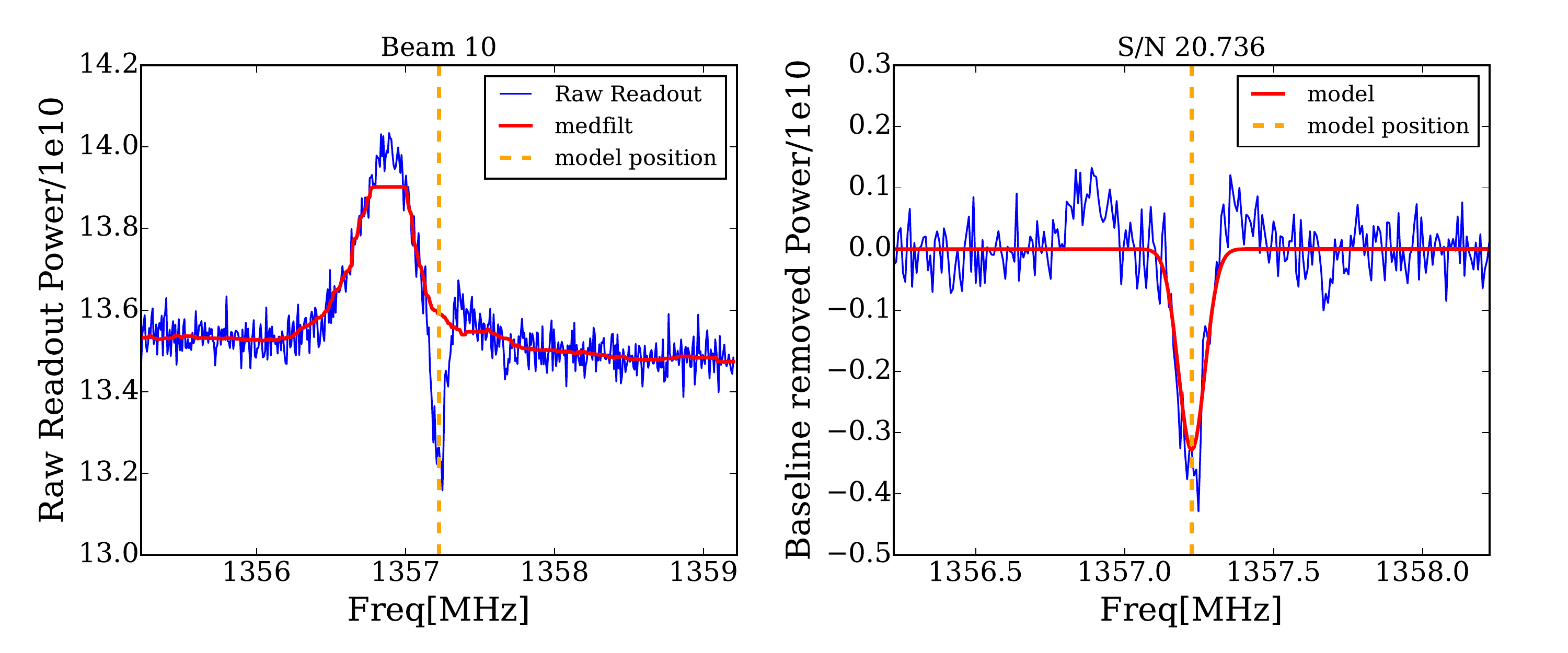}
    \includegraphics[width=8.8cm]{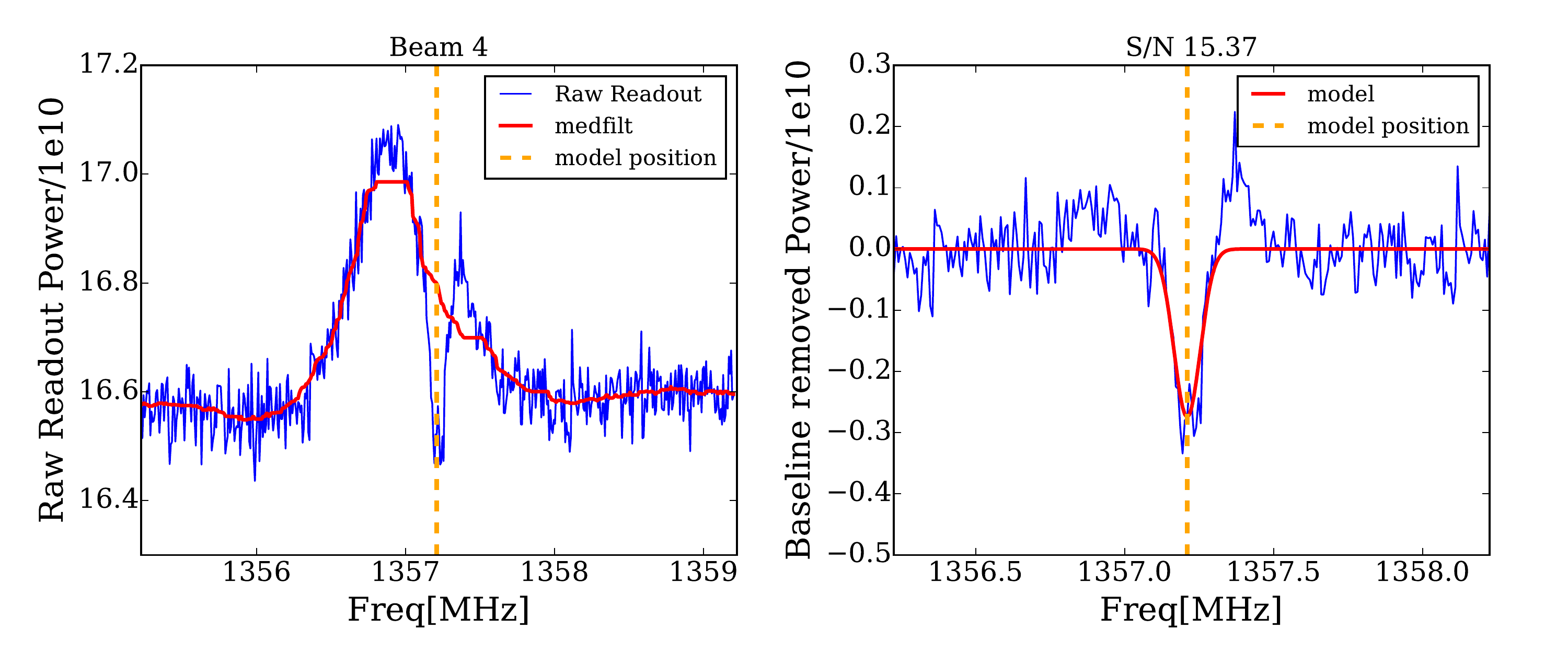}
    \includegraphics[width=8.8cm]{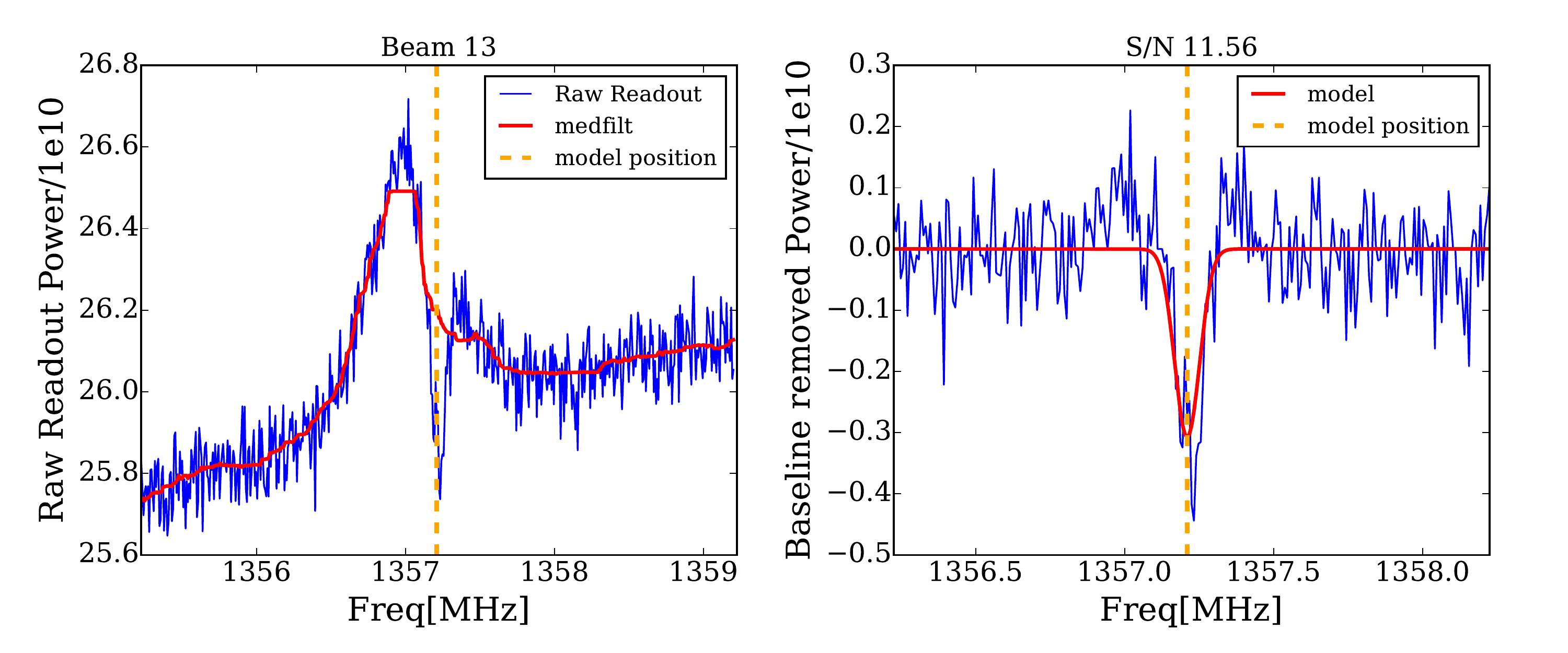}
    
    \includegraphics[width=8.8cm]{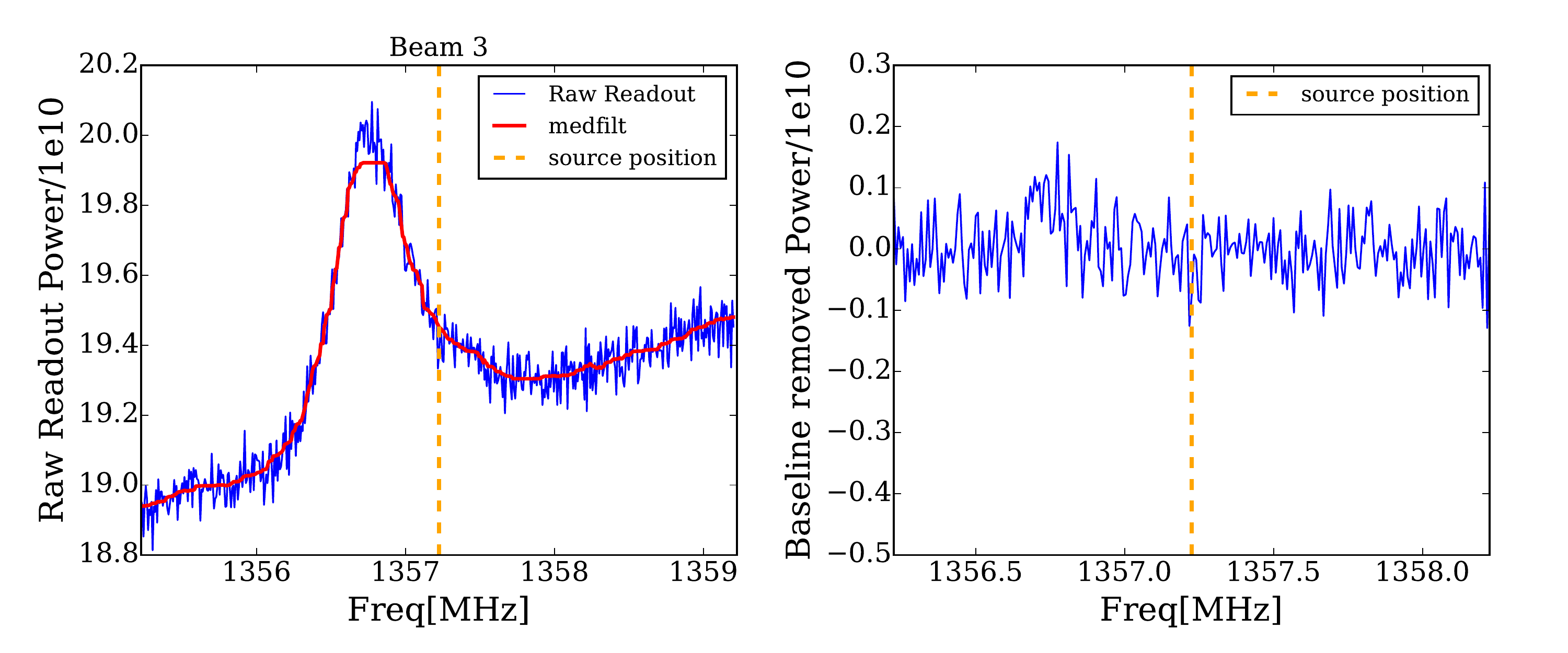}
    \includegraphics[width=8.8cm]{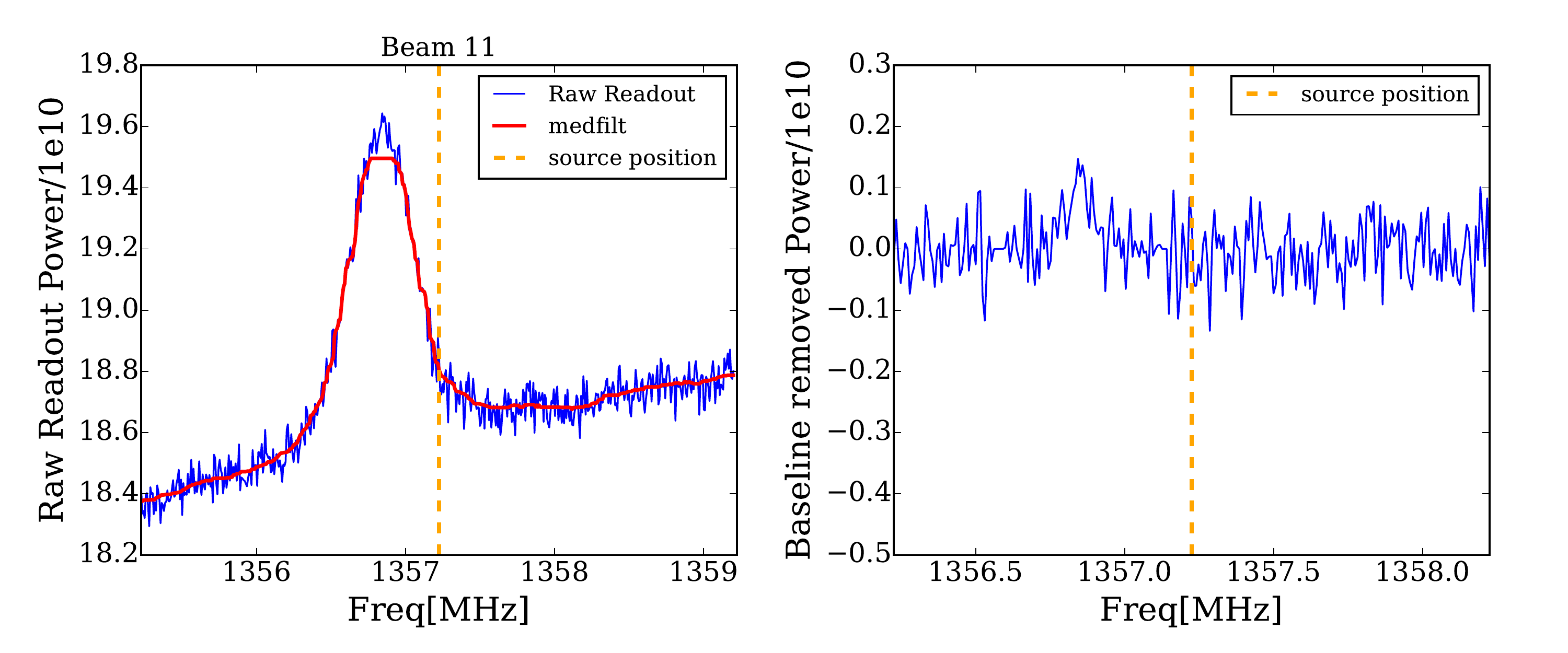}
    \includegraphics[width=8.8cm]{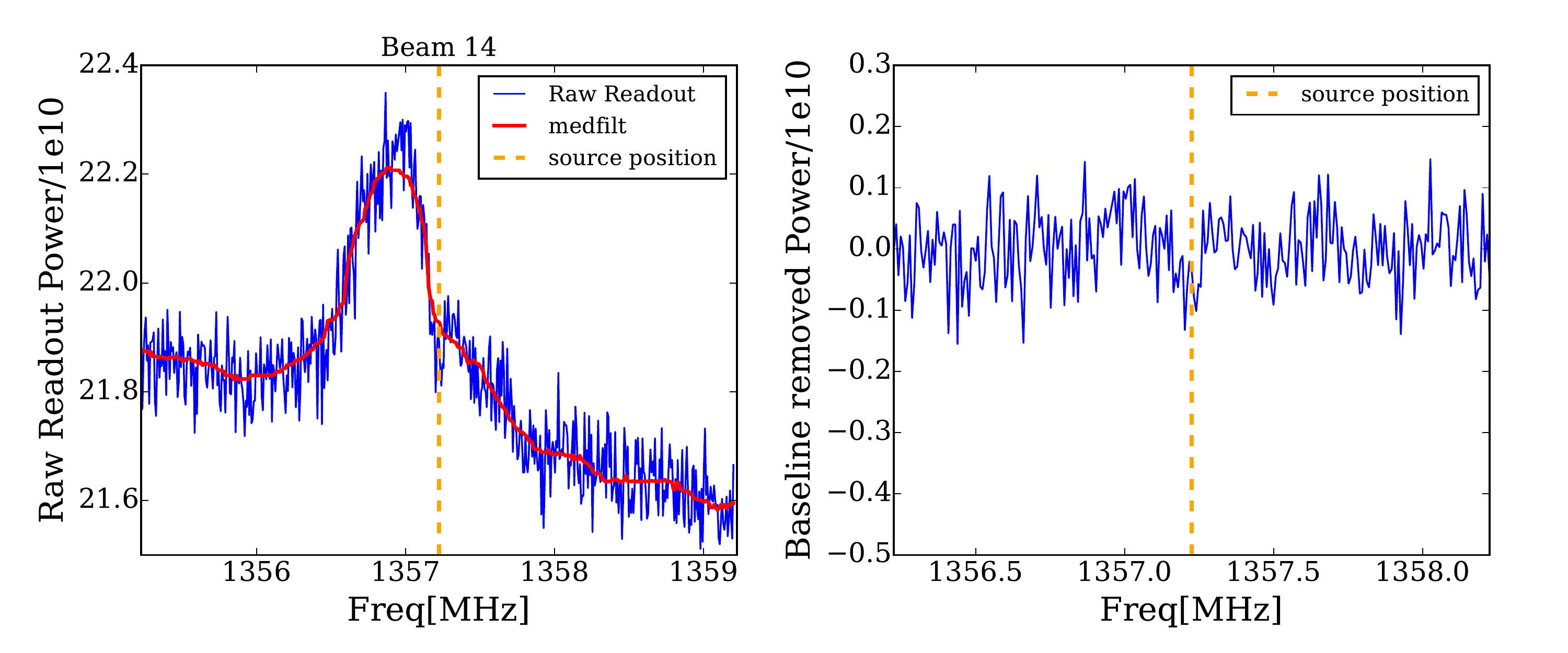}
    \end{multicols}
    \caption{The spectra extracted from different beams at the pointing position closest to UGC\,00613. Left: The spectra of the (previously known) \hi absorption system UGC\,00613. The top, middle and bottom sub-panels show the detection from Beam 10, Beam 4 and Beam 13, respectively. The left column shows the raw data, and the right column shows the baseline removed data. The red-solid lines in the left column indicate the baselines estimated using {\tt medfilt}. The red-solid lines in the right column depict the best template modelling the \hi absorption profile. The vertical orange-dashed lines show the central channel position of the template. The $\mathrm{S/N}$ of the absorption profiles are given as the title of the sub-figures in the right column. Right: Same as the left sub-panels, but for neighbouring Beam 3, Beam 11 and Beam 14, there is no precise detection. The vertical orange-dashed lines refer to the frequency position of UGC\,00613. }
    \label{UGC00613_spec}
\end{figure*}

\begin{figure}
    \centering
    \includegraphics[width=0.45\textwidth]{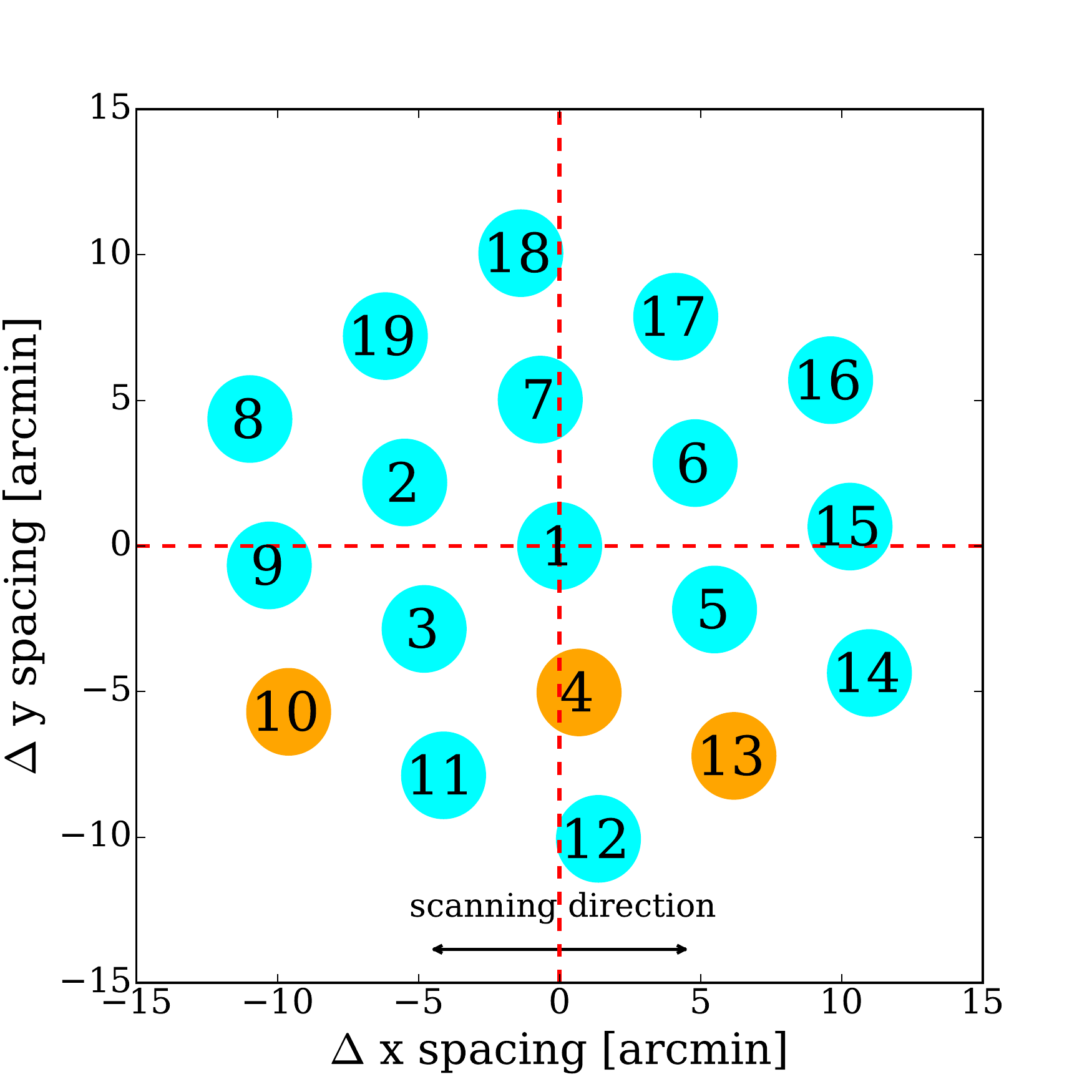}
    \caption{The distribution of the beams from which the \hi absorption system UGC\,00613 was detected. The known \hi absorption system UGC\,00613 can be detected in the data of Beam 10, Beam 4 and Beam 13, which are labelled with orange. These 3 beams distribute in a horizontal line along the scanning direction. The 19 beams have been rotated by 23.4°.}
    \label{UGC00613_beam}
\end{figure}

We take the known \hi absorption system UGC\,00613 as an example of how our candidate selection pipeline works. Using the Matched-Filtering Approach, the absorption profile of UGC\,00613 can be detected in the data from Beam 10, Beam 4 and Beam 13, respectively. In Figure~\ref{UGC00613_spec}, we show the raw spectra (left column) and baseline-removed spectra (right column) from Beam 10, Beam 4 and Beam 13. A high emission peak is beside the absorption profile, which arises from the chronically present RFI. Even though contaminated by the RFI, all detections have significant $\mathrm{S/N}$ (larger than 12). These 3 beams are horizontal along the scanning direction, as shown in Figure\,\ref{UGC00613_beam}. Besides, no other distant beam has detection of this feature. These show that the absorption profile we detected is a true signal, not one from RFI or bandpass variation.

In our data analysis, using criteria (i) and (ii) alone can exclude 73$\%$ and 89$\%$ of the spurious detections, respectively, and combining the spatial and time information together makes our method more effective. From the results, all the excluded candidates are produced by the RFI and standing waves. No true signals are diagnosed as spurious in either criterion (i) or (ii). Using the selection method described above, we excluded most spurious detections ($\sim$ 140,000 cases), and 3 known, 2 new \hi absorption systems and 5 unverified candidates survived. 

\subsection{\hi Absorption Measurement}
\label{sec:theory}
According to the radiative transfer theory, the observed signal ($T_{\rm b}(\nu)$) of an \hi source which is partially covering a background continuum source can be described as \citep{2018A&ARv..26....4M}:
\begin{eqnarray}
    T_{\rm b}(\nu) = c_{\rm f}T_{\rm c}e^{-\tau(\nu)} + (1-c_{\rm f})T_{\rm c} + T_{\rm s}(1-e^{-\tau(\nu)}),
    \label{Tobs}
\end{eqnarray}
where $c_{\rm f}$ is the covering factor which is the fraction of the background source covered by foreground gas clouds, as seen from the centre. $\tau(\nu)$ and $T_{\rm s}$ are the optical depth and spin temperature of the \hi source respectively, and $T_{\rm c}$ is the brightness temperature of the background continuum source. In Eq.\,(\ref{Tobs}), the first term denotes the absorbed emission from the continuum source, the second term is the emission from the part of the continuum source not covered by the cloud, and the last term describes self-absorbed 21-cm line emission from the \hi source. 

The difference between the $T_{\rm b}(\nu)$ and $T_{\rm c}$ is then:
\begin{eqnarray}
    \Delta T(\nu) = T_{\rm b}(\nu) - T_{\rm c} = (T_{\rm s}-c_{\rm f}T_{\rm c})(1-e^{-\tau(\nu)}).
    \label{T_diff}
\end{eqnarray}
If $T_{\rm s}$ > $c_{\rm f}T_{\rm c}$, the spectral line will be seen in emission, while if $T_{\rm s}$ < $c_{\rm f}T_{\rm c}$, the absorption dominates. For most \hi absorption observations, $T_{\rm s} \ll c_{\rm f}T_{\rm c}$, then $\Delta T(\nu) = -c_{\rm f}T_{\rm c}(1-e^{-\tau(\nu)})$. After converting the frequency ($\nu$) to velocity (V), the optical depth can be expressed as :
\begin{eqnarray}
    \tau(V) \approx -\mathrm{ln}(1+\Delta T(V)/(c_{\rm f}T_{\rm c})),
    \label{tau}
\end{eqnarray}
in the limit $T_{\rm s} \ll c_{\rm f}T_{\rm c}$,
where $T_{\rm c}$ can be deduced from the line-free parts of the spectrum or the other database. In this work, we use the $T_{\rm c}$ from the NASA/IPAC Extragalactic Database\footnote{\url{https://ned.ipac.caltech.edu}} (NED). 

The optical depth $\tau$ is proportional to $N_{\hi}/T_{\rm s}$, the \hi column density is given by 
\begin{eqnarray}
    N_{\hi}[\mathrm{cm^{-2}}] = 1.82 \times 10^{18} T_{\rm s}[\mathrm{K}] \int \tau(V)dV[\mathrm{km\,s^{-1}}].
    \label{column_density}
\end{eqnarray}
Throughout this paper we take the assumption that $c_{\rm f} = 1$ \citep{2017A&A...604A..43M} for the calculation of $\tau$ and $N_{\hi}$. 

\section{Results}
\label{sec:Results}

\begin{figure*}
\centering{
\includegraphics[width=8.1cm]{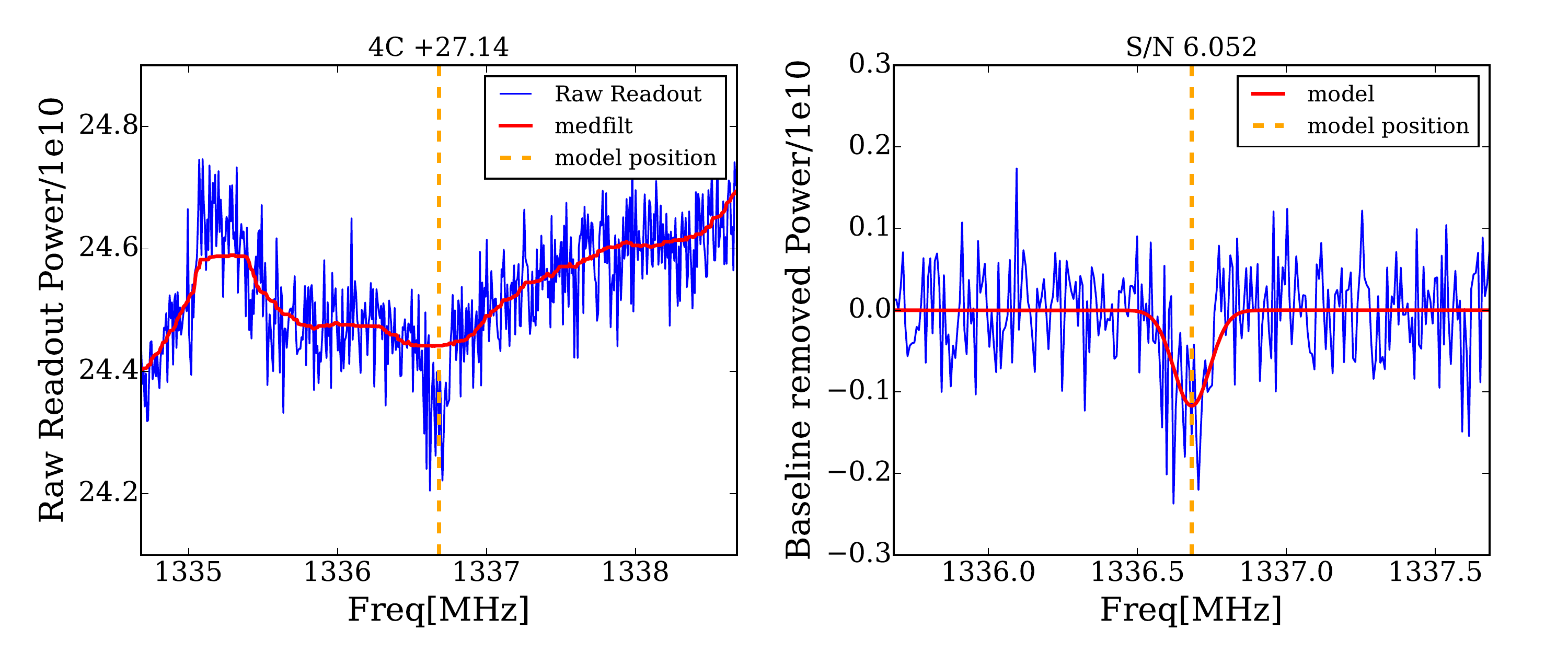}
\includegraphics[width=8.1cm]{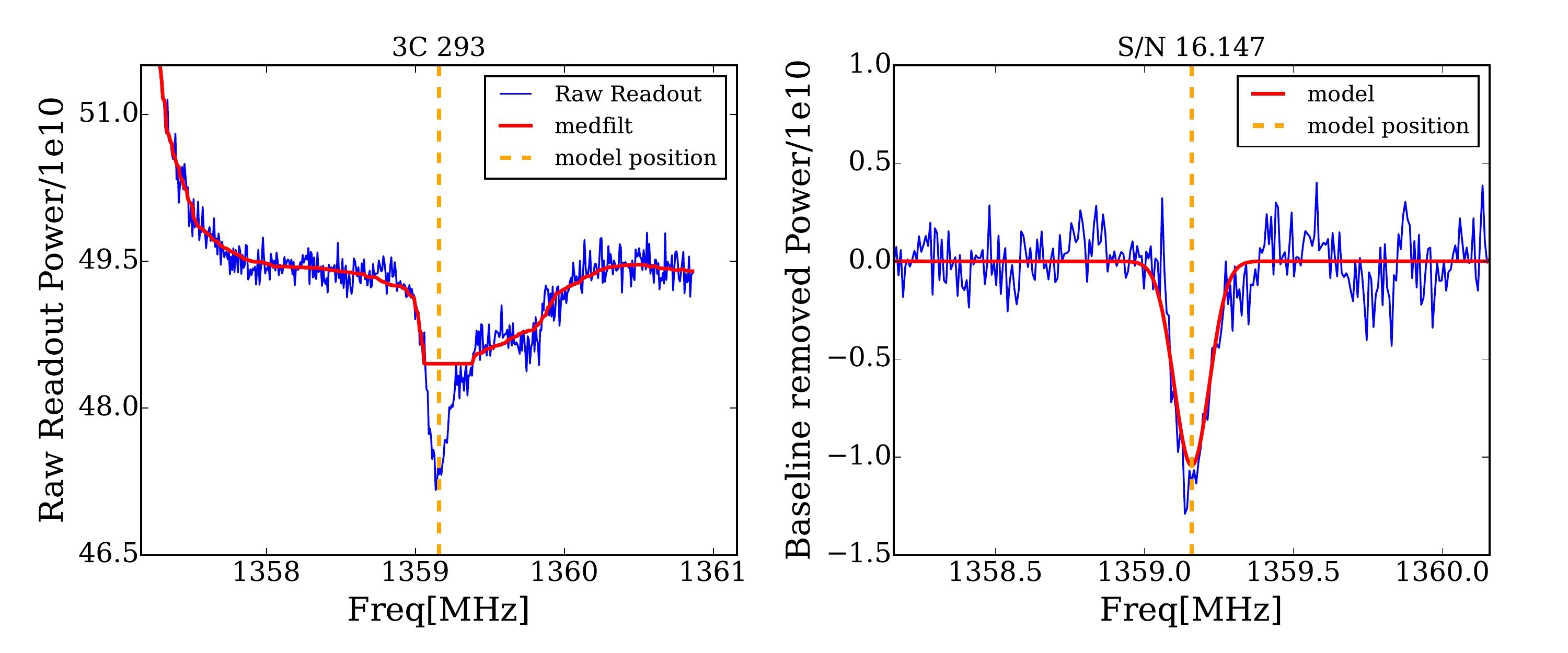}
\includegraphics[width=8.1cm]{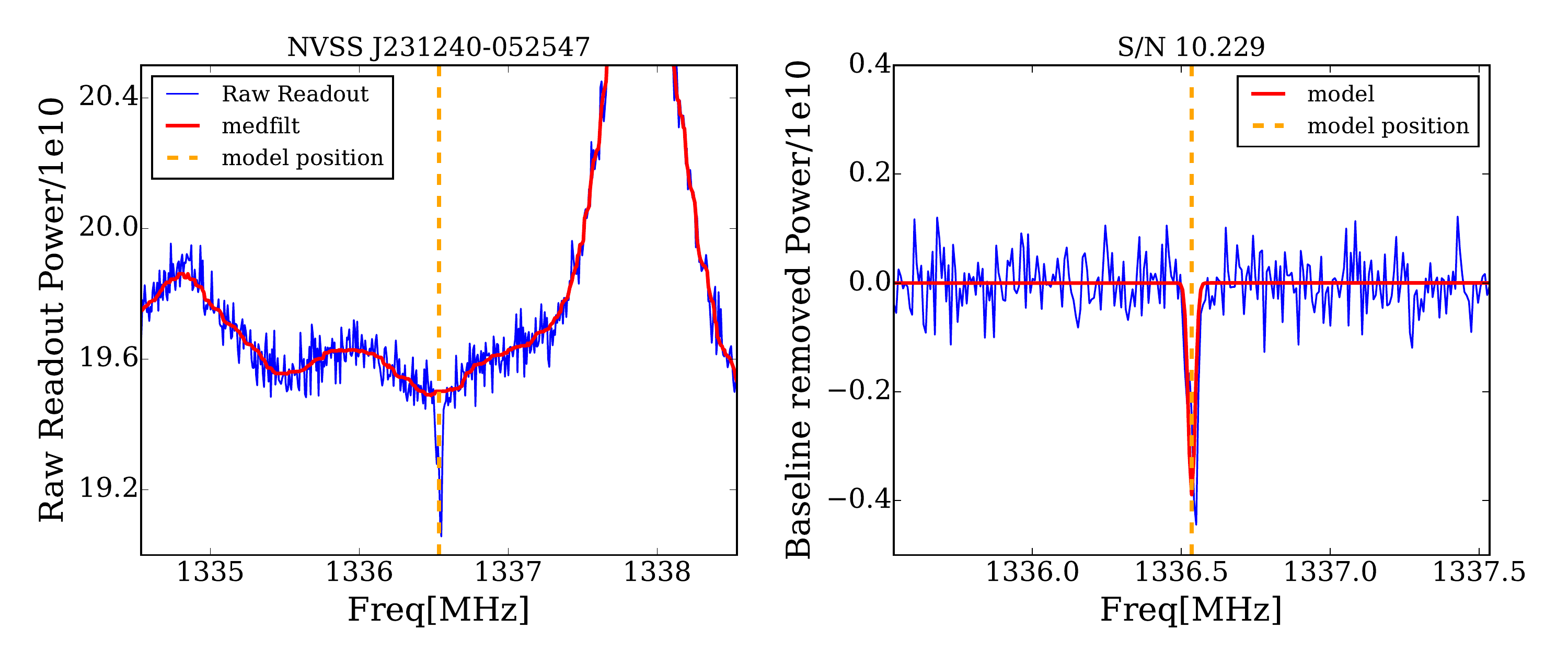}
\includegraphics[width=8.1cm]{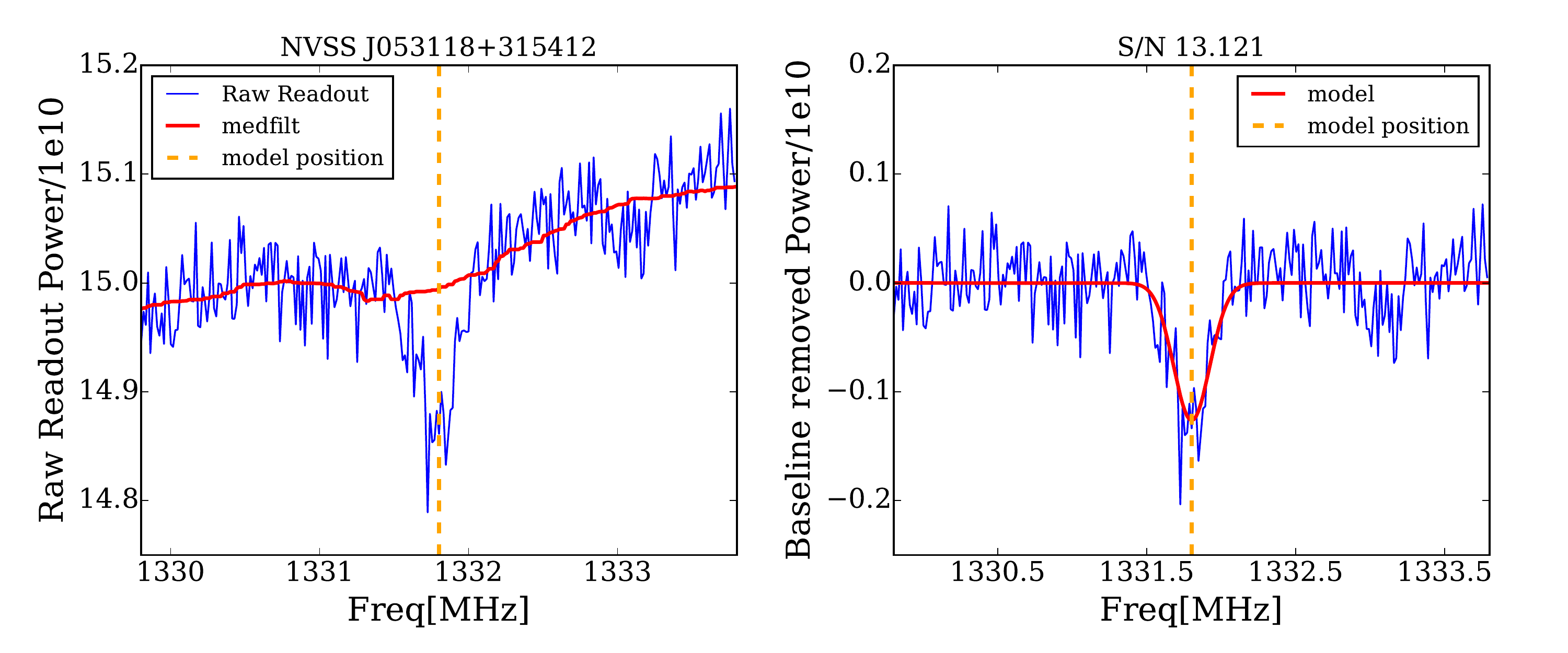}
\includegraphics[width=8.1cm]{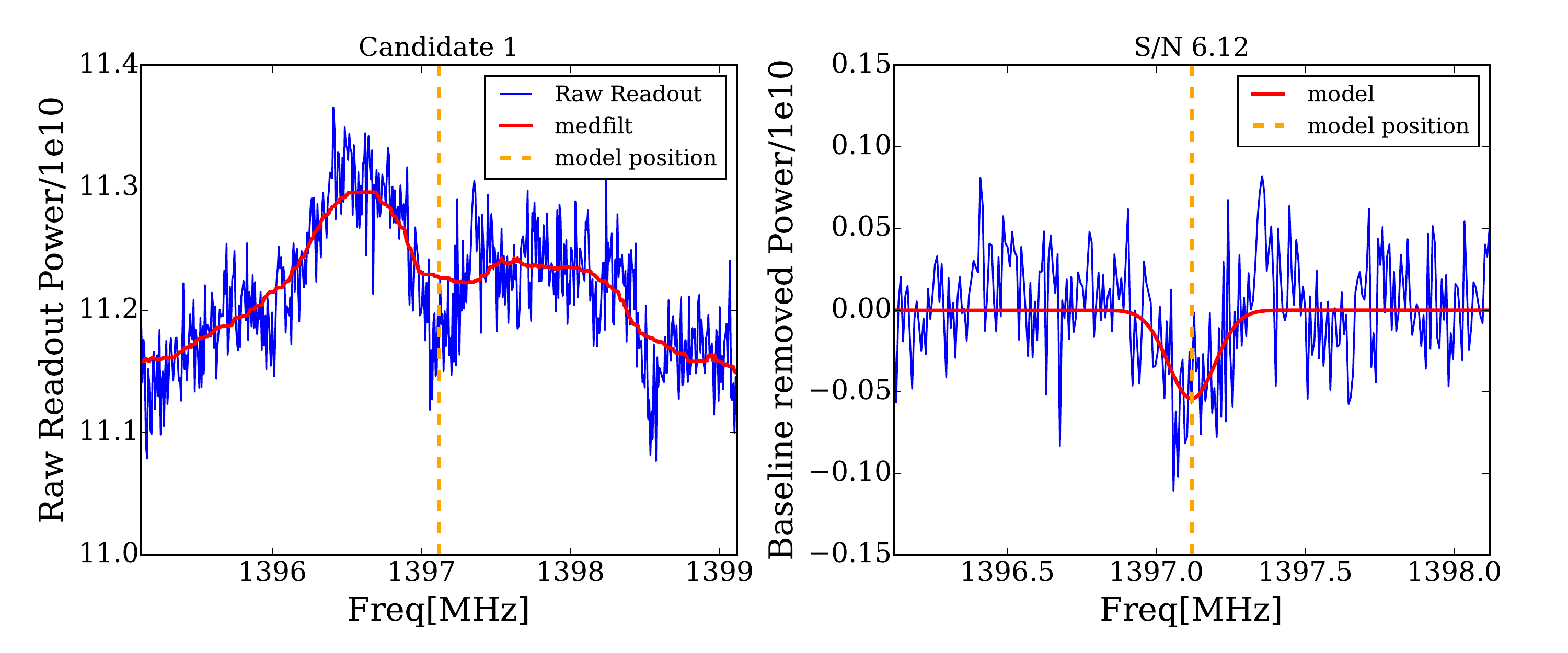}
\includegraphics[width=8.1cm]{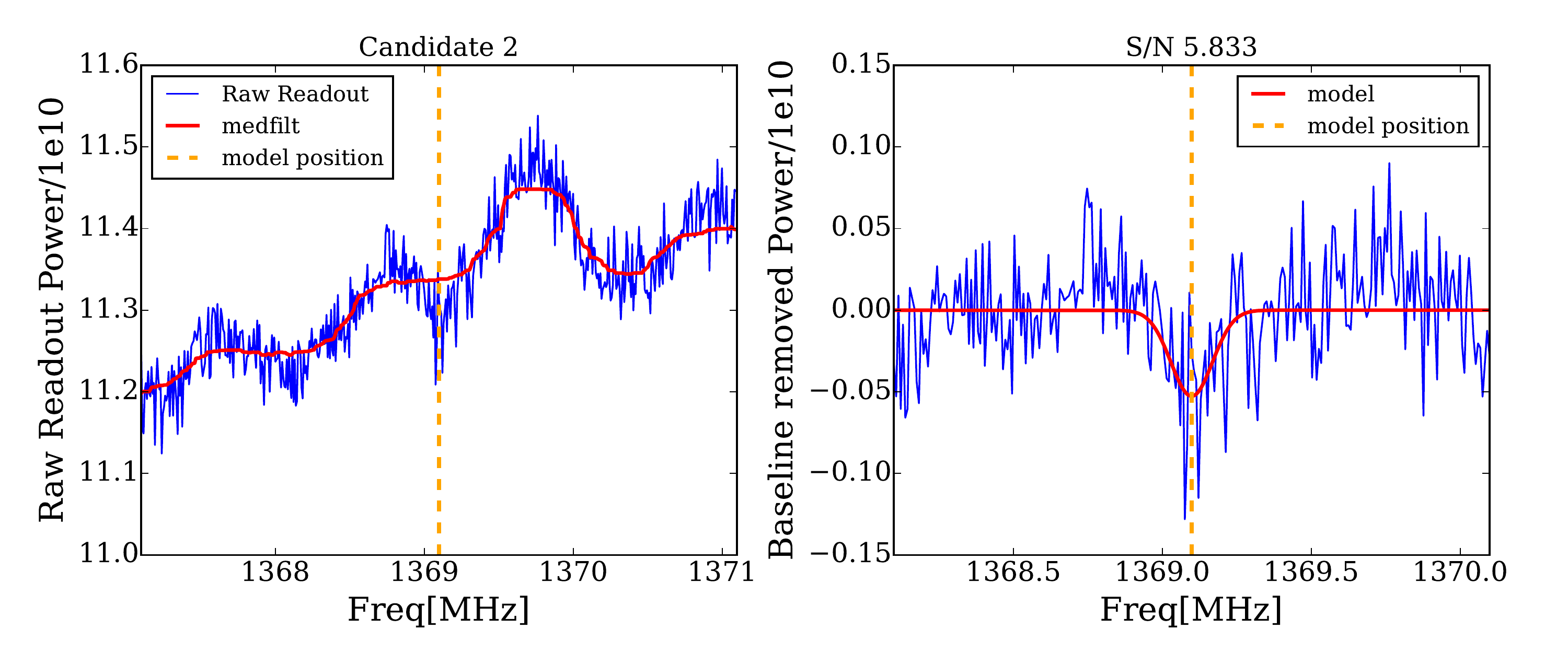}
\includegraphics[width=8.1cm]{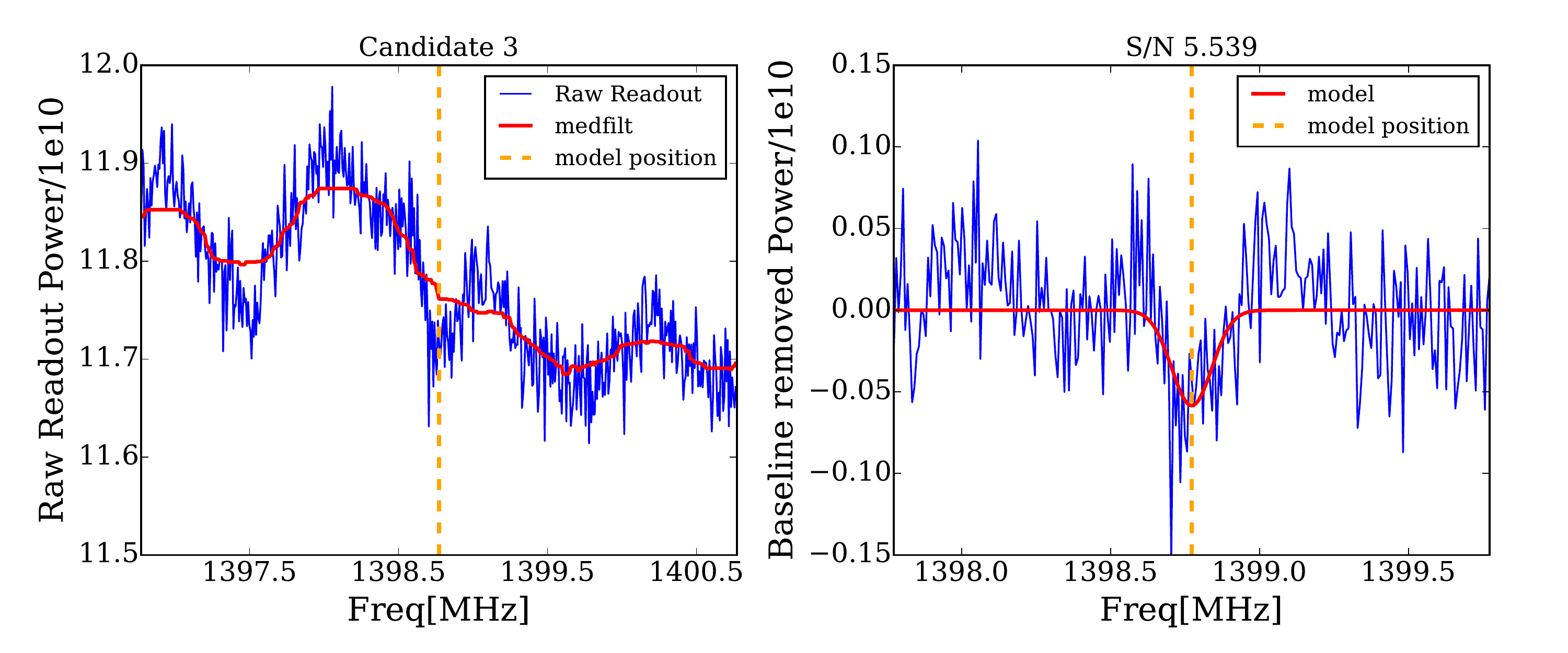}
\includegraphics[width=8.1cm]{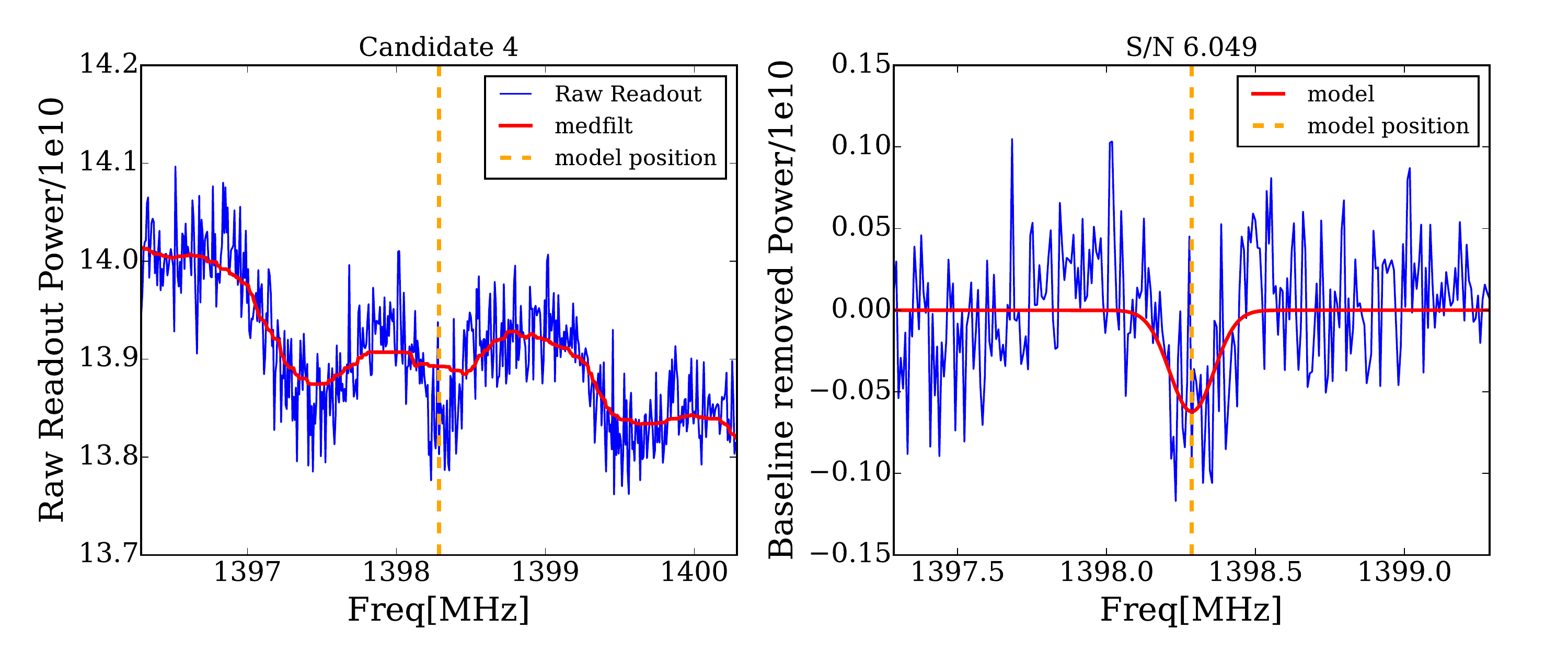}
\includegraphics[width=8.1cm]{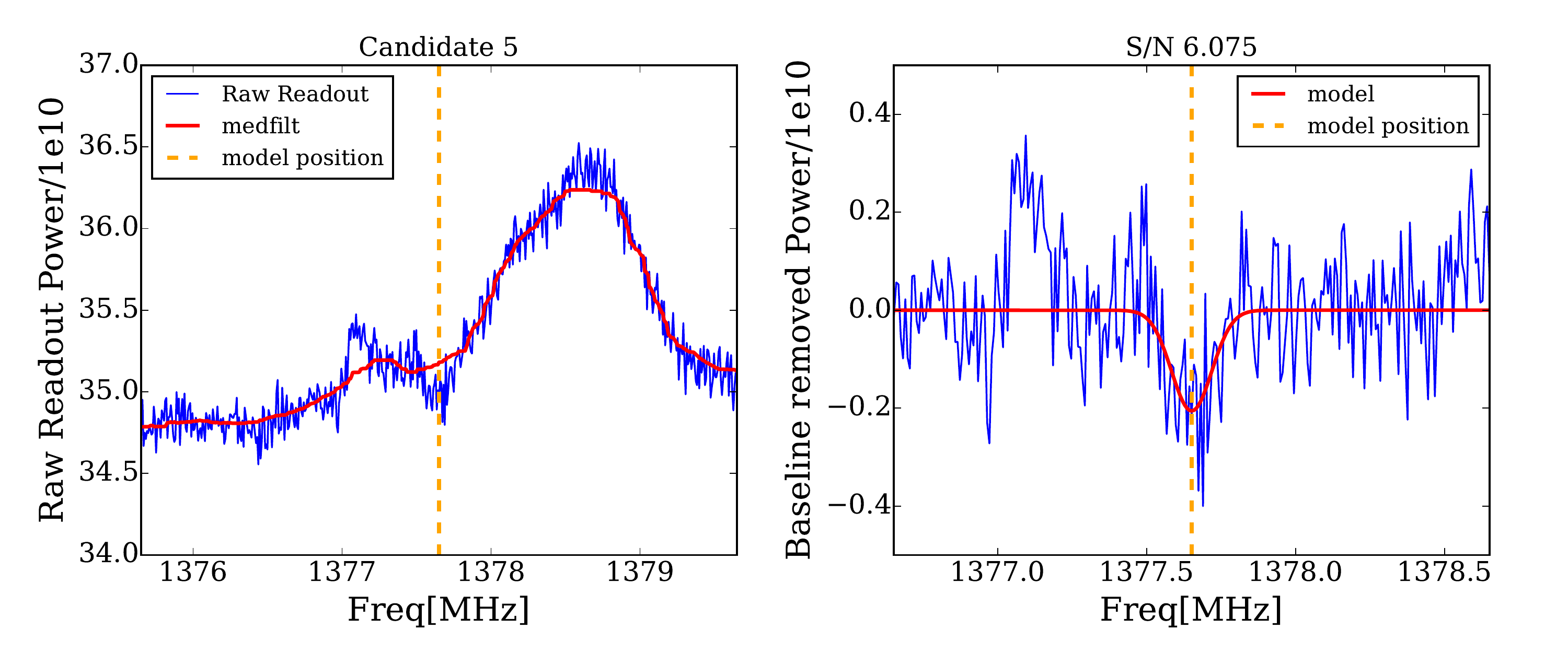}
}
\caption{The CRAFTS spectra of the 5 \hi absorber candidates, 4C\,+27.14 3C\,293, NVSS\,J231240-052547 and NVSS\,J053118+315412. In each row, the odd and even-numbered columns show the raw spectra and the baseline removed spectra, respectively. The vertical lines indicate the frequencies of the absorbers. The red-solid lines in the raw spectra show the baseline. The red-solid lines in the baseline removed spectra show the Gaussian fit absorption profile. Spectra of UGC\,00613 are shown in Figure\,\ref{UGC00613_spec}.}
\label{candidates_plot}
\end{figure*}

By utilizing the matched filter and multi-beam candidate selection pipeline on the collected data, we have identified a total of 10 candidates, among which are three that have been previously detected: UGC\,00613 \citep{2021MNRAS.503.5385Z}, 3C\,293\citep{1981ApJ...243L.143B,2004MNRAS.352...49B} and 4C\,+27.14\citep{2014ApJ...793..132S}. The spectra of these 10 candidates are shown in Figure~\ref{candidates_plot}. All spectra have been corrected for the Doppler shift induced by the Earth's revolution. The high peaks besides the absorption profile (Candidate 5, UGC\,00613 and NVSS\,J231240-052547, for example) are the chronically present RFI. As shown in Figure~\ref{candidates_plot}, the baseline fit in some cases is not good enough. Although the median filter window size is carefully determined, it cannot always provide a perfect estimation of the baseline, especially when factors such as RFI, standing waves (Candidate 4 in Figure~\ref{candidates_plot} for example), or broad absorption lines (3C\,293 in Figure~\ref{candidates_plot} for example) exist. To improve the results, we also tested a smaller median filter window size of 0.45 MHz. However, this did not yield any new \hi absorption detections beyond the five already identified. While spurious detections decreased, the $\mathrm{S/N}$ ratio of the previously identified candidates was reduced. To eliminate the effects of fluctuations and RFI around \hi absorption signals, we will employ a finely-tuned fitting technique using high-order polynomials + multi-component Gaussian functions on the spectra from follow-up observations.

We have made follow-up observations for these candidates in the ON-OFF tracking observation mode, with 990 s integration on each. Besides the three previously known ones, two are confirmed to be bona fide absorbers (along the l.o.s. towards NVSS\,J231240-052547 and NVSS\,J053118+315412), while the other 5 candidates are either produced by features in the bandpass or the combined features from the \hi emission and bandpass ripples (this will be discussed later). We fit the profiles of the true \hi absorptions with multi-components Gaussian functions and presented the basic physical information for each source. The absolute photometric flux is calibrated with 3C\,48 and the FAST built-in noise diode. 
 
\subsection{Previously Known Absorbers}

\subsubsection{UGC 00613}
The \hi absorption feature of UGC\,00613 was firstly reported in \citet{2021MNRAS.503.5385Z} with a flux density depth of $S_{\hi,\rm peak} = -64.28 \pm 6.61$ mJy, FWHM of 26.36 $\pm$ 1.74 \kms, and $\int\tau dv=$ 21.24 $\pm$ 4.10 \kms, from a one-component Gaussian fitting. UGC\,00613 is a flat-spectrum extended radio source \citep{2007ApJS..171...61H} which probably locates in an unrelaxed, merging system \citep{2019MNRAS.488.3416H}. We blindly re-detected its \hi absorption in beam 10, beam 4 and beam 13 in the CRAFTS drift-scan data. As Figure~\ref{UGC00613_spec} shows, the absorption feature overlaps in the frequency band with an emission feature, which is a chronically present RFI. To remove the baseline and RFI, a Gaussian + 2nd-order polynomial function was fitted to the absorption line-free parts of the spectrum. We use the FAST built-in noise diode for flux calibration. The system temperature $T_{\rm sys}$ is estimated as
\begin{eqnarray}
    T_{\rm sys} = \frac{P_{\rm obs}}{P_{\rm noise}}\times T_{\rm noise},
    \label{Tsys}
\end{eqnarray}
where $P_{\rm noise}$ is the recorded noise power from the noise diode, $P_{\rm obs}$ is the CRAFTS recorded observational data, $T_{\rm noise}$ is the noise temperature. $T_{\rm sys}$ was converted to flux density using pre-measured antenna gain \citep{2020RAA....20...64J}: Flux = $T_{\rm sys}$/Gain.

\begin{figure}
    \centering
    \includegraphics[width=0.45\textwidth]{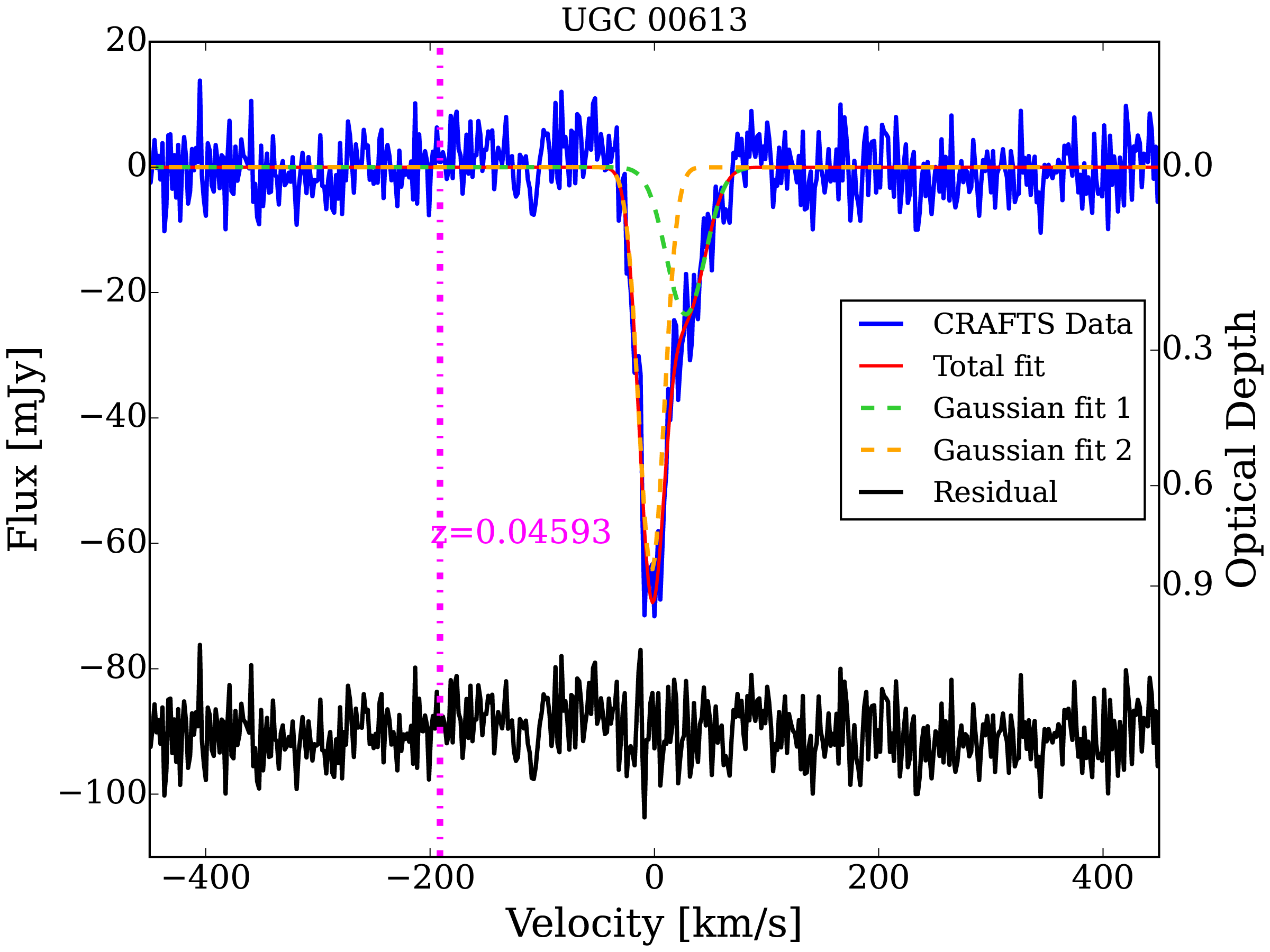}
    \caption{\hi absorption feature of UGC\,00613. 
 The blue solid line depicts the absorption spectrum, and the red solid line shows the fit with a double-component Gaussian model ( components shown as orange and green dashed lines). The magenta dot-dashed vertical line marks the redshift of UGC\,00613 as given in the NASA/IPAC Extragalactic Database. The fitting residual is shown as the black solid line at the bottom. Optical depth value for the \hi absorption is shown on the right scale.}
    \label{UGC00613_fit}
\end{figure}

The absorption spectra of UGC\,00613 and its Gaussian fitting are presented in Figure~\ref{UGC00613_fit}. Two components are found in the absorption feature, and some basic physical information is given in Table\,\ref{radiosource_table} and Table\,\ref{absorption_table}. Our measurements of the \hi absorption of UGC\,00613 with a flux density depth of $S_{\hi,\rm peak} =$ -69.46 $\pm$ 5.56 mJy, FWHM of 31.74 $\pm$ 1.85 \kms, and $\int\tau dv=$ 32.75 $\pm$ 4.39\kms. The measurements of Gaussian component 2 (Table\,\ref{absorption_table}) are consistent with the previous results reported in \citet{2021MNRAS.503.5385Z}.

\subsubsection{3C 293}
3C\,293 is a Fanaroff and Riley type II (FR II) \citep{10.1093/mnras/167.1.31P} radio galaxy, with a compact core of steep spectrum and multiple knots in the radio lobe \citep{1999ApJ...511..730E,2004MNRAS.352...49B}. Both emission and absorption in CO have been detected \citep{1999ApJ...511..730E}, showing that 3C\,293 is a disk galaxy with an optical jet or tidal tail toward its companion galaxy. The outflow of \hi had also been detected, which is probably driven by the radio jet \citep{2013MNRAS.435L..58M}. The radio jets twist by 30°, probably as a result of strong interaction or a recent merger event. Besides, an extremely broad and multi-components \hi absorption feature has been detected \citep{1981ApJ...243L.143B,2004evn..conf..147B,2004MNRAS.352...49B}, indicating a rotating \hi disk.

\begin{figure}
    \centering
    \includegraphics[width=0.45\textwidth]{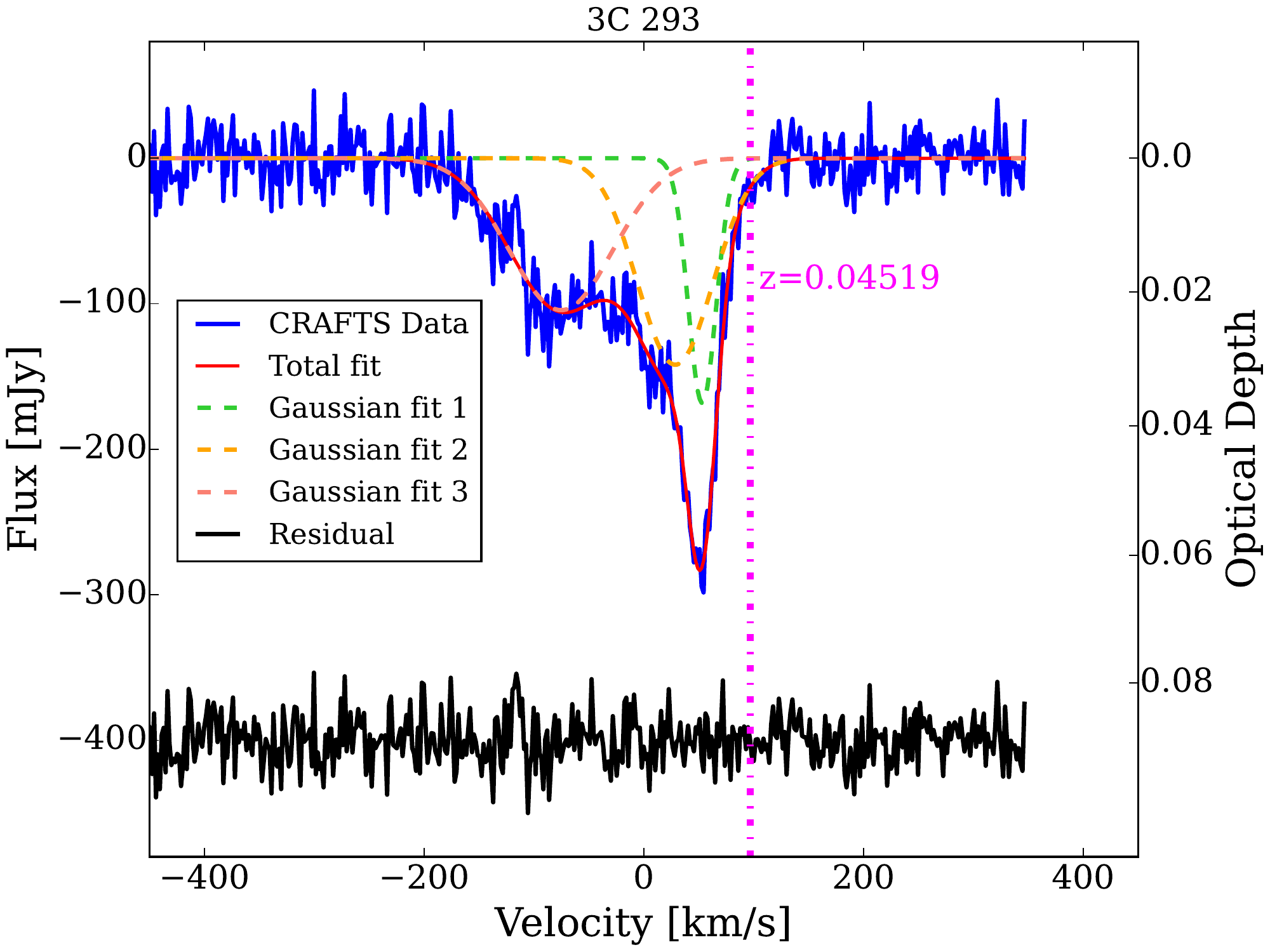}
    \caption{Same as Figure~\ref{UGC00613_fit}, but for the \hi absorption feature of 3C\,293. }
    \label{3C293_fit}
\end{figure}

We detected again the \hi absorption in 3C\,293 in our blind search. Its signal can be detected in beam 10, beam 11, beam 4 and beam 13 with a high signal-to-noise ratio ($\mathrm{S/N} > 10$). We use the built-in noise diode to calibrate the data and measure the basic physical parameters using the method described in Section\,\ref{sec:theory}. The absorption feature is fitted by a triple-component Gaussian function and is presented in Figure~\ref{3C293_fit}, and the physical parameters are given in Table\,\ref{radiosource_table} and Table \,\ref{absorption_table}.

As shown in Figure\,\ref{3C293_fit}, a broad \hi absorption feature with multiple components is obtained. For the total signal, the \hi absorption of 3C\,293 is detected with a flux density depth of $S_{\hi,\rm peak} =$-283.33 $\pm$ 17.24 mJy, FWHM $=61.54 \pm 4.32 \kms$, $\tau_{\rm peak} = 0.062 \pm 0.004$, $\int\tau dv= 6.47 \pm 0.41 \kms$ and $N_{\hi}= 0.12 T_{s} \times 10^{20} \cm^{-2} \K^{-1}$, which are consistent with the previous measurement reported in \citet{1981ApJ...243L.143B} which detected the \hi absorption signal for the first time and gave $\tau_{\rm peak} = 0.085$ and a column density of $0.18 T_{s}\times10^{20} \cm^{-2} \K^{-1}$. The difference is a result of different adoption of flux density of 3C\,293 at 1.4 GHz ($S_{\rm 1.4GHz}$), \citet{1981ApJ...243L.143B} used $S_{\rm 1.4GHz} = 3.76 $ Jy while we adopt recent measurement $S_{\rm 1.4GHz} = 4.69$ Jy \citep{2018AJ....155..188L}.

\begin{figure}
    \centering
    \includegraphics[width=0.45\textwidth]{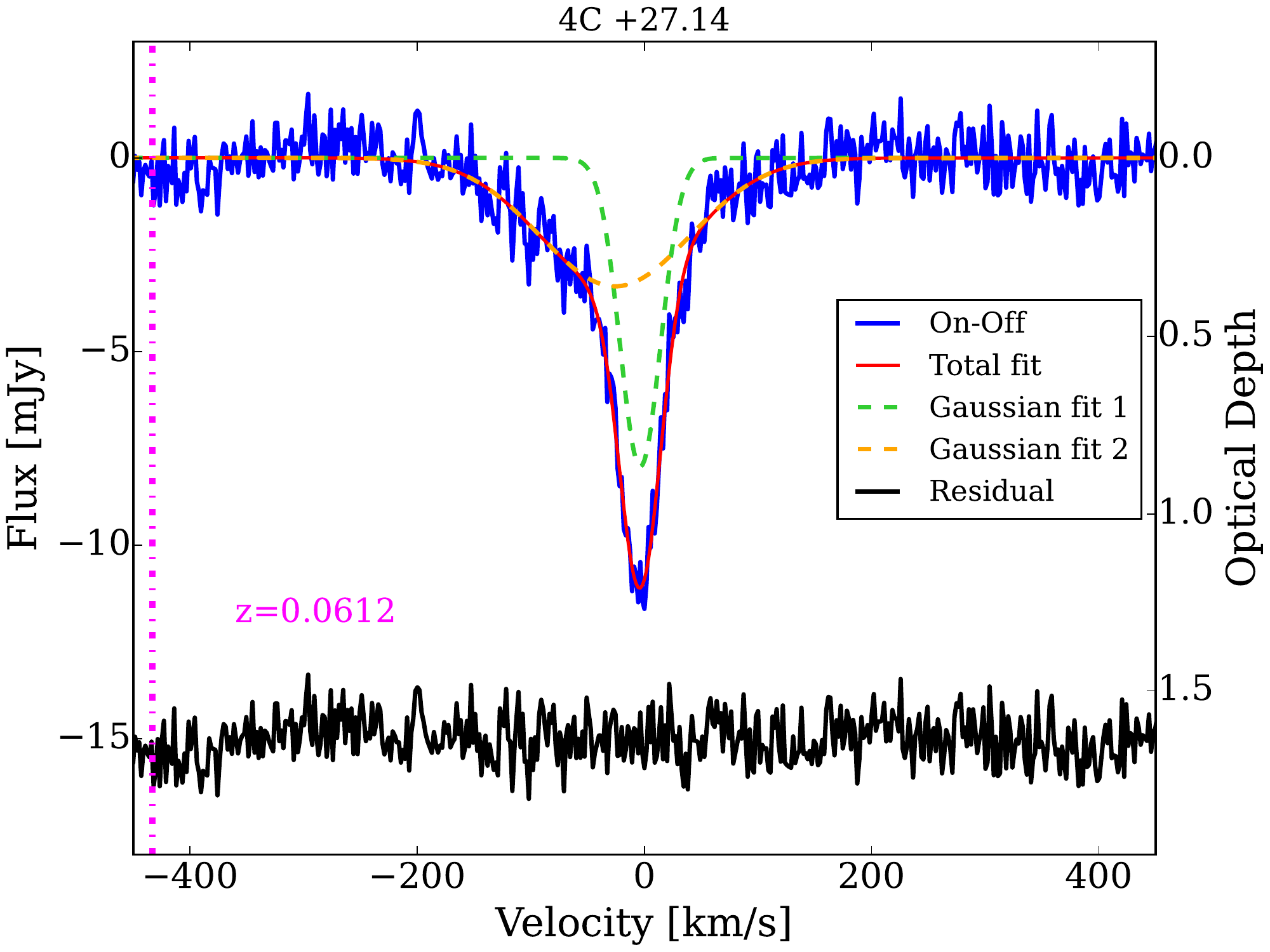}
    \caption{Same as Figure~\ref{UGC00613_fit}, but for the \hi absorption feature of 4C\,+27.14 obtained by processing the follow-up ON-OFF tracking data. }
    \label{4C+27.14_fit}
\end{figure}

\subsubsection{4C +27.14}
4C\,+27.14 is a Type II Seyfert galaxy and displays the properties of a radio-loud AGN \citep{2014A&A...561A..67P}. Broad \hi absorption profile has been observed in the direction of radio continuum source 4C\,+27.14 \citep{2014ApJ...793..132S} for the first time. The \hi absorption lines are detected in the CRAFTS drift-scan data with blind search and the follow-up observation using ON-OFF tracking mode. The absorption signal of 4C\,+27.14 is faint, only detected in beam 11 in the CRAFTS data. The fluctuations in the bandpass are suppressed by subtracting the source-off data from the source-on data. The calibrator 3C\,48 was also observed to calibrate the data. A 2nd order polynomial function was fitted to the absorption line-free parts of the spectrum to remove the baseline of the bandpass. 

The calibrated ON-OFF spectrum and its Gaussian fitting are presented in Figure\,\ref{4C+27.14_fit} (and Figure\,\ref{spectra_followup} for the follow-up observation). Some physical parameters of the system are given in Table\,\ref{absorption_table}. 
Our measurement of the \hi absorption profile of 4C\,+27.14 is consistent with \cite{2014ApJ...793..132S}. 

However, Eq.\,(\ref{tau}) is no more a good assumption because $T_{\rm s} \ll c_{\rm f}T_{\rm c}$ is not satisfied for 4C\,+27.14.  \citet{2003ApJS..145..329H} developed a 17-point observing pattern with the Arecibo telescope and used the least-squares procedure to measure the optical depth and the expected emission profile, which is the profile of \hi source that would be observed at the source position if the background continuum sources were absent. Sixteen off-source measurements and one on-source measurement were applied to each source. The off-source spectra were expressed in a Taylor series expansion of the expected profile and a small contribution from the source intensity attenuated by the optical depth. A least-squares fitting technique is then used to estimate the expected \hi emission and the optical depth.
Using this method, \citet{2014ApJ...793..132S} derived the expected \hi emission profile. The opacity spectrum was then obtained by subtracting the expected \hi emission spectrum from the on-source spectrum. 

We re-scale our optical depth profile in Figure~\ref{4C+27.14_fit} and Table ~\ref{absorption_table} using the data presented in \citet{2014ApJ...793..132S}. our measurements of the \hi absorption associated with 4C\,+27.14 give a flux density depth of $S_{\hi,\rm peak} = -11.11 \pm 0.34 \mJy$, an FWHM of $54.58 \pm 1.99 \kms$, and $\int\tau dv= 3.01 \pm 0.09 \kms$.

\subsection{New Absorbers}
\label{sec:New Absorbers}
Here we report two newly discovered absorbers. We name them by the bright radio sources located very close to the peak position of the absorption seen in our observation, NVSS\,J231240-052547 and NVSS\,J053118+315412. According to the NED and VLASS, these are the only radio sources found within 3 arc mins around these positions where \hi absorptions are detected. 

Is the absorption really produced along the line of sight towards the bright background radio source, or is produced against the blank sky or some fainter, fused sources within the beam? We note that the flux decrements for the two absorbers are quite large. Given that there are bright sources at the position where absorbers are detected, it is much more likely to obtain such a large decrement against a single, bright radio background source since only an absorber of a small area on that line of sight is required, while it is far more difficult to achieve such absorption against fainter, possibly multiple background sources within the same beam, which would require a much larger area for the absorber. 
Combined with the fact that in CRAFTS drift scan data, there is a tight negative relation between the intensity of the absorption signal and the angular distance between the beam centre and the radio sources mentioned above, these two radio sources are very likely related to the newly discovered \hi absorbers. Therefore, we shall proceed under the assumption that these two absorbers are indeed related to that strong radio sources, and this justifies our naming of the absorbers with the radio sources.

\subsubsection{NVSS J231240-052547}

\begin{figure*}
    \centering
    \includegraphics[width=0.33\textwidth]{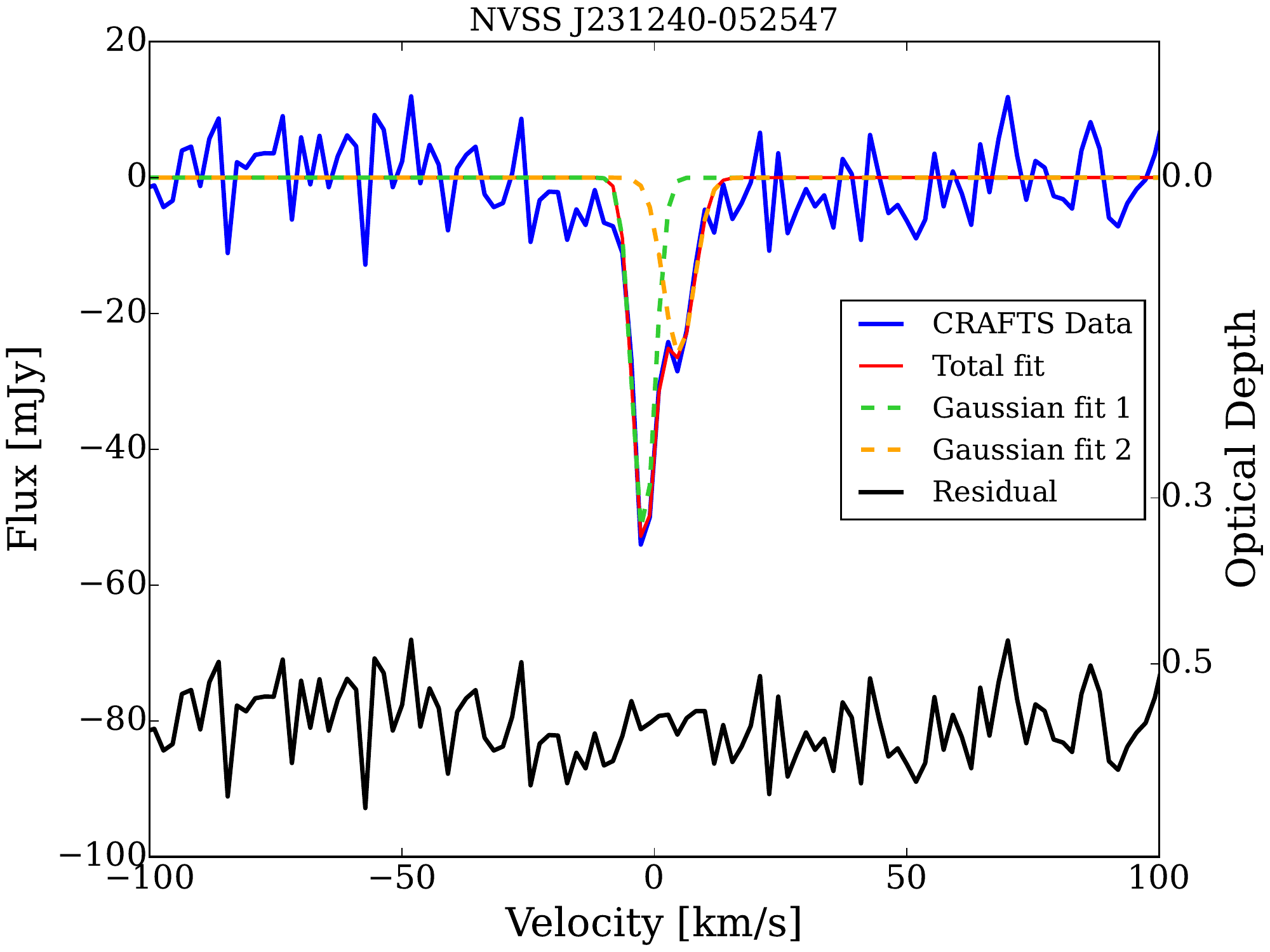}
    \includegraphics[width=0.33\textwidth]{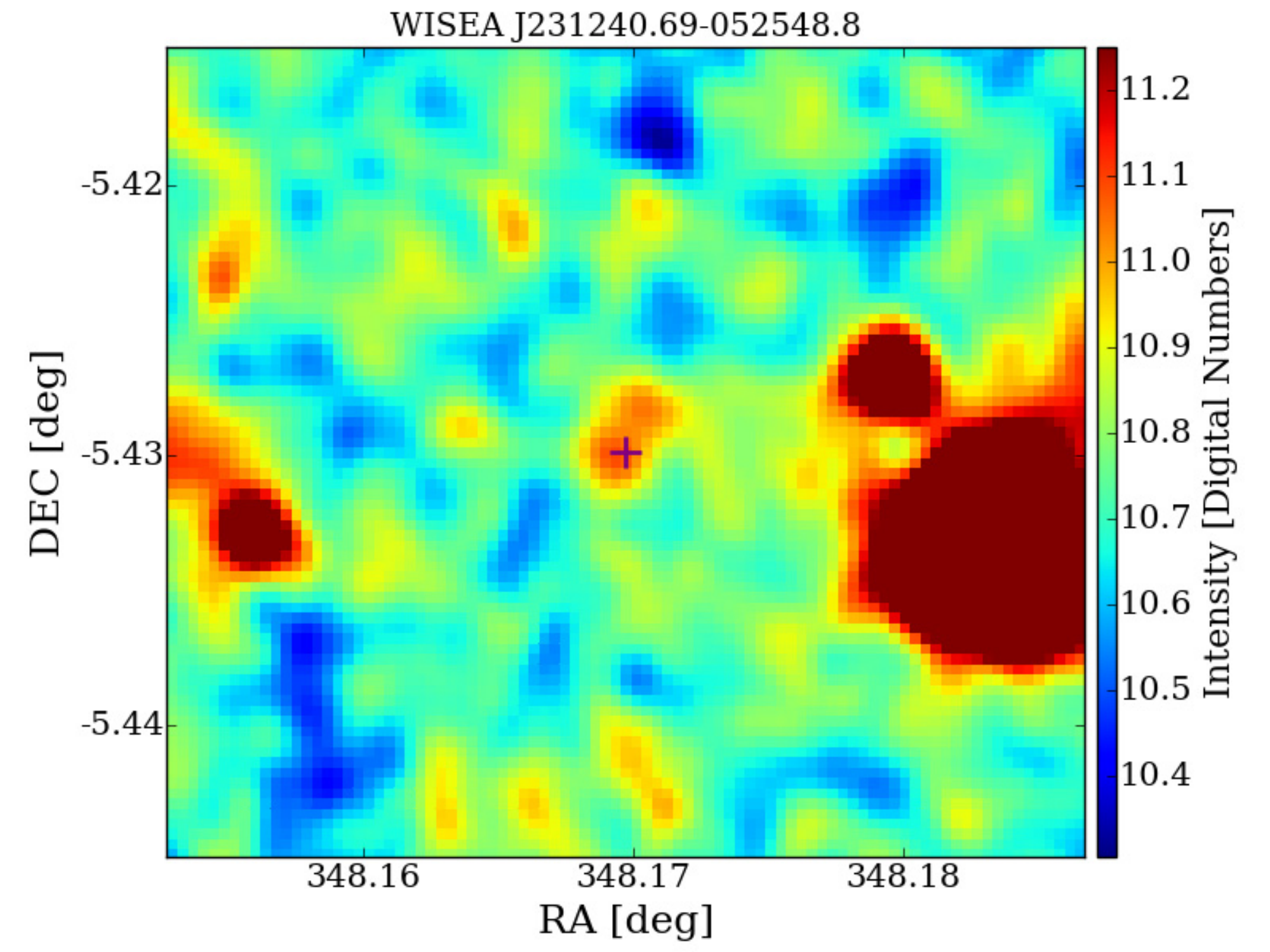}
    \includegraphics[width=0.33\textwidth]{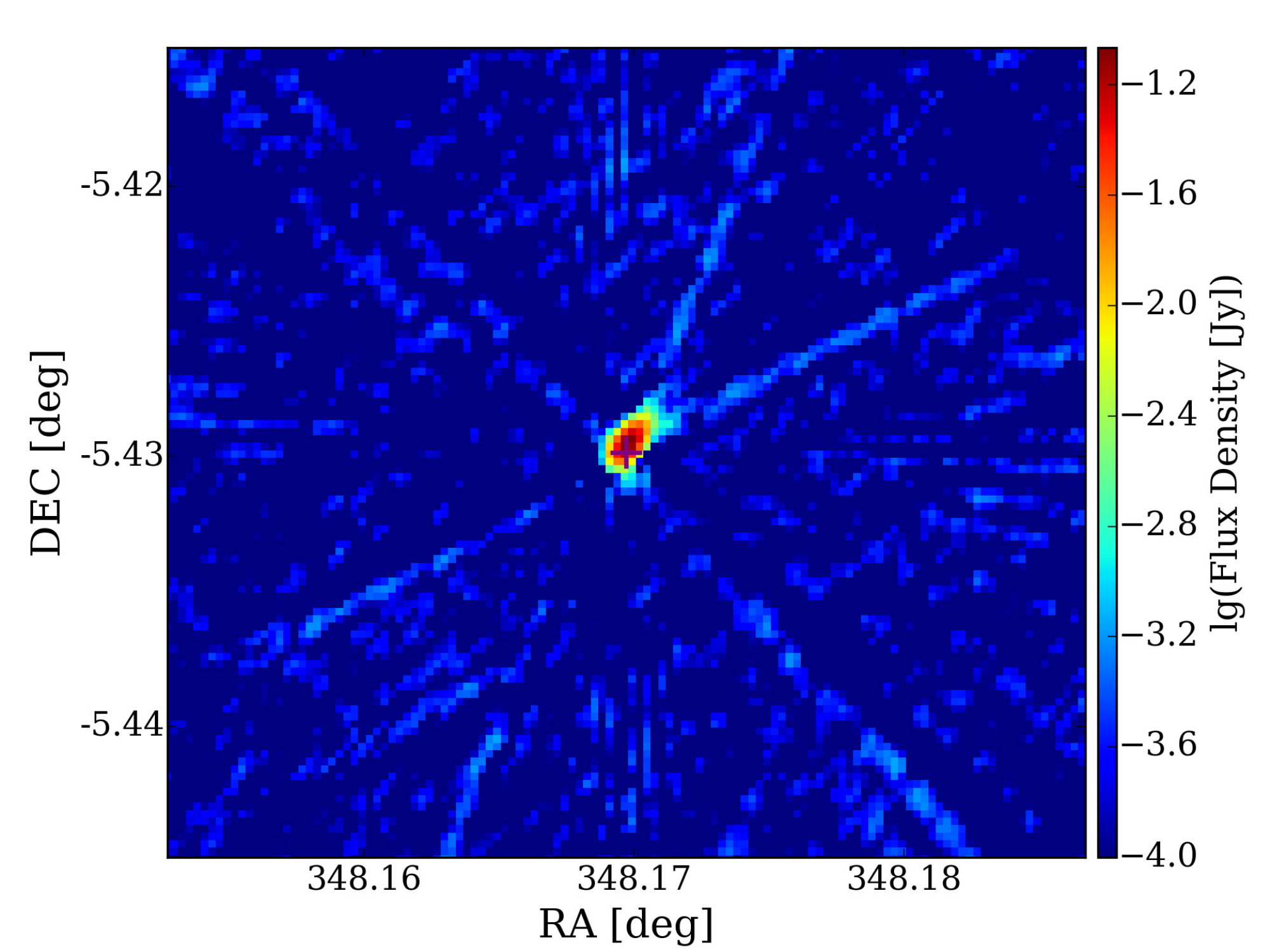}
    \caption{Absorption spectrum and images towards NVSS\,J231240-052547. Left panel: same as Figure~\ref{UGC00613_fit}, but for the \hi absorption feature towards NVSS\,J231240-052547.  Middle panel: image centred at NVSS\,J231240-052547 (or WISEA\,J231240.69-052548.8), using WISE data at 4.618 microns. Right panel: radio image centred at NVSS\,J231240-052547, using VLASS data. The straight cross shows the position of the radio source.}
    \label{NVSS_J231240-052547_fit}
\end{figure*}

\begin{figure*}
    \centering
    \includegraphics[width=0.33\textwidth]{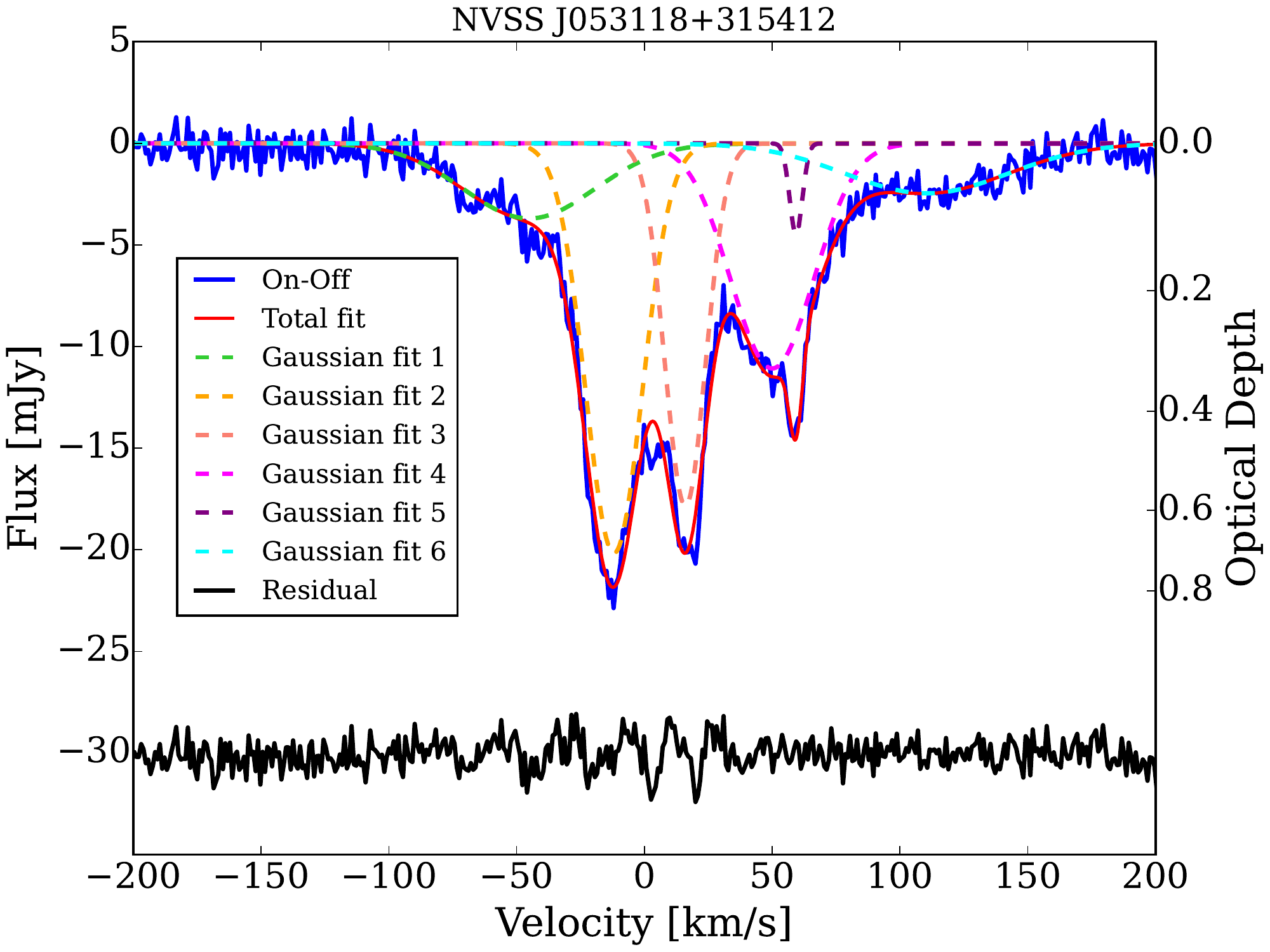}
    \includegraphics[width=0.33\textwidth]{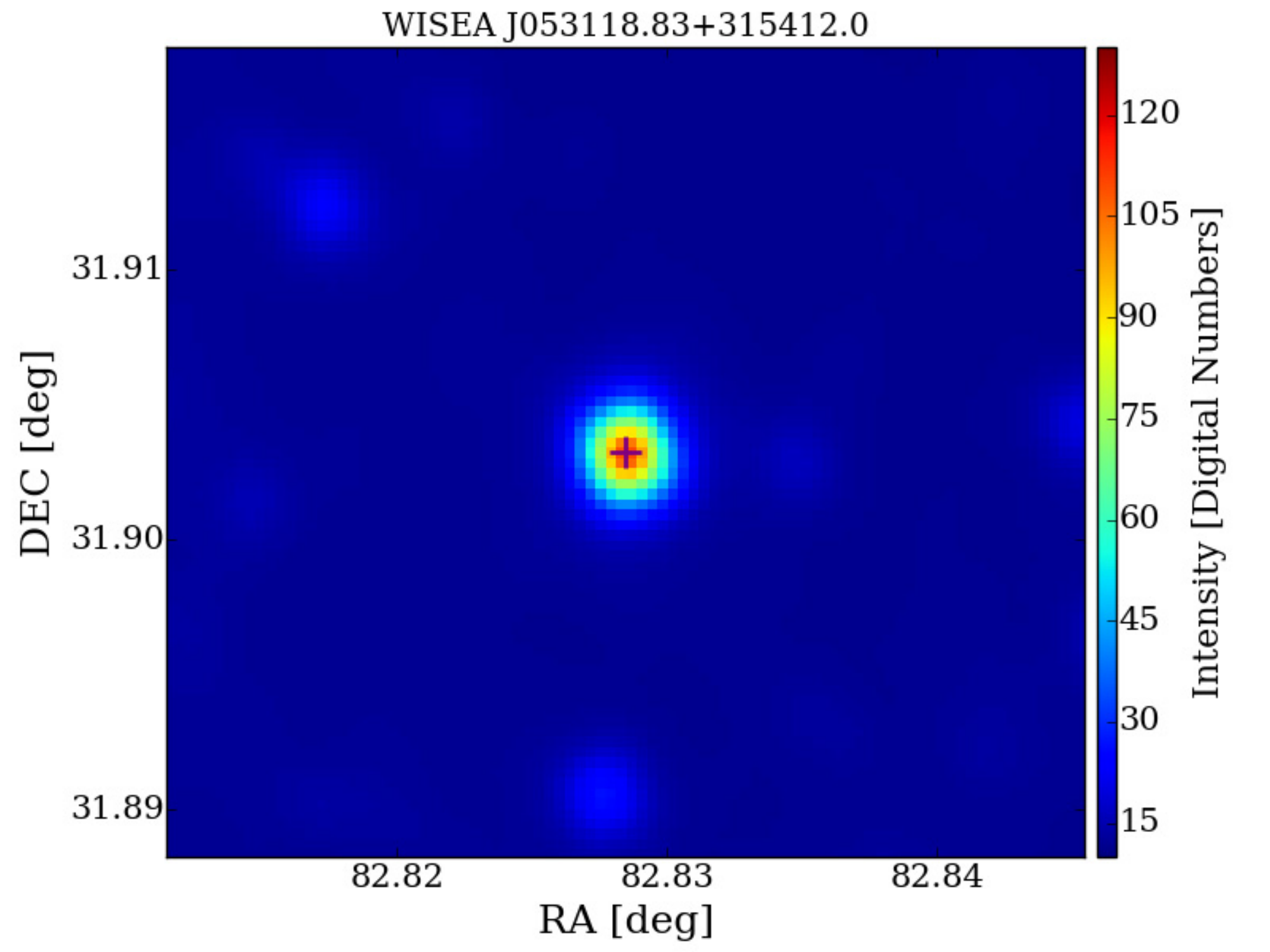}
    \includegraphics[width=0.33\textwidth]{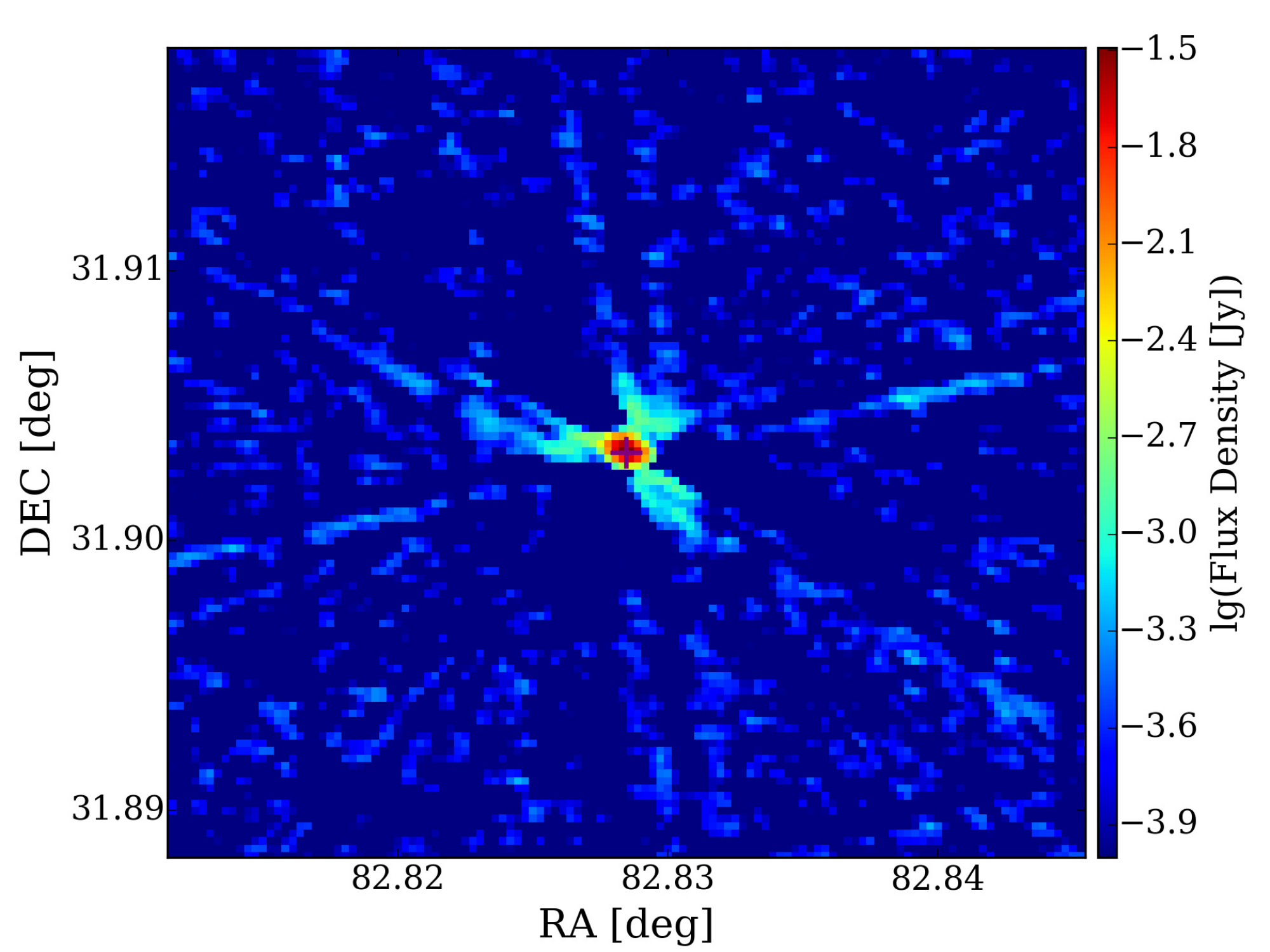}
    \caption{Absorption spectrum and images towards NVSS\,J053118+315412. Left panel: same as Figure~\ref{UGC00613_fit}, but for the \hi absorption feature of NVSS\,J053118+315412. Middle panel: image centred at NVSS\,J053118+315412 (or WISEA\,J053118.83+315412.0), using WISE data at 4.618 microns. Right panel: radio image centred at NVSS\,J231240-052547, using VLASS data. The straight cross shows the position of the radio source.}
    \label{NVSS_J053118+315412_fit}
\end{figure*}

The \hi absorption towards NVSS\,J231240-052547 is here reported for the first time. NVSS\,J231240-052547 is a radio source detected in the NRAO VLA Sky Survey with flux $S_{\rm 1.4GHz} = $ 181.9 mJy. We blindly detected  \hi absorption in beam 10, beam 4 and beam 14 in the CRAFTS drift-scan data. 

The absorption spectra towards NVSS\,J231240-052547 and its Gaussian fitting are shown in the upper panel of Figure~\ref{NVSS_J231240-052547_fit}. Two components are found in the absorption feature, and some basic physical parameters are given in Table ~\ref{radiosource_table} and Table ~\ref{absorption_table}. Under the assumption $T_{\rm s} \ll c_{\rm f}T_{\rm c}$, the \hi absorption towards NVSS\,J231240-052547 has a flux density depth of $S_{\hi,\rm peak} = -54.04 \pm 6.02 \mJy$, an FWHM of $9.12 \pm 1.01 \kms$, $\int\tau dv= 3.01 \pm 0.36 \kms$ and $N_{\hi}= 0.05 T_{s} \times 10^{20} \cm^{-2} \K^{-1}$.

At this same position, a mid-infrared source WISEA\,J231240.69-052548.8 has been detected by the Wide-field Infrared Survey Explorer (WISE, \citealt{2013wise.rept....1C}), and 
cross-identified as the radio source NVSS\,J231240-052547 in the NASA/IPAC Extragalactic Database. The left and right panels of Figure\,\ref{NVSS_J231240-052547_fit} show the image centred at NVSS\,J231240-052547 (or WISEA\,J231240.69-052548.8), constructed using data at 4.618 microns (W2) from WISE and radio data at S-band from the Karl G. Jansky Very Large Array Sky Survey (VLASS) \citep{2020PASP..132c5001L}, respectively. The straight cross marks the position of NVSS\,J231240-052547. The WISE magnitudes for WISEA\,J231240.69-052548.8 are presented in Table ~\ref{wise_table}. The W1-W2 colour of value 0.8 is often used as a simple mid-infrared colour criterion for AGN candidates \citep{2012ApJ...753...30S}. The W1-W2 colour of 0.827 indicates that is an AGN candidate. Combining with the W2-W3 value of 3.752 mag, this source lies in the intersecting region of Seyferts and the Ultra-luminous Infrared Galaxies (ULIRGs) in the WISE colour–colour diagram \citep{2010AJ....140.1868W}. It is probably associated with the background radio source, though there is also the chance that it is located at a lower redshift, and in that case, the absorber could be associated with it.

\begin{figure}
    \centering
    \includegraphics[width=0.33\textwidth]{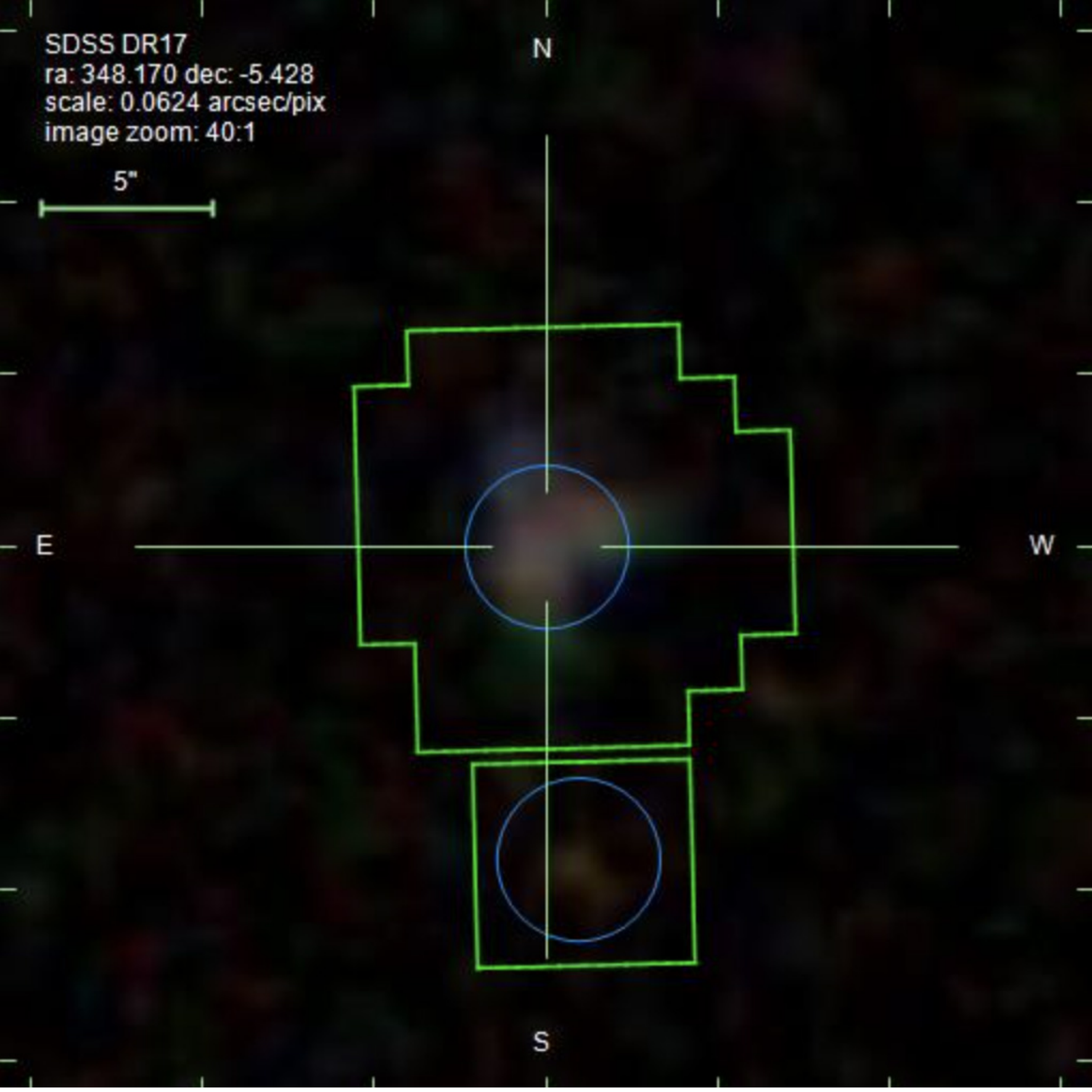}
    \caption{The optical image for SDSS J231240.73-052541.1 (central galaxy). Blue circles mark the photometric objects, and the green squares depict the SDSS outlines.}
    \label{SDSS_J231240.73-052541.1}
\end{figure}

There is also an optical counterpart located at the same position. 
According to the information shown in SDSS DR17 SkyServer website\footnote{\url{https://skyserver.sdss.org/dr17/en/home.aspx}}, WISEA\,J231240.69-052548.8 is cross-identified as SDSS J231240.73-052541.1 in the optical band. SDSS J231240.73-052541.1 is a galaxy with a Petrosian radius of 7.36 arcsec, with a photometric redshift of $photoZ = 0.091 \pm 0.0364$. The magnitudes of SDSS J231240.73-052541.1 are $u=20.66 \pm 0.20, g=19.68 \pm 0.03, r=19.37 \pm 0.03, i=19.16 \pm 0.05$ and $z=19.09 \pm 0.16$. The absorption line towards NVSS\,J231240-052547 is located at $z \sim 0.063$, which is within the error box of the photometric redshift of SDSS J231240.73-052541.1. It could be associated with the absorber, though spectroscopic observations are needed to give a definite answer. The optical image for SDSS J231240.73-052541.1 is extracted from the SDSS SkyServer website and is shown as the central galaxy in Figure~\ref{SDSS_J231240.73-052541.1}.

\subsubsection{NVSS J053118+315412}

The other new absorber is found in the line of sight towards NVSS\,J053118+315412. NVSS\,J053118+315412 has been detected 
at 1.4\,GHz in NVSS with flux $S_{\rm 1.4GHz} = $ 40.2 mJy. Its \hi absorption is only detected in beam 12 of the drift scan covering Declination of +31d57m00s. In order to verify this detection, it was observed in follow-up observation using ON-OFF tracking mode. The absorption spectra towards NVSS\,J053118+315412 and its Gaussian fitting are shown in the upper panel of Figure\,\ref{NVSS_J053118+315412_fit}. Six components are found in the absorption feature, and some basic parameters are given in Table ~\ref{radiosource_table} and Table ~\ref{absorption_table}. Under the assumption $T_{\rm s} \ll c_{\rm f}T_{\rm c}$, the measurements of the \hi absorption towards NVSS\,J053118+315412 give a flux density depth of $S_{\hi,\rm peak} = -21.85 \pm 0.77 \mJy$, an FWHM of $89.53 \pm 2.85 \kms$, $\int\tau dv= 55.81 \pm 2.23 \kms$ and $N_{\hi}= 1.016 T_{s} \times 10^{20} \cm^{-2} \K^{-1}$. 

The radio source has also been cross-identified to a source in the 2MASS Two Micron All Sky Survey (2MASS, \citealt{2006AJ....131.1163S}) and one in the mid-infrared band by WISE by the NASA/IPAC Extragalactic Database. 
We show the image centred at NVSS\,J053118+315412 (or WISEA\,J053118.83+315412.0), constructed using infrared data at 4.618 microns (W2) from WISE and radio data at S-band from VLASS, in the middle and right panel of Figure~\ref{NVSS_J053118+315412_fit}. The straight cross marks the position of NVSS\,J053118+315412. The WISE magnitudes for WISEA\,J053118.83+315412.0 are presented in Table ~\ref{wise_table}. The W1-W2 colour is 1.630, which is higher than the value of 0.8 often used to select AGN candidates and implies that the mid-IR emission comes mainly from the AGN. Combined with the W2-W3 value of 3.598 mag, NVSS\,J053118+315412 lies in the region of QSOs in the WISE colour–colour diagram. The 2MASS counterpart has J-band[\SI{1.235}{\micro\metre}], H-band[\SI{1.662}{\micro\metre}] and K$_{\rm s}$-band[\SI{2.159}{\micro\metre}] magnitudes 15.180 $\pm$ 0.043, 13.943 $\pm$ 0.044 and 13.219 $\pm$ 0.028.

We should note that although the velocity of the absorption is $\sim 1500 \kms$ higher than the velocity of WISEA\,J053118.83+315412.0 (or NVSS\,J053118+315412) as shown in the NED database (given by 2MASS Redshift Catalog, \citealt{2014ApJS..210....9B}), this velocity difference is smaller than the typical precision of 2MASS redshift estimates ($\sim$ 12$\%$), indicating the velocity of absorption is similar to that of NVSS\,J053118+315412.

The complicated structures of the absorption profile indicate the interaction between ISM and host AGN. NVSS\,J053118+315412 exhibits a symmetric absorption spectrum (Gaussian components 2 and 3) relative to the systemic velocity of its host galaxy, suggesting that the \hi gas traces a regular rotating structure. On the other hand, asymmetric absorption wings (Gaussian components 1 and 6) in the absorption profile are generally indicative of unsettled gas structures \citep{2018A&ARv..26....4M}, such as gas outflows that are propelled by the radio jet or tidal gas streams. Additionally, the redshifted absorption components (Gaussian components 4 and 5) could be evidence of accretion onto the SMBH \citep{2010AJ....139...17A,2014A&A...571A..67M}. However, precise identification of infalling \hi can only be achieved via high-resolution observations.

The position of NVSS\,J053118+315412 is outside the footprint of SDSS and other optical surveys. We are applying the telescope time to obtain the optical spectrum of the potential foreground and background sources, which will be investigated in future work.


\subsection{Mid-infrared Colour}

\begin{figure}
    \centering
    \includegraphics[width=0.48\textwidth]{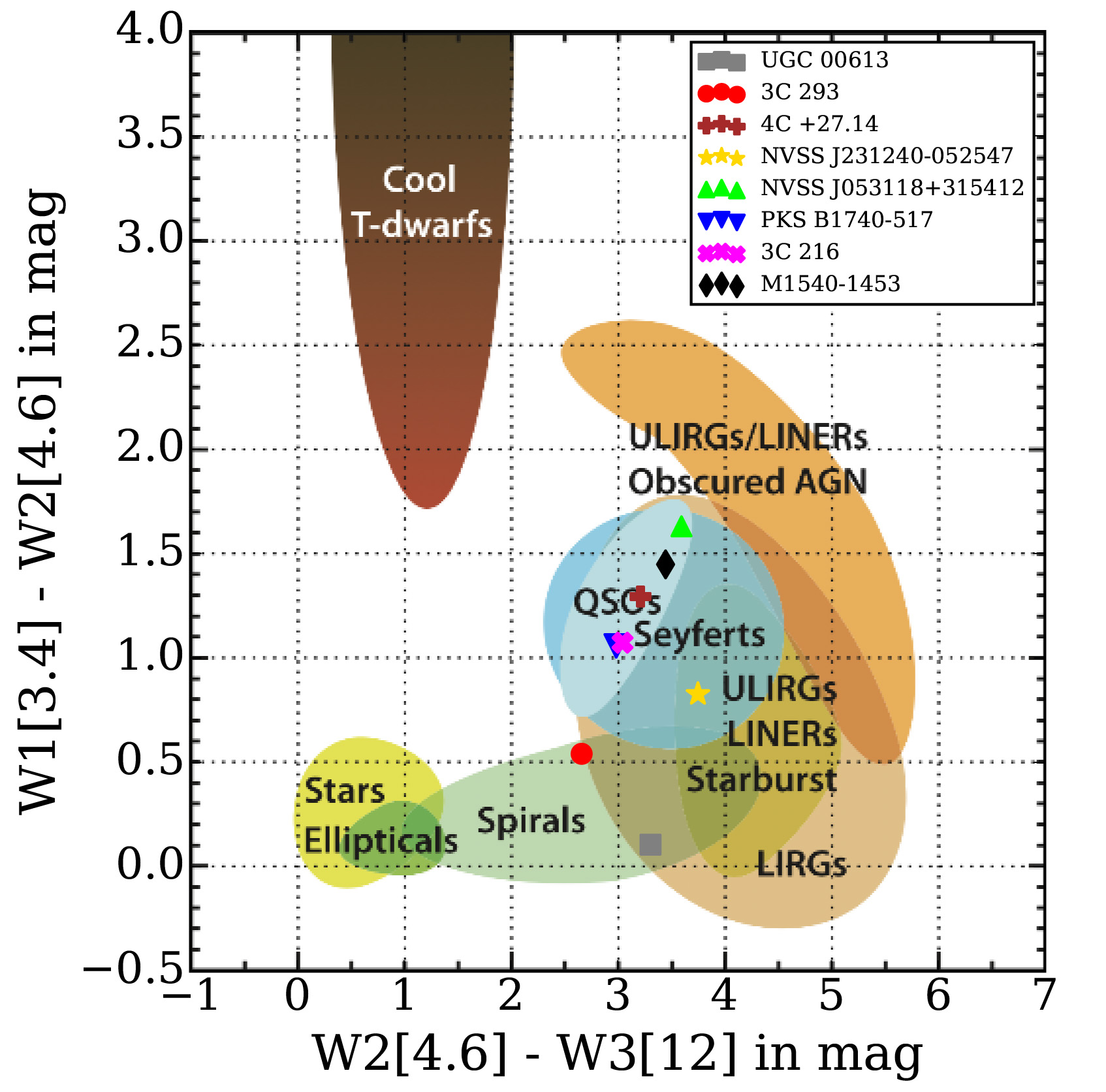}
    \caption{The WISE colour-colour diagram. The WISE colour for the WISE counterparts of UGC\,00613, 3C\,293, 4C\,+27.14, NVSS\,J231240-052547 , NVSS\,J053118+315412 and 3 associated absorption (PKS\,B1740-517, 3C\,216 and M1540-1453) from blind surveys.}
    \label{WISE_color_color}
\end{figure}

The mid-infrared wavelength colour information from the WISE survey for the WISE counterparts of UGC\,00613, 3C\,293, 4C\,+27.14, NVSS\,J231240-052547 and NVSS\,J053118+315412 is shown in Table ~\ref{wise_table}. Figure\,\ref{WISE_color_color} presents the WISE colour-colour diagram from \citet{2010AJ....140.1868W}, overlaid with the WISE counterparts of UGC\,00613, 3C\,293, 4C\,+27.14, NVSS\,J231240-052547, NVSS\,J053118+315412 as well as 3 associated absorption (PKS\,B1740-517, 3C\,216 and M1540-1453) detected in previous blind surveys. In the WISE colour-colour diagram, extragalactic objects are grouped into different classes based on their mid-infrared colour signatures.

From the \hi absorption experiment towards radio AGNs classified as either low-excitation radio galaxies (LERGs) or high-excitation radio galaxies (HERGs), \citet{2020MNRAS.494.5161C} confirmed that compact radio AGNs with WISE colour W2-W3 > 2 have higher detection rates compared to those with W2-W3 < 2 \citep{2017MNRAS.465..997C}. This is because the sources with W2-W3 > 2 are typical of gas-rich systems.

All the 4 associated \hi absorption in our work and 3 associated absorption (PKS\,B1740-517 \citep{2015MNRAS.453.1249A}, 3C\,216 \citep{2020MNRAS.498..883G} and M1540-1453 \citep{2021ApJS..255...28G}) from previous blind surveys reside in the W2-W3 > 2 region, which supports the conclusion given by \citet{2020MNRAS.494.5161C}. More importantly, this associated absorption sample is obtained from a blind survey without any prior selection, indicating a universal relation between mid-infrared wavelength colour and detection rate for all radio sources. However, limited by the small number of unbiased \hi absorption samples, it is necessary to use more \hi absorbers from blind surveys to verify the relationship between mid-infrared colour and detectability for associated and intervening absorption.

\begin{table*}
        \fontsize{9}{10}\selectfont
	\centering
	\caption{WISE magnitudes for the WISE counterparts of two new absorbers NVSS\,J231240-052547 and NVSS\,J053118+315412. Uncertainty is not available if magnitude is a 95\% confidence upper limit (flux measurement has $\mathrm{S/N}$ < 2).}
	\label{wise_table}
	\begin{tabular}{|c|c|c|c|c|c|}
		\hline
		Radio Source & WISE Counterpart & W1[\SI{3.4}{\micro\metre}] &  W2[\SI{4.6}{\micro\metre}] & W3[\SI{12.1}{\micro\metre}] &  W4[\SI{22.2}{\micro\metre}] \\
		\hline
          UGC\,00613 & WISEA\,J005924.42+270332.6 & 11.226 $\pm$ 0.024 & 11.120 $\pm$ 0.020 & 7.812 $\pm$ 0.017 & 5.695 $\pm$ 0.039 \\
          3C\,293 & WISEA\,J135217.81+312646.6 & 11.328 $\pm$ 0.023 & 10.787 $\pm$ 0.021 & 8.124 $\pm$ 0.018 & 5.962 $\pm$ 0.038 \\
          4C\,+27.14 & WISEA\,J045956.08+270602.1 & 10.344 $\pm$ 0.022 & 9.050 $\pm$ 0.019 & 5.836 $\pm$ 0.015 & 3.191 $\pm$ 0.019 \\
		 NVSS\,J231240-052547 & WISEA\,J231240.69-052548.8 & 16.937 $\pm$ 0.162 & 16.110 & 12.358 & 9.016 \\
		 NVSS\,J053118+315412 & WISEA\,J053118.83+315412.0 & 11.607 $\pm$ 0.024 & 9.977 $\pm$ 0.021 & 6.379 $\pm$ 0.014 & 3.613 $\pm$ 0.020 \\
		\hline
	\end{tabular}
\end{table*}

\begin{table*}
        \fontsize{9}{10}\selectfont
	\centering
	\caption{Some basic physical parameters of radio source UGC\,00613, 3C\,293, 4C\,+27.14, NVSS\,J231240-052547 and NVSS\,J053118+315412. The values are retrieved from NASA/IPAC Extragalactic Database \citep{1991ASSL..171...89H}. The redshifts of NVSS\,J231240-052547 and NVSS\,J053118+315412 are SDSS and 2MASS photometric estimates for the nearest optical galaxies cross-matched to the radio sources.}
	\label{radiosource_table}
	\begin{tabular}{|c|c|c|c|c|c|}
		\hline
		Radio Source & Source Type & RA(J2000) & DEC(J2000) & cz (\kms) & $S_{\rm 1.4GHz}$ (mJy)\\
		\hline
		 UGC\,00613 & Diffuse Radio Source & 00h 59m 24.42s & +27d 03m 32.6s & 13770.07 $\pm$ 23.08 & 112.6 $\pm$ 4.0\\
		 3C\,293 & Extended Radio Source & 13h 52m 17.842s & +31d 26m 46.50s & 13547.51 $\pm$ 3.00 & 4690 \\
		 4C\,+27.14 & Compact Radio Source & 04h 59m 56.08s & +27d 06m 02.10s & 18347.30 $\pm$ 0  & 857 \\
		 NVSS\,J231240-052547 & Radio Source & 23h 12m 40.73s & -05d 25m 47.50s & 27281.11 $\pm$ 10912.44 & 181.9 $\pm$ 5.5 \\
		 NVSS\,J053118+315412 & Radio Source & 05h 31m 18.83s & +31d 54m 11.7s & 18359.29 $\pm$ 0 & 40.2 $\pm$ 1.3 \\
		\hline
        
		\hline
	\end{tabular}
\end{table*}

\begin{table*}
        \fontsize{8}{10}\selectfont
	\centering
	\caption{Some basic physical parameters of FAST observations of the \hi absorption profiles for UGC\,00613, 3C\,293 and 4C\,+27.14. The estimations of $\tau$ for 4C\,+27.14 are re-scaled use the data in \citet{2014ApJ...793..132S}.}
	\label{absorption_table}
	\begin{tabular}{|c|c|c|c|c|c|c|c|c|c|}
		\hline
		Radio Source & absorption type & Component & cz$_{\rm peak}$ & FWHM & $S_{\hi, \rm peak}$ & $\int S_{\hi}dv$ & $\tau_{\rm peak}$ & $\int\tau dv$ & $N_{\hi}$\\
		 & & & (\kms) & (\kms) & (mJy) & (mJy\kms) &  & (\kms) & (10$^{20}$cm$^{-2}$K$^{-1}$)\\
		\hline
		  & & 1 & 13987.58 & 40.34 & -23.50 & -1016.51 & 0.23 & 9.78 & 
            0.18$T_{s}$\\
		UGC\,00613  & associated & 2 & 13957.59 & 27.03 & -64.31 & 
            -1826.04 & 0.84 & 21.17 & 0.39$T_{s}$\\
		& & Total & 13957.59 & 31.74 & -69.46 & -2842.55 & 0.96 & 32.75 
            & 0.60$T_{s}$\\
		  \hline
		& & 1 & 13483.28 & 29.93 & -168.09 & -5354.07 & 0.036 & 1.16 & 
            0.02$T_{s}$\\
		3C\,293 & associated & 2 & 13458.66 & 80.60 & -141.90 & 
            -12164.45 & 0.031 & 2.62 & 0.05$T_{s}$\\
		& & 3 & 13354.95 & 110.95 & -104.45 & -12331.98 & 0.023 & 2.65 & 
            0.05$T_{s}$\\
		& & Total & 13481.49 & 61.54 & -283.33 & -29850.50 & 0.062 & 
            6.47 & 0.12$T_{s}$\\
		  \hline
		  & & 1 & 18776.96 & 42.09 & -7.98 & -357.40 & 0.87 & 38.72 &
            0.7$T_{s}$\\
		4C\,+27.14 & associated & 2 & 18755.13 & 155.74 & -3.32 &
            -550.31 & 0.36 & 59.51 & 1.08$T_{s}$ \\
		& & Total & 18776.96 & 54.58 & -11.11 & -907.70 & 1.21 & 98.37 & 
            1.79$T_{s}$ \\
	 	\hline
		  & & 1 & 18790.52 & 5.20 & -53.17 & -294.82 & 0.33 & 1.82 & 
            0.03$T_{s}$\\
		  NVSS\,J231240-052547 & unknown & 2 & 18797.77 & 7.15 & 
            -26.10 & -198.50 & 0.15 & 1.15 & 0.02$T_{s}$ \\
		& & Total & 18797.77 & 9.12 & -54.04 & -493.32 & 0.34 & 3.01 & 
            0.05$T_{s}$ \\
		  \hline
            & & 1 & 19899.88 & 60.81 & -3.70 & -239.20 & 0.097 & 6.19 & 0.113$T_{s}$\\
            & & 2 & 19933.06 & 26.08 & -20.17 & -559.72 & 0.702 & 17.53 & 0.319$T_{s}$\\
            & & 3 & 19960.99 & 18.78 & -17.84 & -356.73 & 0.591 & 10.81 & 0.197$T_{s}$\\
            NVSS\,J053118+315412 & associated & 4 & 19994.99 & 38.18 & -11.08 & -450.01 & 0.324 & 12.55 & 0.229$T_{s}$\\
            & & 5 & 20004.30 & 5.94 & -4.47 & -28.25 & 0.118 & 0.74 & 0.013$T_{s}$\\
            & & 6 & 20055.36 & 75.00 & -2.46 & -196.30 & 0.063 & 5.02 & 0.091$T_{s}$\\
            & & Total & 19944.95 & 89.53 & -21.85 & -1830.21 & 0.790 & 55.81 & 1.016$T_{s}$\\
		\hline
	\end{tabular}
\end{table*}

\subsection{Spurious Detections}

\begin{figure}
\begin{multicols}{2}
\includegraphics[width=4.5cm,height=3.75cm]{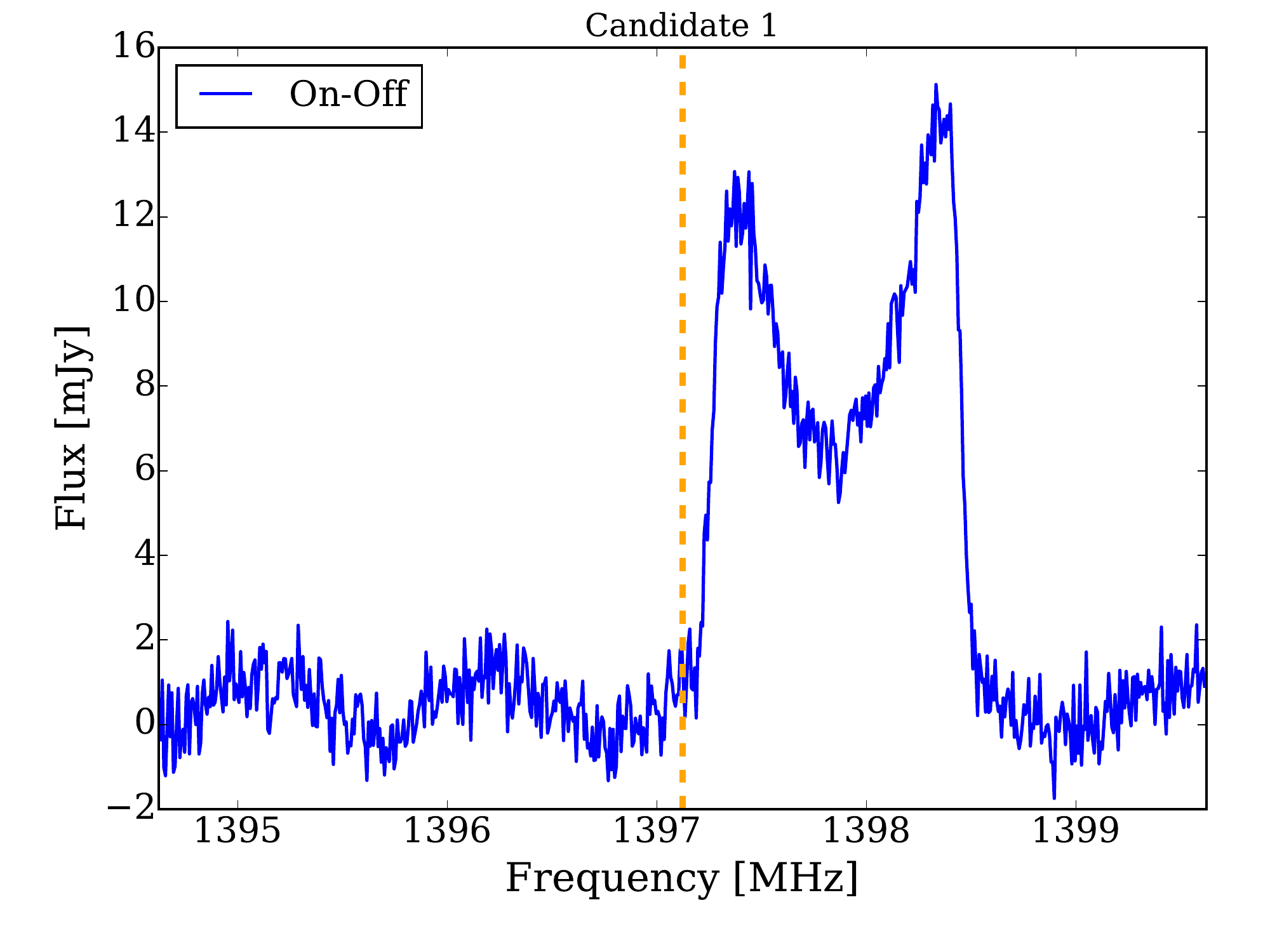}\par
\hspace*{-0.3cm}
\includegraphics[width=4.5cm,height=3.75cm]{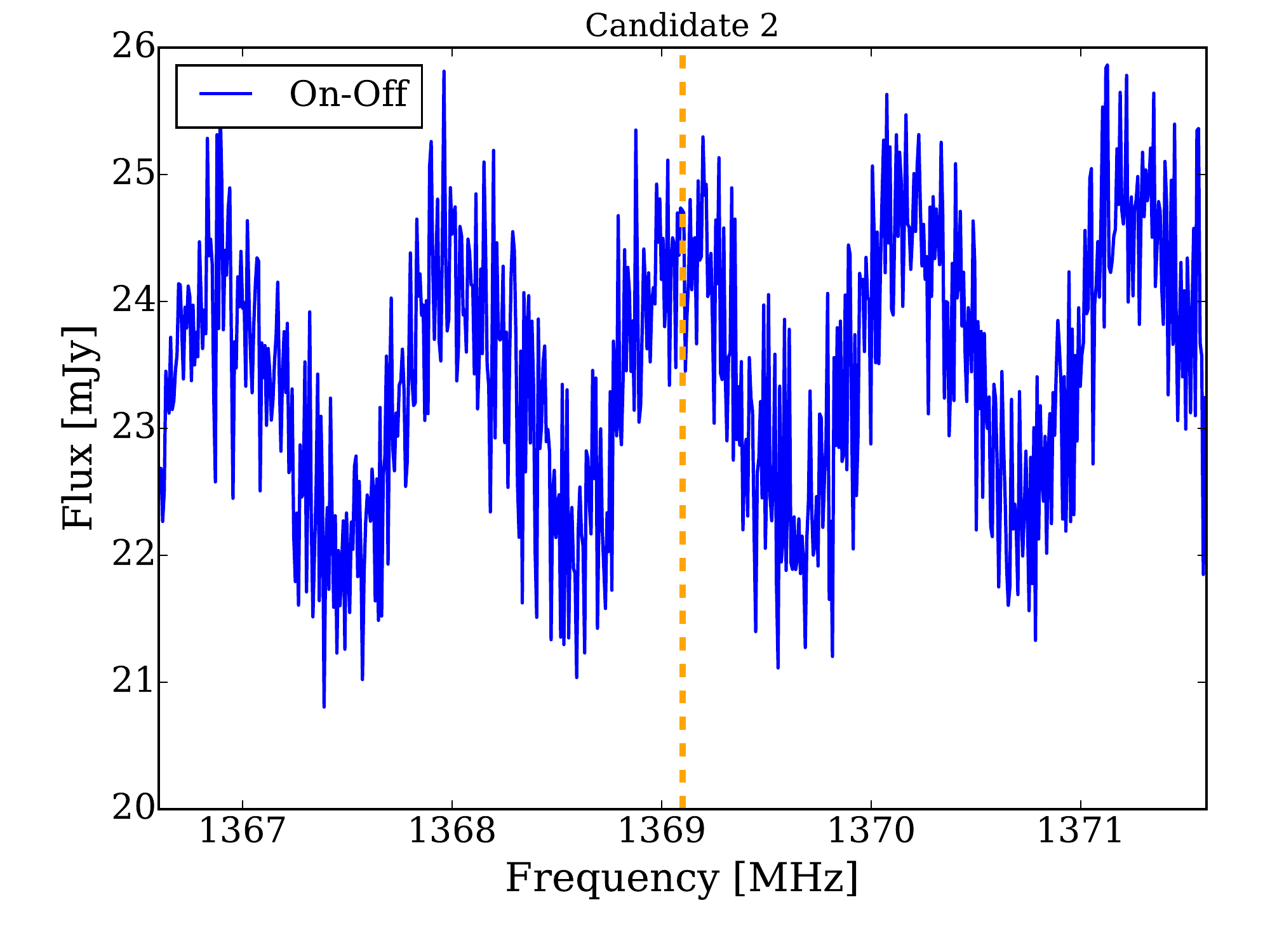}\par
\end{multicols}
\vspace*{-1.05cm}
\begin{multicols}{2}
\includegraphics[width=4.5cm,height=3.75cm]{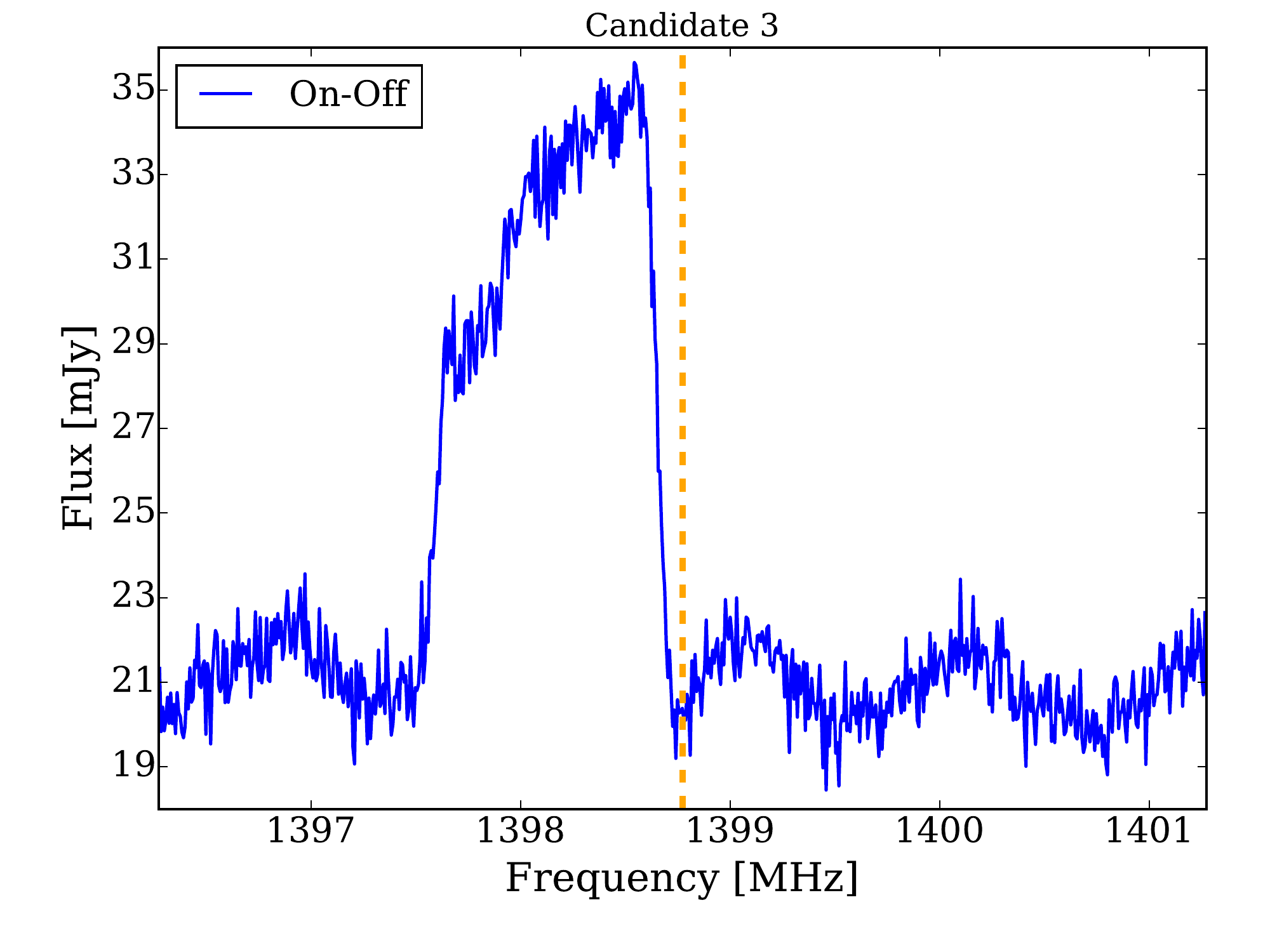}\par
\hspace*{-0.3cm}
\includegraphics[width=4.5cm,height=3.75cm]{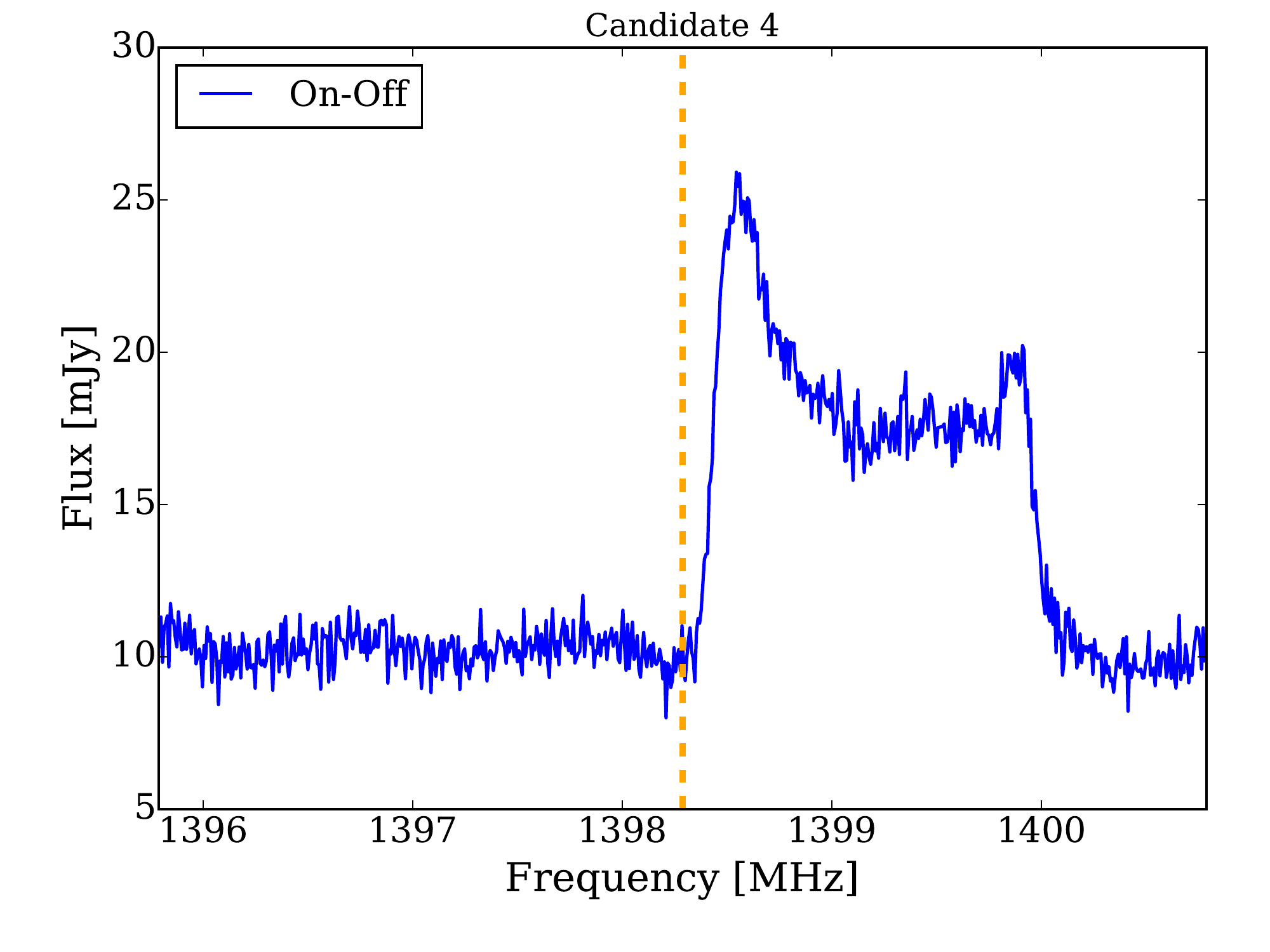}\par
\end{multicols}
\vspace*{-1.05cm}
\begin{multicols}{2}
\includegraphics[width=4.5cm,height=3.75cm]{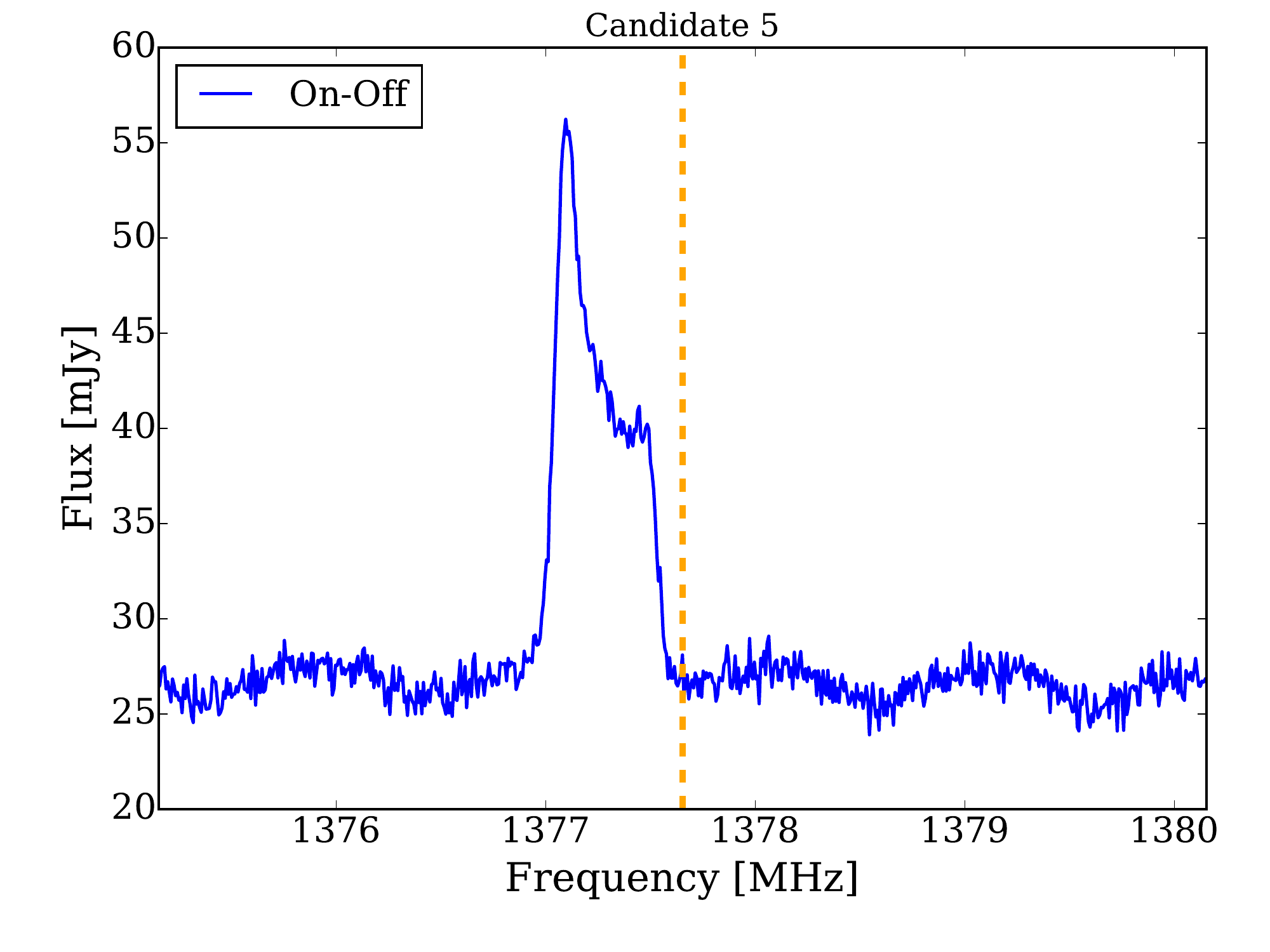}\par
\hspace*{-0.3cm}
\includegraphics[width=4.5cm,height=3.75cm]{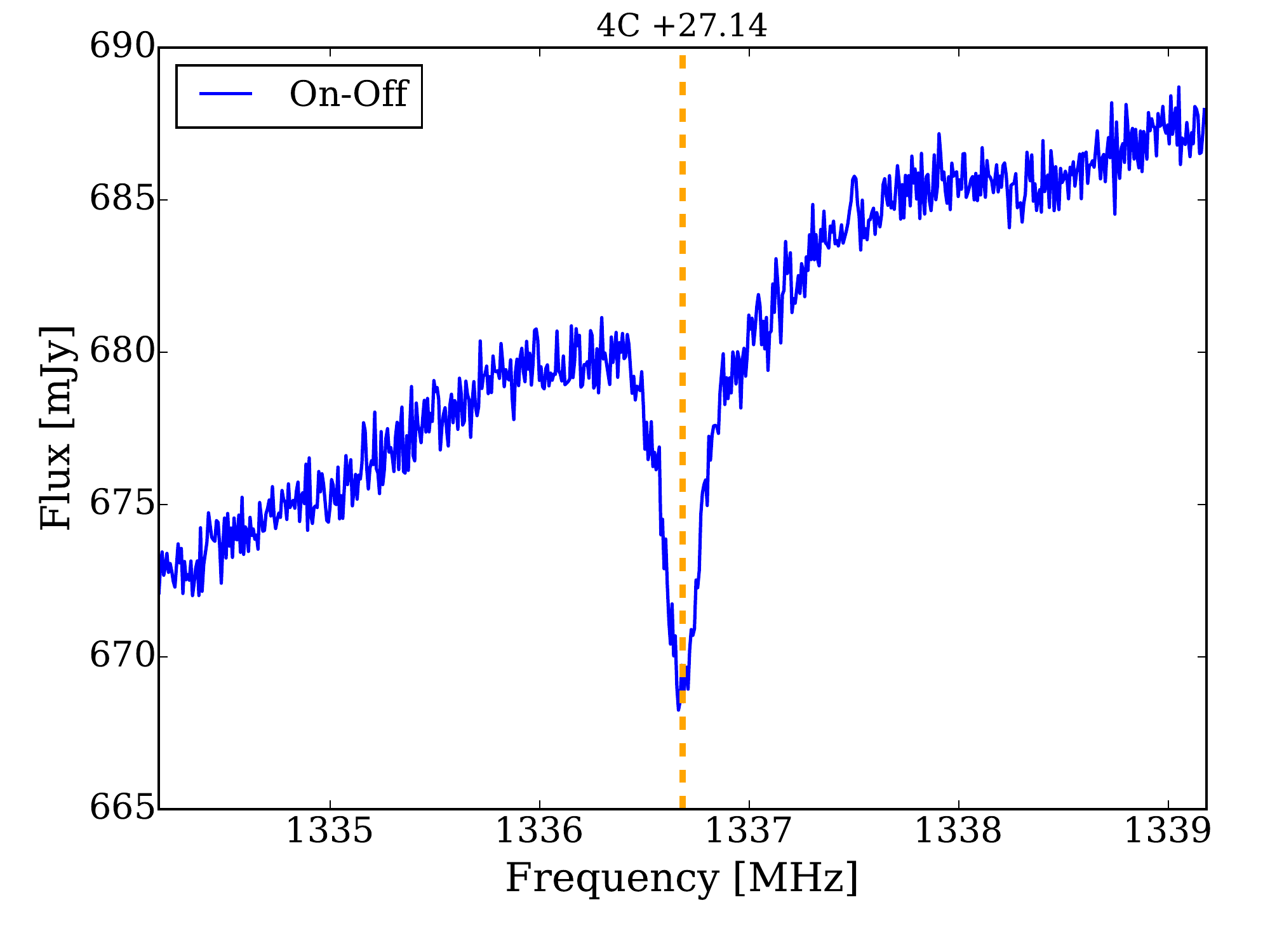}\par
\end{multicols}
\vspace*{-0.8cm}
\caption{The spectra for the candidates 1, 2, 3, 4, 5 and 4C\,+27.14 from the follow-up observation. The orange-dashed lines show the central frequency position of the \hi candidates found in CRAFTS searching (see Figure~\ref{candidates_plot}).}
\label{spectra_followup}
\end{figure}

The other 5 candidates in the follow-up observations are also checked and found to be not true \hi absorptions. The spectra of these sources, together with 4C\,+27.14, are shown in Figure\,\ref{spectra_followup}. The spectra are obtained by subtracting the source off spectra from the source on spectra and are calibrated with 3C 48. The orange-dashed lines show the central frequency position of the absorber candidates (see Figure~\ref{candidates_plot} for comparison). Compared with the spectra in CRAFTS (Figure\,\ref{candidates_plot}), we found that Candidate 1, 3, 4 and 5 arise from the combination of the fluctuations of the bandpass and the \hi emission, while Candidate 2 comes from the fluctuations of the bandpass. 

Although there are significant \hi emission profiles in the source ON - source OFF spectra from follow-up observation, in the CRAFTS data, they are mixed with the bandpass fluctuations induced by standing wave and chronically present RFI, making it difficult to distinguish the \hi emission from the bandpass ripples. The standing wave ripples of the FAST have $c/2f \sim 1.1 \MHz$ (where c denotes the speed of light and $f \sim 138$ m is the FAST focal length), which is not far from the width of \hi emission. 

The spurious absorbers such as Candidate 1, 3, 4 and 5 are the results of the interplay between \hi mission, RFI and bandpass ripples, such spurious absorption features can survive the multi-beam cross-correlation selection and can not be easily excluded without further observation (Section\,\ref{sec:selection}).

\section{Discussions}
\label{sec:Discussion}

\subsection{Correction for Standing Waves}

There have been attempts to suppress the standing waves in radio data. \citet{2019arXiv190911732K} described techniques for modelling and removing the reflections and standing waves in the HERA (the Hydrogen Epoch of Reionization Array) data\citep{2020ApJ...888...70K}. They showed that by combining reflection calibration and cross-coupling subtraction, standing waves could be suppressed down to the integrated noise floor. \citet{2021RAA....21...59L} recognized the standing waves in the Tianlai cylinder array and found the oscillating modulations in the frequency spectrum could be mitigated by removing the corresponding components in the Fourier space. Tests are also being made on correcting the FAST standing waves. By fitting or extracting peaks in Fourier space caused by standing waves components with the period of $\sim 1.1\,$MHz, we can remove them and bring the level of fluctuation down to below 1\,mJy or even lower (W. Yang et al, in preparation). A careful study of the small-scale fluctuations may help improve the calibration. However, the standing waves do vary with time, making it difficult to remove their effect completely.

On the other hand, only a one-pass drift scan has been carried out to date. In the near future, after the 2-pass drift scan is finished, more time-varied RFI and other fluctuations can be excluded by comparing the data covering the same sky at different times.

\subsection{Information from Neighboring Beams}
With so many spurious signals originating from the bandpass fluctuations and RFI being selected as preliminary candidates, it is of vital importance to find a valid and easy-to-use method to distinguish the intrinsic \hi absorption from man-made signals. Fortunately, the 19 beams of the L-band receiver and the drift-scan survey strategy provide an effective solution (described in Section~\ref{sec:selection}). The unique design of FAST makes it a powerful instrument for \hi absorption searching. Considering that only $\sim$ a half of the frequency band from a small part of CRAFTS observable sky has been searched, we expect a larger number of \hi absorbers could be detected with the  2-pass full survey.

In the blind search presented in this work, we run our pipeline in the data from every single beam separately. For drift scan at the meridian circle, $12/\cos\delta$ seconds integration time is obtained in each beam-size pixel. Considering the drift-scan strategy and the positions of the 19 beams, there are overlaps among nearby beams. About $24/\cos\delta$ seconds integration time can be obtained on each beam-size pixel \citep{2020MNRAS.493.5854H} if we combine the data from the 19 beams together. All of the data from the scans of the 19 beams can be utilized fully without losing information by using the linear map-making method \citep{1997ApJ...480L..87T}. It is a well-known technique for turning time-ordered raw data into estimators of the true sky. We have developed an effective map-making pipeline for FAST to combine the data from the 19 beams and make the high signal-to-noise ratio map (Y. Li et al, in preparation). It will be used to process the CRAFTS data in the near future. Predictably more and fainter \hi absorption systems could be detected in the data processed from the full map.

\subsection{Counterparts to the Radio Sources}
Finding the counterpart of the radio source has long been an important yet challenging problem \citep{1983MNRAS.204..151L}. In the above, we have argued that the absorption is very likely against the background radio source seen at the position of the beam if such a radio source is present within the beam. The absorber is then located precisely in the same line of sight. However, there is also the question of the nature of the optical/infrared counterpart, which is found by matching the radio source and the sources in the SDSS/WISE catalogue within a fiducial radius of several arcsecs. These optical/IR sources could be associated with the background radio source or the absorber, or they could be simply coincidentally located near the line of sight.

\begin{figure}
    \centering
    \includegraphics[width=0.45\textwidth]{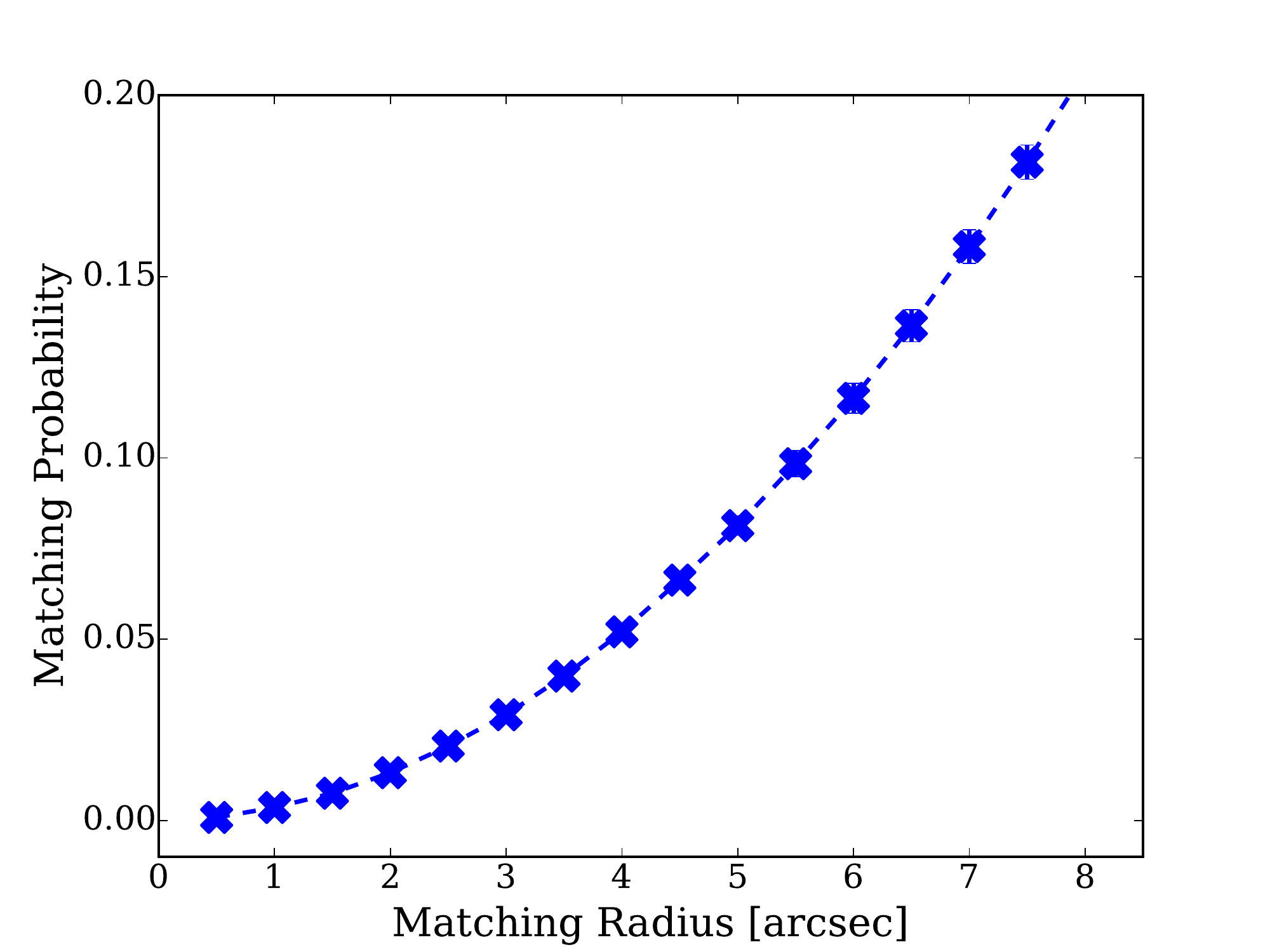}
    \caption{The probability of finding SDSS galaxies within a certain matching radius centred at arbitrary directions.}
    \label{falsematching_probability}
\end{figure}

We quantify the probability of finding coincident optical/infrared sources in the SDSS photometric catalogue, whose surface density is $\sim$ 4.2 per square arcmin, near the line of sight by simulation. We generate 10000 random positions (Ra, Dec) in the surveyed sky region and compute the probability of having galaxies within a fiducial radius to these positions. 
Figure~\ref{falsematching_probability} shows the probability of matching as a function of radius. The probability of finding a galaxy is only 0.0036 at a matching radius of 1 arcsec but increases rapidly with the matching radius. The average separation between the radio sources of our 5 \hi absorption systems and their SDSS counterparts is 0.734 arcsec, within which the probability of having an SDSS galaxy by chance is only 0.0021. For the two newly discovered absorbers towards NVSS\,J231240-052547 and NVSS\,J053118+315412, the separation are 2.351 and 0.387 arcsec, indicating a coincidence probability of 0.018 and 0.0002, respectively. This indicates that the detected SDSS matches are unlikely to be pure coincidence. 

As we argued earlier, the radio background source and the absorber are most likely in the same line of sight. If the background sources are all located at high redshift ($z \sim 1$), for which the optical counterpart would be too faint to be listed in the present SDSS catalogue, then these optical counterparts should be associated with the absorber. However, although most NVSS sources are located at high redshift \citep{2004ApJ...608...10N,2008PhRvD..78d3519H,2015MNRAS.447.3500R}, there is also a fraction of NVSS sources located at low redshift ($z<0.1$) \citep{1998AJ....115.1693C}. A spectroscopic measurement of its redshift is needed to find the final answer.

\subsection{Expected Detection}
\label{sec:expected_detection}
\subsubsection{Comoving Absorption Path}
We estimate the total absorption path length ($\Delta X$) and the $N_{\hi}$ frequency distribution function, f($N_{\hi}$, $X$) for the CRAFTS data survey used here (1300-1450\,MHz), following the method described in \citet{2015MNRAS.453.1249A,2020MNRAS.494.3627A,2021MNRAS.503..985A}. The absorption path length is defined as the comoving interval of the survey that is sensitive to intervening absorbers. It can be obtained by summing over all data elements that are sensitive to a given minimum column density threshold:
\begin{eqnarray}
    \Delta X = \sum^{n_{t}}_{i}\sum^{n_{f}}_{j}C(z_{i,j})w_{i,j}(z_{i,j})\Delta X_{i,j}\delta_{i,j},
    \label{absorption_path}
\end{eqnarray}
where $n_{t}$ and $n_{f}$ are the total number of time points and frequency channels, $\delta_{i,j}$ = 1 when the data element (i,j) is sensitive to a given minimum column density, otherwise $\delta_{i,j}$ = 0. $X_{i,j}$ is the comoving absorption path length for a certain element, $C(z_{i,j})$ denotes the completeness at a certain redshift (frequency), $w_{i,j}$ refers to the redshift weighting function, which gives the probability of the radio source being located beyond $z_{i,j}$, 
\begin{eqnarray}
    w(z) = \frac{\int^{\infty}_{z}N_{\mathrm{src}}(z^{\prime})dz^{\prime}}{\int^{\infty}_{0}N_{\mathrm{src}}(z^{\prime})dz^{\prime}},
    \label{redshift_weighting_function}
\end{eqnarray}
where the redshift distribution of source is \citep{2010A&ARv..18....1D}, 
\begin{eqnarray}
    N_{\mathrm{src}}(z) = 1.29 + 32.37z - 32.89z^{2} + 11.13z^{3} - 1.25z^{4},
    \label{redshift_distribution}
\end{eqnarray}
which is obtained by fitting the bright ($S_{1.4\rm GHz} \geqslant$ 10\,mJy) sources in the Combined EIS-NVSS Survey Of Radio Sources (CENSORS; \citealt{2008MNRAS.385.1297B}). For intervening \hi absorption, we adopt $w_{i,j} = w(z_{i,j}+\Delta z^{\mathrm{asc}}_{i,j})$, where $\Delta z^{\mathrm{asc}}_{i,j} = (1+z_{i,j})\Delta v^{\mathrm{asc}}/c$ and $\Delta v^{\mathrm{asc}} = 3000 \kms $. For associated \hi absorption, we exclude \hi absorption not associated with \hi gas in the host galaxy of the radio source by setting the upper limit of the integral of the denominator to $z_{i,j}+\Delta z^{\mathrm{asc}}_{i,j}$.

The comoving absorption path element for the element i,j in the $\Lambda$CDM model is:
\begin{eqnarray}
    \Delta X_{i,j} = \Delta z_{i,j}(1 + z_{i,j})^2E(z_{i,j})^{-1},
    \label{absorption_path_element}
\end{eqnarray}
where 
\begin{eqnarray}
    E(z) = \sqrt{(1+z)^{3}\Omega_{\rm m} + (1+z)^{2}(1-\Omega_{\rm m}-\Omega_{\Lambda})+\Omega_{\Lambda}}.
    \label{Ez}
\end{eqnarray}

The elements sensitive to a given column density are selected by calculating the column density sensitivity for each element. Assuming the spin temperature of $T_{\rm s} = 100$K, covering factor $c_{\rm f} = 1$,  velocity FWHM $\Delta v_{\rm{FWHM}} = 30$ \kms, the column density sensitivity can be expressed as:
\begin{eqnarray}
    N_{\hi,s} = 5.82 \times 10^{21}\left[\frac{T_{\mathrm{s}}}{100\mathrm{K}}\right]\left[\frac{\Delta v_{\rm{FWHM}}}{30 \mathrm{km\,s^{-1}}}\right]\tau_{\rm{peak}}\mathrm{cm}^{-2},
    \label{column_density_sensitivity}
\end{eqnarray}
where $\tau_{\rm{peak}}$ is the peak optical depth sensitivity estimated from the spectral noise. Adjusting Eq.~(\ref{tau}), $\tau_{\rm{peak}}$ can be expressed as:
\begin{eqnarray}
    \tau_{\rm{peak}} = -\mathrm{ln}\left(1 - (\mathrm{S/N})_{\rm{peak}}\left[\frac{\sigma_{\mathrm{rms}}}{S_{c}c_{\mathrm{f}}}  \right]\right),
    \label{tau_peak}
\end{eqnarray}
where $(\mathrm{S/N})_{\rm{peak}}$ is the minimum peak signal-to-noise ratio required to reach high completeness. $(\mathrm{S/N})_{\rm{peak}}$ is adopted as 5.5 in the comoving path calculation, corresponding to a $(\mathrm{S/N})_{\rm{int}}$ of $\sim$ 12, for the Gaussian narrow line absorption template with an FWHM of 30 \kms and the frequency resolution of 15.26 kHz. From Figure\,\ref{completeness}, the detection of mock absorption lines with a $(\mathrm{S/N})_{\rm{int}}$ of 12 can hit the completeness of 0.9. In Eq.~(\ref{tau_peak}), $\sigma_{\mathrm{rms}}$ is the RMS noise, $S_{c}$ is the background continuum flux density, which is estimated by calibrating the CRAFTS data using the FAST built-in noise diode and then removing the system temperature and Galactic foreground signal. The system temperature for each beam and observations pointing at different Declination is estimated using the models and parameters given by \cite{2020RAA....20...64J}. The Galactic foreground signal is estimated from the global sky model (GSM) \citep{2008MNRAS.388..247D,2017MNRAS.464.3486Z,Huang:2018ral}. In order to make sure all RFI are masked, the calibrated spectra are then flagged with a 3 sigma criterion.

The completeness correction employed in comoving absorption path estimation is determined using mock absorption with $\mathrm{S/N}$ = 12 and FWHM = 30 $\kms$. Nonetheless, at a fixed velocity-integrated sensitivity, absorptions with broader lines exhibit lower completeness when considering. Figure\,\ref{completeness_moreparameters} demonstrates that when the velocity-integrated sensitivity is low (e.g., $\mathrm{S/N}$ = 5.5), the completeness decreases rapidly with increasing velocity widths. On the other hand, for a fixed high velocity-integrated sensitivity (e.g., $\mathrm{S/N}$ = 12), there is little disparity in completeness among absorptions with velocity widths of 15 $\kms$, 30 $\kms$, and 60 $\kms$. Consequently, the selection of velocity width has a minor impact on our estimation of the comoving absorption path length.

\begin{figure}
    \centering
    \includegraphics[width=8.8cm]{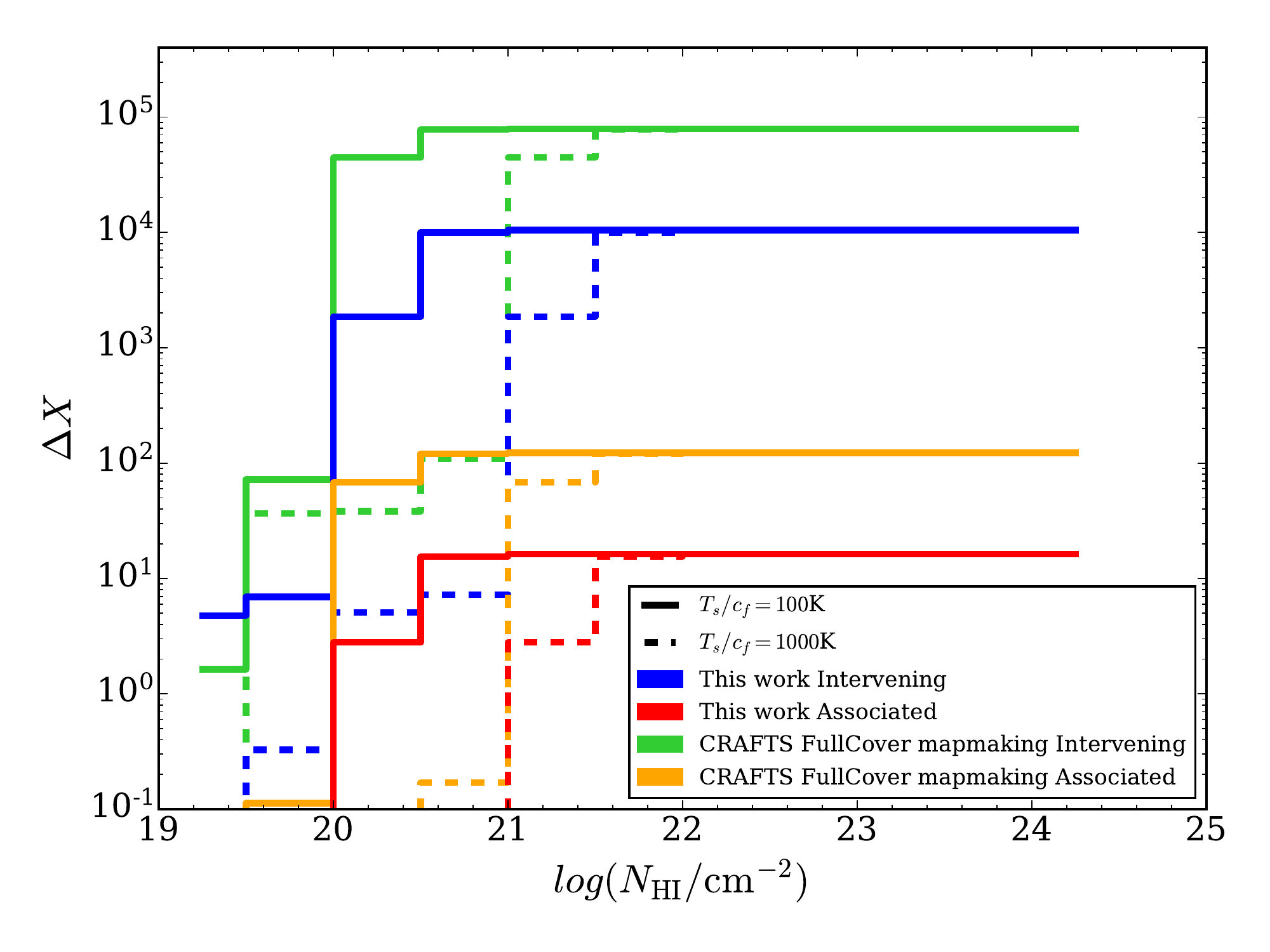}
    \caption{The comoving absorption path length ($\Delta X$) as a function of \hi column density sensitivity. The results for the spin temperature to source covering fraction ratios of 100 and 1000 K are depicted as solid and dashed lines, respectively. The comoving absorption paths spanned by our data are labelled as blue (intervening absorption) and red (associated absorption) colour. The predicted results from \hi absorption searching in map-making data from the full coverage of the FAST observable sky are presented as green (intervening absorption) and orange (associated absorption) colour.}
    \label{comoving_absorption_path}
\end{figure}

We show the comoving absorption path length for our current data as a function of \hi column density in Figure~\ref{comoving_absorption_path}. We present the results for $T_{s}/c_{f}$ = 100 and 1000 K, which are the typical spin temperature of the cold neutral medium (CNM) and warm neutral medium (WNM) \citep{2018ApJS..238...14M}. The comoving absorption paths for $T_{s}/c_{f}$ = 100 K and 1000 K are labelled as solid and dashed lines in Figure~\ref{comoving_absorption_path}. Under the assumption of $T_{s}/c_{f}$ = 100 K, the total comoving absorption path length spanned by our data is $\Delta X^{\mathrm{inv}}$ = 1.05$\times10^{4}$ ($\Delta z^{\mathrm{inv}} = 9.80\times10^{3}$) and $\Delta X^{\mathrm{asc}}$ = 1.63$\times10^{1}$ ($\Delta z^{\mathrm{asc}} = 1.51\times10^{1}$). The absorption path sensitive to all DLAs ($N_{\hi} \geqslant 2\times10^{20} \cm^{-2}$) and super-DLAs ($N_{\hi} \geqslant 2\times10^{21} \cm^{-2}$) are $\Delta X^{\mathrm{inv}}_{\mathrm{all}}$ = 8.33$\times10^3$ ($\Delta z^{\mathrm{inv}}_{\mathrm{all}} = 7.81\times10^{3}$) and $\Delta X^{\mathrm{inv}}_{\mathrm{super}}$ = 1.05$\times10^4$ ($\Delta z^{\mathrm{inv}}_{\mathrm{super}} = 9.80\times10^{3}$) for intervening absorption and $\Delta X^{\mathrm{asc}}_{\mathrm{all}}$ = 1.28$\times10^1$ ($\Delta z^{\mathrm{asc}}_{\mathrm{all}} = 1.19\times10^{1}$) and $\Delta X^{\mathrm{asc}}_{\mathrm{super}}$ = 1.63$\times10^1$ ($\Delta z^{\mathrm{asc}}_{\mathrm{super}} = 1.51\times10^{1}$) for associated absorption, respectively.

The maximum zenith angle for FAST is $\sim 40^\circ$, allowing the full cover of the FAST observable sky $\sim$ 23800 deg$^{2}$. The predicted results for \hi absorption searching at $z \leqslant 0.09$ in map-making data from the full coverage of the FAST observable sky are also presented in Figure~\ref{comoving_absorption_path}. Under the assumption of $T_{s}/c_{f}$ = 100\, K, for full cover searching in map-making data, the total path sensitive to all DLAs and super-DLAs increases to $\Delta X^{\mathrm{inv}}_{\mathrm{all}}$ = 7.37$\times10^4$ and $\Delta X^{\mathrm{inv}}_{\mathrm{super}}$ = 7.90$\times10^4$ for intervening absorption and to $\Delta X^{\mathrm{asc}}_{\mathrm{all}}$ = 1.14$\times10^2$ and $\Delta X^{\mathrm{asc}}_{\mathrm{super}}$ = 1.23$\times10^2$ for associated absorption, respectively.

\subsubsection{$N_{\hi}$ Frequency Distribution Function}
\begin{figure}
    \centering
    \includegraphics[width=8.8cm]{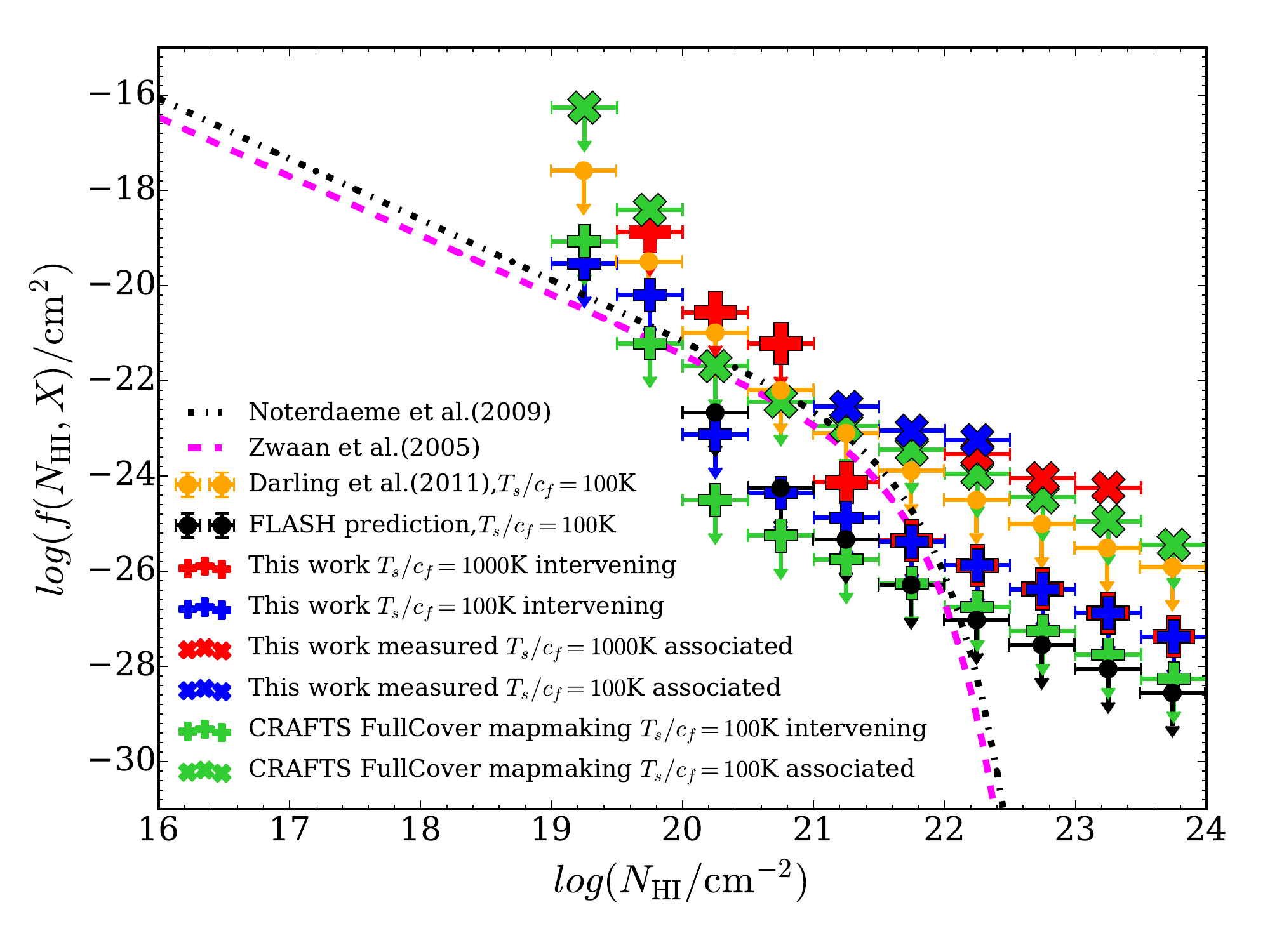}
    \caption{The measured \hi column density frequency distribution ($f(N_{\hi},X)$) or the 95 per cent upper limits on $f(N_{\hi},X)$ for CRAFTS data and previous works. 
    For data used in this work, $f(N_{\hi}, X)$ for associated \hi absorption (measured using the 4 associated absorption) and intervening \hi absorption (the 95 per cent upper limits) are presented as oblique and straight crosses, the distributions for the spin temperature to source covering fraction ratio of 100\, K and 1000\, K are shown as blue and red crosses. The predicted $f(N_{\hi}, X)$ for the \hi absorption searching in map-making data from the full coverage of the FAST observable sky are presented as green oblique (associated) and straight (intervening) crosses. The predicted distribution from planned FLASH survey \citep{2020MNRAS.494.3627A} (integration time 2 hours, 36 antennas, sky area of $\delta$ = +10 deg and redshift coverage of z = 0.4 to 1.0), the ALFALFA absorption pilot survey \citep{2011ApJ...742...60D} and model fits to \hi emission-line observations with WSRT (Westerbork Synthesis Radio Telescope) at z=0 \citep{2005MNRAS.364.1467Z} and DLA systems in SDSS-DR7 at z$\sim$3 \citep{2009A&A...505.1087N} are also presented.}
    \label{f_HI_X}
\end{figure}

With the comoving absorption path estimated above, we compute the $N_{\hi}$ frequency distribution function \citep{2005ApJ...635..123P,2010ApJ...708..868C,2011ApJ...742...60D,2020MNRAS.494.3627A,2020MNRAS.498..883G}, $f(N_{\hi}, X)$, which is the number of absorbers per column density $N_{\hi}$ bin per absorption path length sensitive to $N_{\hi}$, it is given by:
\begin{eqnarray}
    f(N_{\hi},X) = \frac{\lambda_{\mathrm{absorption}}}{\Delta N_{\hi}\Delta X},
    \label{NHI_frequency_distribution}
\end{eqnarray}
where $\lambda_{\mathrm{absorption}}$ is the number of \hi absorption systems within each column density interval, $\Delta N_{\hi}$ = 0.5\,dex here and $\Delta X$ is the comoving absorption path length searched with $N_{\hi}$. Making use of the five detected \hi absorption systems, we calculate the measured $f(N_{\hi}, X)$ for the analysed data in this work and present it in Figure~\ref{f_HI_X} as oblique crosses. 

We also compare our results with the upper limits on the $f(N_{\hi}, X)$ for the other surveys. For the case of upper limits, $N_{\hi}$ frequency distribution function is expressed as $f(N_{\hi},X) < \frac{\lambda_{\mathrm{max}}}{\Delta N_{\hi}\Delta X},$ where $\lambda_{\mathrm{max}}$ is the Poisson upper limit on the detection rate of absorbers with column density $N_{\hi}$ in the interval $\Delta N_{\hi}$. The 95\% confidence upper limit on the Poisson rate is $\lambda_{\mathrm{max}}$ = 3.0 when non-detection is assumed \citep{1986ApJ...303..336G}.

Considering 4 associated and only 1 tentative intervening \hi absorption are detected in this work, we present measured $f(N_{\hi}, X)$ for associated absorption and the 95 \% upper limits of $f(N_{\hi}, X)$ for intervening absorption for $T_{s}/c_{f}$ = 100 and 1000 K in Figure~\ref{f_HI_X}, together with the distribution from planned FLASH survey \citep{2020MNRAS.494.3627A} (integration time 2 hours, 36 antennas, sky area of $\delta$ = +10 deg and redshift coverage from 0.4 to 1.0), the ALFALFA absorption pilot survey \citep{2011ApJ...742...60D} and model fits to \hi emission-line observations with WSRT (Westerbork Synthesis Radio Telescope) at z=0 \citep{2005MNRAS.364.1467Z} and DLA systems in SDSS-DR7 at z $\sim$ 3 \citep{2009A&A...505.1087N}. The predicted $f(N_{\hi}, X)$ for the \hi absorption searching in map-making data from the full coverage of the FAST observable sky is comparable with FLASH $f(N_{\hi}, X)$ upper limits prediction at super-DLAs ($N_{\hi} \geqslant 2\times 10^{20} \cm^{-2}$) regions.

We should note that limited by the RFIs, computing resources and authorized follow-up observation time, only the 1300-1450\, MHz band are used for absorption lines search and sensitivity calculation. With more comprehensive data processing methods and more frequency bands (1000-1300\, MHz) and larger sky searched in the near future, a much larger comoving absorption path and higher sensitivity to N$_{\hi}$ frequency distribution function could be obtained with CRAFTS.

\subsubsection{Comparing Forecast with real Blind Searching}
Utilising the number density distribution of \hi clouds and radio sources over the sky, \citet{2017RAA....17...49Y} predicted that a one-month drift scan around the celestial equator by FAST could detect $\sim$ 200 absorption systems with $\mathrm{S/N}$ of 10\,$\sigma$ in redshift range 0 < z < 0.39.  \citet{2021MNRAS.503.5385Z} considered the NVSS bright source within the FAST observable sky and forecasted that $\sim$ 6000 extragalactic \hi absorbers can be found with a detection rate of $\sim$ 5.5 per cent (detection rate of 40$\%$ of the ALFALFA data). However, we only blindly find 5 absorption systems in the 643.8-hour drift scan survey. This difference between forecast and real data processing is attributed to the RFI and unstable bandpass (standing waves), which can contaminate the spectrum profile of true absorption systems and produce a spurious absorption line shape. It is expected that with standing waves corrected and the map-making method applied in the near future, more \hi absorbers could be unveiled with the FAST extra-galactic sky survey data.

\section{Summary}
\label{sec:summary}
In this paper, we present the early science results of a purely blind search for extragalactic \hi 21-cm absorption lines in the CRAFTS drift-scan data from FAST. Nearly 643.8\, hours and 3155\,deg$^{2}$ survey (covers 44827 radio sources with a flux density greater than 12 mJy) from the Commensal Radio Astronomy FAST Survey (CRAFTS) are searched for \hi absorber. The total comoving absorption path length spanned by our data for intervening and associated absorption is $\Delta X^{\mathrm{inv}}$ = 1.05$\times10^{4}$ ($\Delta z^{\mathrm{inv}} = 9.80\times10^{3}$) and $\Delta X^{\mathrm{asc}}$ = 1.63$\times10^{1}$ ($\Delta z^{\mathrm{asc}} = 1.51\times10^{1}$, assuming $T_{s}/c_{f}$ = 100 K). The completeness correction, determined using mock absorption signals with FWHM = 30 $\kms$, has been applied to the estimates of comoving absorption path. Both polarizations data of all 19 beams are processed. Due to the RFI, only the relatively clean data in the frequency range of 1.3 GHz to 1.45 GHz are searched.

We use a matched-filtering approach to detect the \hi absorption profiles at each time point of the drift scan. By cross-correlating information from different beams, we effectively eliminate spurious signals and identify three previously known \hi absorbers (UGC\,00613, 3C\,293 and 4C\,+27.14) as well as two newly discovered ones (NVSS\, J231240-052547 and NVSS\, J053118+315412). We fit the \hi profiles with multi-components Gaussian functions and calculate the redshift, FWHM, flux density, optical depth and \hi column densities for each source. The newly discovered \hi absorption feature in NVSS\,J053118+315412 displays associated absorption with both blueshifted and redshifted wings, as well as redshifted components. These characteristics suggest the presence of unsettled gas structures and possible accretion of gas onto the SMBH. Our results demonstrate the power of FAST in \hi absorption blind searching. The forecast for the FAST \hi absorption searching indicates a long comoving absorption path and high sensitivity to N$_{\hi}$ frequency distribution function could be obtained with CRAFTS.

Discovery of intervening \hi absorbers in blind radio search may enable us to unravel the nature of the DLAs. As most DLAs were discovered in the observation of optically bright background quasars, it is often difficult to observe the absorber itself. In the present case, \hi absorbers with column density comparable to the DLAs are detected in the radio, it is hopeful that the absorber could also be observed optically. We have found optical/IR counterparts in data archives along the line of sight where the two absorptions are detected, and follow-up observation with redshift measurement will determine whether NVSS\, J231240-052547 is associated with the \hi absorbers and the nature of the counterparts. 

\section*{Acknowledgements}
We acknowledge the support of the National SKA Program of China, No.2022SKA0110100. This work is supported by the National Key R\&D Program 2017YFA0402603, the Ministry of Science and Technology (MoST) inter-government cooperation program China-South Africa Cooperation Flagship project 2018YFE0120800, the MoST grant 2016YFE0100300, the National Natural Science Foundation of China (NSFC) key project grant 11633004, the Chinese Academy of Sciences (CAS) Frontier Science Key Project QYZDJ-SSW-SLH017 and the CAS Interdisciplinary Innovation Team grant (JCTD-2019-05), the NSFC-ISF joint research program No. 11761141012, the CAS Strategic Priority Research Program XDA15020200, and the NSFC grant 11773034. 

Wenkai Hu and Guilaine Lagache are supported by the European Research Council (ERC) under the European Union's Horizon 2020 research and innovation programme (project CONCERTO, grant agreement No 788212) and from the Excellence Initiative of Aix-Marseille University-A*Midex, a French "Investissements d'Avenir" programme. Zheng Zheng is supported by NSFC grant No. 11988101, U1931110 and 12041302. Zheng Zheng is also supported by CAS Interdisciplinary Innovation Team (JCTD-2019-05). The authors thank Sne$\rm\check{z}$ana Stanimirovi$\rm\acute{c}$, Elizabeth K. Mahony, Stephen J. Curran and  James R. Allison for helpful discussion.
Ue-Li Pen receives support from Ontario Research Fund—Research Excellence Program (ORF-RE), Natural Sciences and Engineering Research Council of Canada (NSERC) [funding reference number RGPIN-2019-067, CRD 523638-18, 555585-20], Canadian Institute for Advanced Research (CIFAR), the National Science Foundation of China (Grants No. 11929301), Thoth Technology Inc, Alexander von Humboldt Foundation, and the National Science and Technology Council (NSTC) (111-2123-M-001 -008 -, and 111-2811-M-001 -040 -). Computations were performed on the SOSCIP Consortium’s [Blue Gene/Q, Cloud Data Analytics, Agile and/or Large Memory System] computing platform(s). SOSCIP is funded by the Federal Economic Development Agency of Southern Ontario, the Province of Ontario, IBM Canada Ltd., Ontario Centres of Excellence, Mitacs and 15 Ontario academic member institutions.

The radio data analyzed in this work can be accessed by sending a request to the FAST Data Centre or the corresponding authors of this paper.


\bibliographystyle{mnras}
\bibliography{absorption}

\begin{thebibliography}{}
\makeatletter
\relax
\def\mn@urlcharsother{\let\do\@makeother \do\$\do\&\do\#\do\^\do\_\do\%\do\~}
\def\mn@doi{\begingroup\mn@urlcharsother \@ifnextchar [ {\mn@doi@}
  {\mn@doi@[]}}
\def\mn@doi@[#1]#2{\def\@tempa{#1}\ifx\@tempa\@empty \href
  {http://dx.doi.org/#2} {doi:#2}\else \href {http://dx.doi.org/#2} {#1}\fi
  \endgroup}
\def\mn@eprint#1#2{\mn@eprint@#1:#2::\@nil}
\def\mn@eprint@arXiv#1{\href {http://arxiv.org/abs/#1} {{\tt arXiv:#1}}}
\def\mn@eprint@dblp#1{\href {http://dblp.uni-trier.de/rec/bibtex/#1.xml}
  {dblp:#1}}
\def\mn@eprint@#1:#2:#3:#4\@nil{\def\@tempa {#1}\def\@tempb {#2}\def\@tempc
  {#3}\ifx \@tempc \@empty \let \@tempc \@tempb \let \@tempb \@tempa \fi \ifx
  \@tempb \@empty \def\@tempb {arXiv}\fi \@ifundefined
  {mn@eprint@\@tempb}{\@tempb:\@tempc}{\expandafter \expandafter \csname
  mn@eprint@\@tempb\endcsname \expandafter{\@tempc}}}

\bibitem[\protect\citeauthoryear{{Allison}}{{Allison}}{2021}]{2021MNRAS.503..985A}
{Allison} J.~R.,  2021, \mn@doi [\mnras] {10.1093/mnras/stab518}, \href
  {https://ui.adsabs.harvard.edu/abs/2021MNRAS.503..985A} {503, 985}

\bibitem[\protect\citeauthoryear{{Allison} et~al.,}{{Allison}
  et~al.}{2015}]{2015MNRAS.453.1249A}
{Allison} J.~R.,  et~al., 2015, \mn@doi [\mnras] {10.1093/mnras/stv1532}, \href
  {https://ui.adsabs.harvard.edu/abs/2015MNRAS.453.1249A} {453, 1249}

\bibitem[\protect\citeauthoryear{{Allison}, {Zwaan}, {Duchesne}  \&
  {Curran}}{{Allison} et~al.}{2016}]{2016MNRAS.462.1341A}
{Allison} J.~R.,  {Zwaan} M.~A.,  {Duchesne} S.~W.,   {Curran} S.~J.,  2016,
  \mn@doi [\mnras] {10.1093/mnras/stw1722}, \href
  {https://ui.adsabs.harvard.edu/abs/2016MNRAS.462.1341A} {462, 1341}

\bibitem[\protect\citeauthoryear{{Allison} et~al.,}{{Allison}
  et~al.}{2020}]{2020MNRAS.494.3627A}
{Allison} J.~R.,  et~al., 2020, \mn@doi [\mnras] {10.1093/mnras/staa949}, \href
  {https://ui.adsabs.harvard.edu/abs/2020MNRAS.494.3627A} {494, 3627}

\bibitem[\protect\citeauthoryear{{Allison} et~al.,}{{Allison}
  et~al.}{2022}]{2022PASA...39...10A}
{Allison} J.~R.,  et~al., 2022, \mn@doi [\pasa] {10.1017/pasa.2022.3}, \href
  {https://ui.adsabs.harvard.edu/abs/2022PASA...39...10A} {39, e010}

\bibitem[\protect\citeauthoryear{{Araya}, {Rodr{\'\i}guez}, {Pihlstr{\"o}m},
  {Taylor}, {Tremblay}  \& {Vermeulen}}{{Araya}
  et~al.}{2010}]{2010AJ....139...17A}
{Araya} E.~D.,  {Rodr{\'\i}guez} C.,  {Pihlstr{\"o}m} Y.,  {Taylor} G.~B.,
  {Tremblay} S.,   {Vermeulen} R.~C.,  2010, \mn@doi [\aj]
  {10.1088/0004-6256/139/1/17}, \href
  {https://ui.adsabs.harvard.edu/abs/2010AJ....139...17A} {139, 17}

\bibitem[\protect\citeauthoryear{{Baan} \& {Haschick}}{{Baan} \&
  {Haschick}}{1981}]{1981ApJ...243L.143B}
{Baan} W.~A.,  {Haschick} A.~D.,  1981, \mn@doi [\apjl] {10.1086/183461}, \href
  {https://ui.adsabs.harvard.edu/abs/1981ApJ...243L.143B} {243, L143}

\bibitem[\protect\citeauthoryear{{Beswick}, {Peck}, {Taylor}, {Giovannini}  \&
  {Pedlar}}{{Beswick} et~al.}{2004a}]{2004evn..conf..147B}
{Beswick} R.~J.,  {Peck} A.~B.,  {Taylor} G.~B.,  {Giovannini} G.,   {Pedlar}
  A.,  2004a, in European VLBI Network on New Developments in VLBI Science and
  Technology. pp 147--150 (\mn@eprint {arXiv} {astro-ph/0501540})

\bibitem[\protect\citeauthoryear{{Beswick}, {Peck}, {Taylor}  \&
  {Giovannini}}{{Beswick} et~al.}{2004b}]{2004MNRAS.352...49B}
{Beswick} R.~J.,  {Peck} A.~B.,  {Taylor} G.~B.,   {Giovannini} G.,  2004b,
  \mn@doi [\mnras] {10.1111/j.1365-2966.2004.07892.x}, \href
  {https://ui.adsabs.harvard.edu/abs/2004MNRAS.352...49B} {352, 49}

\bibitem[\protect\citeauthoryear{{Bilicki}, {Jarrett}, {Peacock}, {Cluver}  \&
  {Steward}}{{Bilicki} et~al.}{2014}]{2014ApJS..210....9B}
{Bilicki} M.,  {Jarrett} T.~H.,  {Peacock} J.~A.,  {Cluver} M.~E.,   {Steward}
  L.,  2014, \mn@doi [\apjs] {10.1088/0067-0049/210/1/9}, \href
  {https://ui.adsabs.harvard.edu/abs/2014ApJS..210....9B} {210, 9}

\bibitem[\protect\citeauthoryear{{Bordoloi} et~al.,}{{Bordoloi}
  et~al.}{2022}]{Bordoloi2022}
{Bordoloi} R.,  et~al., 2022, \mn@doi [\nat] {10.1038/s41586-022-04616-1},
  \href {https://ui.adsabs.harvard.edu/abs/2022Natur.606...59B} {606, 59}

\bibitem[\protect\citeauthoryear{{Brookes}, {Best}, {Peacock}, {R{\"o}ttgering}
   \& {Dunlop}}{{Brookes} et~al.}{2008}]{2008MNRAS.385.1297B}
{Brookes} M.~H.,  {Best} P.~N.,  {Peacock} J.~A.,  {R{\"o}ttgering} H.~J.~A.,
  {Dunlop} J.~S.,  2008, \mn@doi [\mnras] {10.1111/j.1365-2966.2008.12786.x},
  \href {https://ui.adsabs.harvard.edu/abs/2008MNRAS.385.1297B} {385, 1297}

\bibitem[\protect\citeauthoryear{{Brown} \& {Roberts}}{{Brown} \&
  {Roberts}}{1973}]{1973ApJ...184L...7B}
{Brown} R.~L.,  {Roberts} M.~S.,  1973, \mn@doi [\apjl] {10.1086/181276}, \href
  {https://ui.adsabs.harvard.edu/abs/1973ApJ...184L...7B} {184, L7}

\bibitem[\protect\citeauthoryear{{Catinella}, {Haynes}, {Giovanelli}, {Gardner}
   \& {Connolly}}{{Catinella} et~al.}{2008}]{2008ApJ...685L..13C}
{Catinella} B.,  {Haynes} M.~P.,  {Giovanelli} R.,  {Gardner} J.~P.,
  {Connolly} A.~J.,  2008, \mn@doi [\apjl] {10.1086/592328}, \href
  {http://adsabs.harvard.edu/abs/2008ApJ...685L..13C} {685, L13}

\bibitem[\protect\citeauthoryear{{Chandola} \& {Saikia}}{{Chandola} \&
  {Saikia}}{2017}]{2017MNRAS.465..997C}
{Chandola} Y.,  {Saikia} D.~J.,  2017, \mn@doi [\mnras]
  {10.1093/mnras/stw2705}, \href
  {https://ui.adsabs.harvard.edu/abs/2017MNRAS.465..997C} {465, 997}

\bibitem[\protect\citeauthoryear{{Chandola}, {Sirothia}  \&
  {Saikia}}{{Chandola} et~al.}{2011}]{2011MNRAS.418.1787C}
{Chandola} Y.,  {Sirothia} S.~K.,   {Saikia} D.~J.,  2011, \mn@doi [\mnras]
  {10.1111/j.1365-2966.2011.19607.x}, \href
  {https://ui.adsabs.harvard.edu/abs/2011MNRAS.418.1787C} {418, 1787}

\bibitem[\protect\citeauthoryear{{Chandola}, {Saikia}  \& {Li}}{{Chandola}
  et~al.}{2020}]{2020MNRAS.494.5161C}
{Chandola} Y.,  {Saikia} D.~J.,   {Li} D.,  2020, \mn@doi [\mnras]
  {10.1093/mnras/staa1029}, \href
  {https://ui.adsabs.harvard.edu/abs/2020MNRAS.494.5161C} {494, 5161}

\bibitem[\protect\citeauthoryear{{Chang}, {Pen}, {Peterson}  \&
  {McDonald}}{{Chang} et~al.}{2008}]{2008PhRvL.100i1303C}
{Chang} T.-C.,  {Pen} U.-L.,  {Peterson} J.~B.,   {McDonald} P.,  2008, \mn@doi
  [\prl] {10.1103/PhysRevLett.100.091303}, \href
  {https://ui.adsabs.harvard.edu/abs/2008PhRvL.100i1303C} {100, 091303}

\bibitem[\protect\citeauthoryear{{Combes} et~al.,}{{Combes}
  et~al.}{2021}]{2021A&A...648A.116C}
{Combes} F.,  et~al., 2021, \mn@doi [\aap] {10.1051/0004-6361/202040167}, \href
  {https://ui.adsabs.harvard.edu/abs/2021A&A...648A.116C} {648, A116}

\bibitem[\protect\citeauthoryear{{Condon}, {Cotton}, {Greisen}, {Yin},
  {Perley}, {Taylor}  \& {Broderick}}{{Condon}
  et~al.}{1998}]{1998AJ....115.1693C}
{Condon} J.~J.,  {Cotton} W.~D.,  {Greisen} E.~W.,  {Yin} Q.~F.,  {Perley}
  R.~A.,  {Taylor} G.~B.,   {Broderick} J.~J.,  1998, \mn@doi [\aj]
  {10.1086/300337}, \href
  {https://ui.adsabs.harvard.edu/abs/1998AJ....115.1693C} {115, 1693}

\bibitem[\protect\citeauthoryear{{Cooksey}, {Thom}, {Prochaska}  \&
  {Chen}}{{Cooksey} et~al.}{2010}]{2010ApJ...708..868C}
{Cooksey} K.~L.,  {Thom} C.,  {Prochaska} J.~X.,   {Chen} H.-W.,  2010, \mn@doi
  [\apj] {10.1088/0004-637X/708/1/868}, \href
  {https://ui.adsabs.harvard.edu/abs/2010ApJ...708..868C} {708, 868}

\bibitem[\protect\citeauthoryear{{Curran}, {Duchesne}, {Divoli}  \&
  {Allison}}{{Curran} et~al.}{2016}]{2016MNRAS.462.4197C}
{Curran} S.~J.,  {Duchesne} S.~W.,  {Divoli} A.,   {Allison} J.~R.,  2016,
  \mn@doi [\mnras] {10.1093/mnras/stw1938}, \href
  {https://ui.adsabs.harvard.edu/abs/2016MNRAS.462.4197C} {462, 4197}

\bibitem[\protect\citeauthoryear{{Cutri} et~al.,}{{Cutri}
  et~al.}{2013}]{2013wise.rept....1C}
{Cutri} R.~M.,  et~al., 2013, {Explanatory Supplement to the AllWISE Data
  Release Products}, Explanatory Supplement to the AllWISE Data Release
  Products

\bibitem[\protect\citeauthoryear{{Darling}, {Giovanelli}, {Haynes}, {Bolatto}
  \& {Bower}}{{Darling} et~al.}{2004}]{2004ApJ...613L.101D}
{Darling} J.,  {Giovanelli} R.,  {Haynes} M.~P.,  {Bolatto} A.~D.,   {Bower}
  G.~C.,  2004, \mn@doi [\apjl] {10.1086/425143}, \href
  {https://ui.adsabs.harvard.edu/abs/2004ApJ...613L.101D} {613, L101}

\bibitem[\protect\citeauthoryear{{Darling}, {Macdonald}, {Haynes}  \&
  {Giovanelli}}{{Darling} et~al.}{2011}]{2011ApJ...742...60D}
{Darling} J.,  {Macdonald} E.~P.,  {Haynes} M.~P.,   {Giovanelli} R.,  2011,
  \mn@doi [\apj] {10.1088/0004-637X/742/1/60}, \href
  {https://ui.adsabs.harvard.edu/abs/2011ApJ...742...60D} {742, 60}

\bibitem[\protect\citeauthoryear{Dunning et~al.,}{Dunning
  et~al.}{2017}]{8105012}
Dunning A.,  et~al., 2017, in 2017 XXXIInd General Assembly and Scientific
  Symposium of the International Union of Radio Science (URSI GASS). pp~1--4,
  \mn@doi{10.23919/URSIGASS.2017.8105012}

\bibitem[\protect\citeauthoryear{{Emonts} et~al.,}{{Emonts}
  et~al.}{2010}]{2010MNRAS.406..987E}
{Emonts} B.~H.~C.,  et~al., 2010, \mn@doi [\mnras]
  {10.1111/j.1365-2966.2010.16706.x}, \href
  {https://ui.adsabs.harvard.edu/abs/2010MNRAS.406..987E} {406, 987}

\bibitem[\protect\citeauthoryear{{Evans}, {Sanders}, {Surace}  \&
  {Mazzarella}}{{Evans} et~al.}{1999}]{1999ApJ...511..730E}
{Evans} A.~S.,  {Sanders} D.~B.,  {Surace} J.~A.,   {Mazzarella} J.~M.,  1999,
  \mn@doi [\apj] {10.1086/306717}, \href
  {https://ui.adsabs.harvard.edu/abs/1999ApJ...511..730E} {511, 730}

\bibitem[\protect\citeauthoryear{Fanaroff \& Riley}{Fanaroff \&
  Riley}{1974}]{10.1093/mnras/167.1.31P}
Fanaroff B.~L.,  Riley J.~M.,  1974, \mn@doi [Monthly Notices of the Royal
  Astronomical Society] {10.1093/mnras/167.1.31P}, 167, 31P

\bibitem[\protect\citeauthoryear{{Fern{\'a}ndez} et~al.,}{{Fern{\'a}ndez}
  et~al.}{2016}]{2016ApJ...824L...1F}
{Fern{\'a}ndez} X.,  et~al., 2016, \mn@doi [\apjl]
  {10.3847/2041-8205/824/1/L1}, \href
  {http://adsabs.harvard.edu/abs/2016ApJ...824L...1F} {824, L1}

\bibitem[\protect\citeauthoryear{{Gehrels}}{{Gehrels}}{1986}]{1986ApJ...303..336G}
{Gehrels} N.,  1986, \mn@doi [\apj] {10.1086/164079}, \href
  {https://ui.adsabs.harvard.edu/abs/1986ApJ...303..336G} {303, 336}

\bibitem[\protect\citeauthoryear{{Ger{\'e}b}, {Maccagni}, {Morganti}  \&
  {Oosterloo}}{{Ger{\'e}b} et~al.}{2015}]{2015A&A...575A..44G}
{Ger{\'e}b} K.,  {Maccagni} F.~M.,  {Morganti} R.,   {Oosterloo} T.~A.,  2015,
  \mn@doi [\aap] {10.1051/0004-6361/201424655}, \href
  {https://ui.adsabs.harvard.edu/abs/2015A&A...575A..44G} {575, A44}

\bibitem[\protect\citeauthoryear{{Giovanelli} et~al.,}{{Giovanelli}
  et~al.}{2005}]{2005AJ....130.2598G}
{Giovanelli} R.,  et~al., 2005, \mn@doi [\aj] {10.1086/497431}, \href
  {http://adsabs.harvard.edu/abs/2005AJ....130.2598G} {130, 2598}

\bibitem[\protect\citeauthoryear{{Giovanelli} et~al.,}{{Giovanelli}
  et~al.}{2007}]{2007AJ....133.2569G}
{Giovanelli} R.,  et~al., 2007, \mn@doi [\aj] {10.1086/516635}, \href
  {http://adsabs.harvard.edu/abs/2007AJ....133.2569G} {133, 2569}

\bibitem[\protect\citeauthoryear{{Grasha}, {Darling}, {Leroy}  \&
  {Bolatto}}{{Grasha} et~al.}{2020}]{2020MNRAS.498..883G}
{Grasha} K.,  {Darling} J.,  {Leroy} A.~K.,   {Bolatto} A.~D.,  2020, \mn@doi
  [\mnras] {10.1093/mnras/staa2521}, \href
  {https://ui.adsabs.harvard.edu/abs/2020MNRAS.498..883G} {498, 883}

\bibitem[\protect\citeauthoryear{Gupta et~al.,}{Gupta
  et~al.}{2018}]{Gupta:2018A7}
Gupta N.,  et~al., 2018, \mn@doi [PoS] {10.22323/1.277.0014}, MeerKAT2016, 014

\bibitem[\protect\citeauthoryear{{Gupta} et~al.,}{{Gupta}
  et~al.}{2021a}]{2021ApJS..255...28G}
{Gupta} N.,  et~al., 2021a, \mn@doi [\apjs] {10.3847/1538-4365/ac03b5}, \href
  {https://ui.adsabs.harvard.edu/abs/2021ApJS..255...28G} {255, 28}

\bibitem[\protect\citeauthoryear{{Gupta} et~al.,}{{Gupta}
  et~al.}{2021b}]{2021ApJ...907...11G}
{Gupta} N.,  et~al., 2021b, \mn@doi [\apj] {10.3847/1538-4357/abcb85}, \href
  {https://ui.adsabs.harvard.edu/abs/2021ApJ...907...11G} {907, 11}

\bibitem[\protect\citeauthoryear{{Hardcastle} et~al.,}{{Hardcastle}
  et~al.}{2019}]{2019MNRAS.488.3416H}
{Hardcastle} M.~J.,  et~al., 2019, \mn@doi [\mnras] {10.1093/mnras/stz1910},
  \href {https://ui.adsabs.harvard.edu/abs/2019MNRAS.488.3416H} {488, 3416}

\bibitem[\protect\citeauthoryear{{Healey}, {Romani}, {Taylor}, {Sadler},
  {Ricci}, {Murphy}, {Ulvestad}  \& {Winn}}{{Healey}
  et~al.}{2007}]{2007ApJS..171...61H}
{Healey} S.~E.,  {Romani} R.~W.,  {Taylor} G.~B.,  {Sadler} E.~M.,  {Ricci} R.,
   {Murphy} T.,  {Ulvestad} J.~S.,   {Winn} J.~N.,  2007, \mn@doi [\apjs]
  {10.1086/513742}, \href
  {https://ui.adsabs.harvard.edu/abs/2007ApJS..171...61H} {171, 61}

\bibitem[\protect\citeauthoryear{{Heiles} \& {Troland}}{{Heiles} \&
  {Troland}}{2003}]{2003ApJS..145..329H}
{Heiles} C.,  {Troland} T.~H.,  2003, \mn@doi [\apjs] {10.1086/367785}, \href
  {https://ui.adsabs.harvard.edu/abs/2003ApJS..145..329H} {145, 329}

\bibitem[\protect\citeauthoryear{{Helou}, {Madore}, {Schmitz}, {Bicay}, {Wu}
  \& {Bennett}}{{Helou} et~al.}{1991}]{1991ASSL..171...89H}
{Helou} G.,  {Madore} B.~F.,  {Schmitz} M.,  {Bicay} M.~D.,  {Wu} X.,
  {Bennett} J.,  1991, {The NASA/IPAC extragalactic database.}.
pp 89--106, \mn@doi{10.1007/978-94-011-3250-3\_10}

\bibitem[\protect\citeauthoryear{{Ho}, {Hirata}, {Padmanabhan}, {Seljak}  \&
  {Bahcall}}{{Ho} et~al.}{2008}]{2008PhRvD..78d3519H}
{Ho} S.,  {Hirata} C.,  {Padmanabhan} N.,  {Seljak} U.,   {Bahcall} N.,  2008,
  \mn@doi [\prd] {10.1103/PhysRevD.78.043519}, \href
  {https://ui.adsabs.harvard.edu/abs/2008PhRvD..78d3519H} {78, 043519}

\bibitem[\protect\citeauthoryear{{Hu} et~al.,}{{Hu}
  et~al.}{2019}]{2019MNRAS.489.1619H}
{Hu} W.,  et~al., 2019, \mn@doi [\mnras] {10.1093/mnras/stz2038}, \href
  {https://ui.adsabs.harvard.edu/abs/2019MNRAS.489.1619H} {489, 1619}

\bibitem[\protect\citeauthoryear{{Hu}, {Wang}, {Wu}, {Wang}, {Zhang}  \&
  {Chen}}{{Hu} et~al.}{2020}]{2020MNRAS.493.5854H}
{Hu} W.,  {Wang} X.,  {Wu} F.,  {Wang} Y.,  {Zhang} P.,   {Chen} X.,  2020,
  \mn@doi [\mnras] {10.1093/mnras/staa650}, \href
  {https://ui.adsabs.harvard.edu/abs/2020MNRAS.493.5854H} {493, 5854}

\bibitem[\protect\citeauthoryear{{Hu} et~al.,}{{Hu}
  et~al.}{2021}]{2021MNRAS.508.2897H}
{Hu} W.,  et~al., 2021, \mn@doi [\mnras] {10.1093/mnras/stab2728}, \href
  {https://ui.adsabs.harvard.edu/abs/2021MNRAS.508.2897H} {508, 2897}

\bibitem[\protect\citeauthoryear{Huang, Wu  \& Chen}{Huang
  et~al.}{2019}]{Huang:2018ral}
Huang Q.,  Wu F.,   Chen X.,  2019, \mn@doi [Sci. China Phys. Mech. Astron.]
  {10.1007/s11433-018-9333-1}, 62, 989511

\bibitem[\protect\citeauthoryear{{Jarvis} et~al.,}{{Jarvis}
  et~al.}{2014}]{2014arXiv1401.4018J}
{Jarvis} M.~J.,  et~al., 2014, preprint, \href
  {http://adsabs.harvard.edu/abs/2014arXiv1401.4018J} {} (\mn@eprint {arXiv}
  {1401.4018})

\bibitem[\protect\citeauthoryear{{Jiang} et~al.,}{{Jiang}
  et~al.}{2020}]{2020RAA....20...64J}
{Jiang} P.,  et~al., 2020, \mn@doi [Research in Astronomy and Astrophysics]
  {10.1088/1674-4527/20/5/64}, \href
  {https://ui.adsabs.harvard.edu/abs/2020RAA....20...64J} {20, 064}

\bibitem[\protect\citeauthoryear{{Kern}, {Parsons}, {Dillon}, {Lanman},
  {Fagnoni}  \& {de Lera Acedo}}{{Kern} et~al.}{2019}]{2019arXiv190911732K}
{Kern} N.~S.,  {Parsons} A.~R.,  {Dillon} J.~S.,  {Lanman} A.~E.,  {Fagnoni}
  N.,   {de Lera Acedo} E.,  2019, arXiv e-prints, \href
  {https://ui.adsabs.harvard.edu/abs/2019arXiv190911732K} {p. arXiv:1909.11732}

\bibitem[\protect\citeauthoryear{{Kern} et~al.,}{{Kern}
  et~al.}{2020}]{2020ApJ...888...70K}
{Kern} N.~S.,  et~al., 2020, \mn@doi [\apj] {10.3847/1538-4357/ab5e8a}, \href
  {https://ui.adsabs.harvard.edu/abs/2020ApJ...888...70K} {888, 70}

\bibitem[\protect\citeauthoryear{{Koribalski} et~al.,}{{Koribalski}
  et~al.}{2020}]{2020Ap&SS.365..118K}
{Koribalski} B.~S.,  et~al., 2020, \mn@doi [\apss]
  {10.1007/s10509-020-03831-4}, \href
  {https://ui.adsabs.harvard.edu/abs/2020Ap&SS.365..118K} {365, 118}

\bibitem[\protect\citeauthoryear{{Lacy} et~al.,}{{Lacy}
  et~al.}{2020}]{2020PASP..132c5001L}
{Lacy} M.,  et~al., 2020, \mn@doi [\pasp] {10.1088/1538-3873/ab63eb}, \href
  {https://ui.adsabs.harvard.edu/abs/2020PASP..132c5001L} {132, 035001}

\bibitem[\protect\citeauthoryear{{Laing}, {Riley}  \& {Longair}}{{Laing}
  et~al.}{1983}]{1983MNRAS.204..151L}
{Laing} R.~A.,  {Riley} J.~M.,   {Longair} M.~S.,  1983, \mn@doi [\mnras]
  {10.1093/mnras/204.1.151}, \href
  {https://ui.adsabs.harvard.edu/abs/1983MNRAS.204..151L} {204, 151}

\bibitem[\protect\citeauthoryear{{Li} et~al.,}{{Li}
  et~al.}{2018}]{2018IMMag..19..112L}
{Li} D.,  et~al., 2018, \mn@doi [IEEE Microwave Magazine]
  {10.1109/MMM.2018.2802178}, \href
  {https://ui.adsabs.harvard.edu/abs/2018IMMag..19..112L} {19, 112}

\bibitem[\protect\citeauthoryear{{Li} et~al.,}{{Li}
  et~al.}{2021}]{2021RAA....21...59L}
{Li} J.-X.,  et~al., 2021, \mn@doi [Research in Astronomy and Astrophysics]
  {10.1088/1674-4527/21/3/059}, \href
  {https://ui.adsabs.harvard.edu/abs/2021RAA....21...59L} {21, 059}

\bibitem[\protect\citeauthoryear{{Lin}, {Huang}  \& {Chen}}{{Lin}
  et~al.}{2018}]{2018AJ....155..188L}
{Lin} Y.-T.,  {Huang} H.-J.,   {Chen} Y.-C.,  2018, \mn@doi [\aj]
  {10.3847/1538-3881/aab5b4}, \href
  {https://ui.adsabs.harvard.edu/abs/2018AJ....155..188L} {155, 188}

\bibitem[\protect\citeauthoryear{{Maccagni}, {Morganti}, {Oosterloo}  \&
  {Mahony}}{{Maccagni} et~al.}{2014}]{2014A&A...571A..67M}
{Maccagni} F.~M.,  {Morganti} R.,  {Oosterloo} T.~A.,   {Mahony} E.~K.,  2014,
  \mn@doi [\aap] {10.1051/0004-6361/201424334}, \href
  {https://ui.adsabs.harvard.edu/abs/2014A&A...571A..67M} {571, A67}

\bibitem[\protect\citeauthoryear{{Maccagni}, {Morganti}, {Oosterloo},
  {Ger{\'e}b}  \& {Maddox}}{{Maccagni} et~al.}{2017}]{2017A&A...604A..43M}
{Maccagni} F.~M.,  {Morganti} R.,  {Oosterloo} T.~A.,  {Ger{\'e}b} K.,
  {Maddox} N.,  2017, \mn@doi [\aap] {10.1051/0004-6361/201730563}, \href
  {https://ui.adsabs.harvard.edu/abs/2017A&A...604A..43M} {604, A43}

\bibitem[\protect\citeauthoryear{{Mahony}, {Morganti}, {Emonts}, {Oosterloo}
  \& {Tadhunter}}{{Mahony} et~al.}{2013}]{2013MNRAS.435L..58M}
{Mahony} E.~K.,  {Morganti} R.,  {Emonts} B.~H.~C.,  {Oosterloo} T.~A.,
  {Tadhunter} C.,  2013, \mn@doi [\mnras] {10.1093/mnrasl/slt094}, \href
  {https://ui.adsabs.harvard.edu/abs/2013MNRAS.435L..58M} {435, L58}

\bibitem[\protect\citeauthoryear{{Mahony} et~al.,}{{Mahony}
  et~al.}{2022}]{2022MNRAS.509.1690M}
{Mahony} E.~K.,  et~al., 2022, \mn@doi [\mnras] {10.1093/mnras/stab3041}, \href
  {https://ui.adsabs.harvard.edu/abs/2022MNRAS.509.1690M} {509, 1690}

\bibitem[\protect\citeauthoryear{{Maina} et~al.,}{{Maina}
  et~al.}{2022}]{2022MNRAS.516.2050M}
{Maina} E.~K.,  et~al., 2022, \mn@doi [\mnras] {10.1093/mnras/stac1752}, \href
  {https://ui.adsabs.harvard.edu/abs/2022MNRAS.516.2050M} {516, 2050}

\bibitem[\protect\citeauthoryear{{Meyer} et~al.,}{{Meyer}
  et~al.}{2004}]{2004MNRAS.350.1195M}
{Meyer} M.~J.,  et~al., 2004, \mn@doi [\mnras]
  {10.1111/j.1365-2966.2004.07710.x}, \href
  {http://adsabs.harvard.edu/abs/2004MNRAS.350.1195M} {350, 1195}

\bibitem[\protect\citeauthoryear{{Morganti} \& {Oosterloo}}{{Morganti} \&
  {Oosterloo}}{2018}]{2018A&ARv..26....4M}
{Morganti} R.,  {Oosterloo} T.,  2018, \mn@doi [\aapr]
  {10.1007/s00159-018-0109-x}, \href
  {https://ui.adsabs.harvard.edu/abs/2018A&ARv..26....4M} {26, 4}

\bibitem[\protect\citeauthoryear{{Morganti}, {Oosterloo}, {Emonts}, {van der
  Hulst}  \& {Tadhunter}}{{Morganti} et~al.}{2003}]{2003ApJ...593L..69M}
{Morganti} R.,  {Oosterloo} T.~A.,  {Emonts} B.~H.~C.,  {van der Hulst} J.~M.,
   {Tadhunter} C.~N.,  2003, \mn@doi [\apjl] {10.1086/378219}, \href
  {https://ui.adsabs.harvard.edu/abs/2003ApJ...593L..69M} {593, L69}

\bibitem[\protect\citeauthoryear{{Murray}, {Stanimirovi{\'c}}, {Goss},
  {Heiles}, {Dickey}, {Babler}  \& {Kim}}{{Murray}
  et~al.}{2018}]{2018ApJS..238...14M}
{Murray} C.~E.,  {Stanimirovi{\'c}} S.,  {Goss} W.~M.,  {Heiles} C.,  {Dickey}
  J.~M.,  {Babler} B.,   {Kim} C.-G.,  2018, \mn@doi [\apjs]
  {10.3847/1538-4365/aad81a}, \href
  {https://ui.adsabs.harvard.edu/abs/2018ApJS..238...14M} {238, 14}

\bibitem[\protect\citeauthoryear{{Nan} et~al.,}{{Nan}
  et~al.}{2011}]{2011IJMPD..20..989N}
{Nan} R.,  et~al., 2011, \mn@doi [International Journal of Modern Physics D]
  {10.1142/S0218271811019335}, \href
  {http://adsabs.harvard.edu/abs/2011IJMPD..20..989N} {20, 989}

\bibitem[\protect\citeauthoryear{{Nolta} et~al.,}{{Nolta}
  et~al.}{2004}]{2004ApJ...608...10N}
{Nolta} M.~R.,  et~al., 2004, \mn@doi [\apj] {10.1086/386536}, \href
  {https://ui.adsabs.harvard.edu/abs/2004ApJ...608...10N} {608, 10}

\bibitem[\protect\citeauthoryear{{Noterdaeme}, {Petitjean}, {Ledoux}  \&
  {Srianand}}{{Noterdaeme} et~al.}{2009}]{2009A&A...505.1087N}
{Noterdaeme} P.,  {Petitjean} P.,  {Ledoux} C.,   {Srianand} R.,  2009, \mn@doi
  [\aap] {10.1051/0004-6361/200912768}, \href
  {https://ui.adsabs.harvard.edu/abs/2009A&A...505.1087N} {505, 1087}

\bibitem[\protect\citeauthoryear{{Parisi} et~al.,}{{Parisi}
  et~al.}{2014}]{2014A&A...561A..67P}
{Parisi} P.,  et~al., 2014, \mn@doi [\aap] {10.1051/0004-6361/201322409}, \href
  {https://ui.adsabs.harvard.edu/abs/2014A&A...561A..67P} {561, A67}

\bibitem[\protect\citeauthoryear{{Prochaska}, {Herbert-Fort}  \&
  {Wolfe}}{{Prochaska} et~al.}{2005}]{2005ApJ...635..123P}
{Prochaska} J.~X.,  {Herbert-Fort} S.,   {Wolfe} A.~M.,  2005, \mn@doi [\apj]
  {10.1086/497287}, \href
  {https://ui.adsabs.harvard.edu/abs/2005ApJ...635..123P} {635, 123}

\bibitem[\protect\citeauthoryear{{Rahman}, {M{\'e}nard}, {Scranton}, {Schmidt}
  \& {Morrison}}{{Rahman} et~al.}{2015}]{2015MNRAS.447.3500R}
{Rahman} M.,  {M{\'e}nard} B.,  {Scranton} R.,  {Schmidt} S.~J.,   {Morrison}
  C.~B.,  2015, \mn@doi [\mnras] {10.1093/mnras/stu2636}, \href
  {https://ui.adsabs.harvard.edu/abs/2015MNRAS.447.3500R} {447, 3500}

\bibitem[\protect\citeauthoryear{{Sadler} et~al.,}{{Sadler}
  et~al.}{2020}]{2020MNRAS.499.4293S}
{Sadler} E.~M.,  et~al., 2020, \mn@doi [\mnras] {10.1093/mnras/staa2390}, \href
  {https://ui.adsabs.harvard.edu/abs/2020MNRAS.499.4293S} {499, 4293}

\bibitem[\protect\citeauthoryear{{Saintonge}}{{Saintonge}}{2007}]{2007AJ....133.2087S}
{Saintonge} A.,  2007, \mn@doi [\aj] {10.1086/513515}, \href
  {https://ui.adsabs.harvard.edu/abs/2007AJ....133.2087S} {133, 2087}

\bibitem[\protect\citeauthoryear{{Schulz}, {Morganti}, {Nyland}, {Paragi},
  {Mahony}  \& {Oosterloo}}{{Schulz} et~al.}{2021}]{2021A&A...647A..63S}
{Schulz} R.,  {Morganti} R.,  {Nyland} K.,  {Paragi} Z.,  {Mahony} E.~K.,
  {Oosterloo} T.,  2021, \mn@doi [\aap] {10.1051/0004-6361/202037677}, \href
  {https://ui.adsabs.harvard.edu/abs/2021A&A...647A..63S} {647, A63}

\bibitem[\protect\citeauthoryear{{Skrutskie} et~al.,}{{Skrutskie}
  et~al.}{2006}]{2006AJ....131.1163S}
{Skrutskie} M.~F.,  et~al., 2006, \mn@doi [\aj] {10.1086/498708}, \href
  {https://ui.adsabs.harvard.edu/abs/2006AJ....131.1163S} {131, 1163}

\bibitem[\protect\citeauthoryear{Smith, Dunning, Smart, Shaw, Mackay, Bowen  \&
  Hayman}{Smith et~al.}{2017}]{8073111}
Smith S.~L.,  Dunning A.,  Smart K.~W.,  Shaw R.,  Mackay S.,  Bowen M.,
  Hayman D.,  2017, in 2017 IEEE International Symposium on Antennas and
  Propagation USNC/URSI National Radio Science Meeting. pp 2137--2138,
  \mn@doi{10.1109/APUSNCURSINRSM.2017.8073111}

\bibitem[\protect\citeauthoryear{{Stanimirovi{\'c}}, {Murray}, {Lee}, {Heiles}
  \& {Miller}}{{Stanimirovi{\'c}} et~al.}{2014}]{2014ApJ...793..132S}
{Stanimirovi{\'c}} S.,  {Murray} C.~E.,  {Lee} M.-Y.,  {Heiles} C.,   {Miller}
  J.,  2014, \mn@doi [\apj] {10.1088/0004-637X/793/2/132}, \href
  {https://ui.adsabs.harvard.edu/abs/2014ApJ...793..132S} {793, 132}

\bibitem[\protect\citeauthoryear{{Stern} et~al.,}{{Stern}
  et~al.}{2012}]{2012ApJ...753...30S}
{Stern} D.,  et~al., 2012, \mn@doi [\apj] {10.1088/0004-637X/753/1/30}, \href
  {https://ui.adsabs.harvard.edu/abs/2012ApJ...753...30S} {753, 30}

\bibitem[\protect\citeauthoryear{{Struve} \& {Conway}}{{Struve} \&
  {Conway}}{2010}]{2010A&A...513A..10S}
{Struve} C.,  {Conway} J.~E.,  2010, \mn@doi [\aap]
  {10.1051/0004-6361/200913572}, \href
  {https://ui.adsabs.harvard.edu/abs/2010A&A...513A..10S} {513, A10}

\bibitem[\protect\citeauthoryear{{Su} et~al.,}{{Su}
  et~al.}{2022}]{2022MNRAS.516.2947S}
{Su} R.,  et~al., 2022, \mn@doi [\mnras] {10.1093/mnras/stac2257}, \href
  {https://ui.adsabs.harvard.edu/abs/2022MNRAS.516.2947S} {516, 2947}

\bibitem[\protect\citeauthoryear{{Taylor}}{{Taylor}}{1996}]{1996ApJ...470..394T}
{Taylor} G.~B.,  1996, \mn@doi [\apj] {10.1086/177874}, \href
  {https://ui.adsabs.harvard.edu/abs/1996ApJ...470..394T} {470, 394}

\bibitem[\protect\citeauthoryear{{Tegmark}}{{Tegmark}}{1997}]{1997ApJ...480L..87T}
{Tegmark} M.,  1997, \mn@doi [\apjl] {10.1086/310631}, \href
  {https://ui.adsabs.harvard.edu/abs/1997ApJ...480L..87T} {480, L87}

\bibitem[\protect\citeauthoryear{Van~Rossum \& Drake~Jr}{Van~Rossum \&
  Drake~Jr}{1995}]{van1995python}
Van~Rossum G.,  Drake~Jr F.~L.,  1995, Python reference manual.
Centrum voor Wiskunde en Informatica Amsterdam

\bibitem[\protect\citeauthoryear{Verheijen, van Gorkom, Szomoru, Dwarakanath,
  Poggianti  \& Schiminovich}{Verheijen et~al.}{2007}]{1538-4357-668-1-L9}
Verheijen M.,  van Gorkom J.~H.,  Szomoru A.,  Dwarakanath K.~S.,  Poggianti
  B.~M.,   Schiminovich D.,  2007, The Astrophysical Journal Letters, 668, L9

\bibitem[\protect\citeauthoryear{Virtanen et~al.,}{Virtanen
  et~al.}{2020}]{2020SciPy-NMeth}
Virtanen P.,  et~al., 2020, \mn@doi [Nature Methods]
  {10.1038/s41592-019-0686-2}, \href {https://rdcu.be/b08Wh} {17, 261}

\bibitem[\protect\citeauthoryear{{Vreeswijk, P. M.} et~al.,}{{Vreeswijk, P. M.}
  et~al.}{2004}]{vreeswijk}
{Vreeswijk, P. M.} et~al., 2004, \mn@doi [A\&A] {10.1051/0004-6361:20040086},
  419, 927

\bibitem[\protect\citeauthoryear{{Wolfe}, {Gawiser}  \& {Prochaska}}{{Wolfe}
  et~al.}{2005}]{2005ARA&A..43..861W}
{Wolfe} A.~M.,  {Gawiser} E.,   {Prochaska} J.~X.,  2005, \mn@doi [\araa]
  {10.1146/annurev.astro.42.053102.133950}, \href
  {https://ui.adsabs.harvard.edu/abs/2005ARA&A..43..861W} {43, 861}

\bibitem[\protect\citeauthoryear{{Wright} et~al.,}{{Wright}
  et~al.}{2010}]{2010AJ....140.1868W}
{Wright} E.~L.,  et~al., 2010, \mn@doi [\aj] {10.1088/0004-6256/140/6/1868},
  \href {https://ui.adsabs.harvard.edu/abs/2010AJ....140.1868W} {140, 1868}

\bibitem[\protect\citeauthoryear{{Yu}, {Pen}, {Zhang}, {Li}  \& {Chen}}{{Yu}
  et~al.}{2017}]{2017RAA....17...49Y}
{Yu} H.-R.,  {Pen} U.-L.,  {Zhang} T.-J.,  {Li} D.,   {Chen} X.,  2017, \mn@doi
  [Research in Astronomy and Astrophysics] {10.1088/1674-4527/17/6/49}, \href
  {https://ui.adsabs.harvard.edu/abs/2017RAA....17...49Y} {17, 049}

\bibitem[\protect\citeauthoryear{{Zhang}, {Zhu}, {Wu}, {Yu}, {Jiang}, {Yue},
  {Huang}  \& {Hao}}{{Zhang} et~al.}{2021}]{2021MNRAS.503.5385Z}
{Zhang} B.,  {Zhu} M.,  {Wu} Z.-Z.,  {Yu} Q.-Z.,  {Jiang} P.,  {Yue} Y.-L.,
  {Huang} M.-L.,   {Hao} Q.-L.,  2021, \mn@doi [\mnras]
  {10.1093/mnras/stab754}, \href
  {https://ui.adsabs.harvard.edu/abs/2021MNRAS.503.5385Z} {503, 5385}

\bibitem[\protect\citeauthoryear{{Zheng} et~al.,}{{Zheng}
  et~al.}{2017}]{2017MNRAS.464.3486Z}
{Zheng} H.,  et~al., 2017, \mn@doi [\mnras] {10.1093/mnras/stw2525}, \href
  {https://ui.adsabs.harvard.edu/abs/2017MNRAS.464.3486Z} {464, 3486}

\bibitem[\protect\citeauthoryear{{Zheng}, {Li}, {Sadler}, {Allison}  \&
  {Tang}}{{Zheng} et~al.}{2020}]{2020MNRAS.499.3085Z}
{Zheng} Z.,  {Li} D.,  {Sadler} E.~M.,  {Allison} J.~R.,   {Tang} N.,  2020,
  \mn@doi [\mnras] {10.1093/mnras/staa3033}, \href
  {https://ui.adsabs.harvard.edu/abs/2020MNRAS.499.3085Z} {499, 3085}

\bibitem[\protect\citeauthoryear{{Zwaan}, {van Dokkum}  \& {Verheijen}}{{Zwaan}
  et~al.}{2001}]{2001Sci...293.1800Z}
{Zwaan} M.~A.,  {van Dokkum} P.~G.,   {Verheijen} M.~A.~W.,  2001, \mn@doi
  [Science] {10.1126/science.1063034}, \href
  {http://adsabs.harvard.edu/abs/2001Sci...293.1800Z} {293, 1800}

\bibitem[\protect\citeauthoryear{{Zwaan} et~al.,}{{Zwaan}
  et~al.}{2004}]{2004MNRAS.350.1210Z}
{Zwaan} M.~A.,  et~al., 2004, \mn@doi [\mnras]
  {10.1111/j.1365-2966.2004.07782.x}, \href
  {http://adsabs.harvard.edu/abs/2004MNRAS.350.1210Z} {350, 1210}

\bibitem[\protect\citeauthoryear{{Zwaan}, {van der Hulst}, {Briggs},
  {Verheijen}  \& {Ryan-Weber}}{{Zwaan} et~al.}{2005}]{2005MNRAS.364.1467Z}
{Zwaan} M.~A.,  {van der Hulst} J.~M.,  {Briggs} F.~H.,  {Verheijen} M.~A.~W.,
   {Ryan-Weber} E.~V.,  2005, \mn@doi [\mnras]
  {10.1111/j.1365-2966.2005.09698.x}, \href
  {https://ui.adsabs.harvard.edu/abs/2005MNRAS.364.1467Z} {364, 1467}

\bibitem[\protect\citeauthoryear{{de Oliveira-Costa}, {Tegmark}, {Gaensler},
  {Jonas}, {Landecker}  \& {Reich}}{{de Oliveira-Costa}
  et~al.}{2008}]{2008MNRAS.388..247D}
{de Oliveira-Costa} A.,  {Tegmark} M.,  {Gaensler} B.~M.,  {Jonas} J.,
  {Landecker} T.~L.,   {Reich} P.,  2008, \mn@doi [\mnras]
  {10.1111/j.1365-2966.2008.13376.x}, \href
  {https://ui.adsabs.harvard.edu/abs/2008MNRAS.388..247D} {388, 247}

\bibitem[\protect\citeauthoryear{{de Zotti}, {Massardi}, {Negrello}  \&
  {Wall}}{{de Zotti} et~al.}{2010}]{2010A&ARv..18....1D}
{de Zotti} G.,  {Massardi} M.,  {Negrello} M.,   {Wall} J.,  2010, \mn@doi
  [\aapr] {10.1007/s00159-009-0026-0}, \href
  {https://ui.adsabs.harvard.edu/abs/2010A&ARv..18....1D} {18, 1}

\makeatother
\end{thebibliography}

\end{document}